\documentclass[onecolumn,a4paper,fleqn,usenatbib]{mnras}

\usepackage{newtxtext,newtxmath}
\usepackage[T1]{fontenc}
\usepackage{ae,aecompl}

\usepackage{amsmath}
\usepackage{mathtools}
\usepackage{graphicx}
\usepackage{float}
\usepackage{amsfonts}
\usepackage{amssymb}
\usepackage{makeidx}
\usepackage{mathrsfs}
\usepackage{subcaption}
\usepackage{natbib}
\usepackage{multirow}

\newcommand{\rs}{r_{_{\rm S}}}
\newcommand{\agsq}{a_{\rm th}^2}
\newcommand{\arsq}{a_{\rm rel}^2}
\newcommand{\agcsq}{a_{\rm th,c}^2}
\newcommand{\arcsq}{a_{\rm rel,c}^2}

\newcommand{\rc}{r_c}
\newcommand{\vcsq}{\upsilon_c^2}
\newcommand{\OmegaK}{\Omega_{\rm K}}

\newcommand{\msun}{M_\odot}

\newcommand\greens{f_{\rm G}}

\newcommand{\sgr}{Sgr~A*~}
\newcommand{\vel}{\upsilon}

\title[A Two-Fluid Model for Black-Hole Accretion Flows]{A Two-Fluid Model for Black-Hole Accretion Flows: Particle Acceleration and Disc Structure}

\author[J. Lee and P. Becker]{Jason P. Lee,$^{1}$\thanks{Email: je@gmu.edu} Peter A. Becker,$^{1}$\thanks{Email: pbecker@gmu.edu}
\\
$^{1}$Department of Physics \& Astronomy,
George Mason University, Fairfax, VA 22030-4444, USA
}

\date{Accepted . Received ; in original form }

\pubyear{2016}

\begin{document}
\label{firstpage}
\pagerange{\pageref{firstpage}--\pageref{lastpage}}
\maketitle

\begin{abstract}
Hot, tenuous advection-dominated accretion flows around black holes are ideal sites for the Fermi acceleration of relativistic particles at standing shock waves in the accretion disc. Previous work has demonstrated that the shock-acceleration process can be efficient enough to power the observed, strong outflows in radio-loud active galaxies such as M87. However, the dynamical effect (back-reaction) on the flow, exerted by the pressure of the relativistic particles, has not been previously considered, and this effect can have a significant influence on the disc structure. We reexamine the problem by developing a new, two-fluid model for the structure of the accretion disc that includes the dynamical effect of the relativistic particle pressure, combined with the pressure of the background (thermal) gas. The new model is analogous to the two-fluid model of cosmic-ray acceleration in supernova-driven shock waves. As part of the model, we also develop a new set of shock jump conditions, which are solved along with the hydrodynamic conservation equations to determine the structure of the accretion disc. The solutions include the formation of a mildly relativistic outflow (jet) at the shock radius, driven by the relativistic particles accelerated in the disc. One of our main conclusions is that in the context of the new two-fluid accretion model, global smooth (shock-free) solutions do not exist, and the disc must always contain a standing shock wave, at least in the inviscid case considered here.
\end{abstract}




\section{INTRODUCTION}

The two-temperature advection-dominated accretion flow (ADAF) model seems to provide a good description of the physics occurring around radio-loud accreting supermassive black holes such as M87 and Sgr~A*. In these discs, radiative cooling is inefficient, and the ions and electrons reach nearly virial temperatures. These radio-loud sources are often associated with strong relativistic outflows (jets). On the other hand, jets are not usually detected in X-ray luminous sources, which are well described by the thin-disc model (Shakura \& Sunyaev 1973; Shapiro, Lightman, \& Eardley 1976, hereafter SLE; Blandford \& Begelman 1999). The low radiative efficiency in the underfed black holes, with accretion rates far below the Eddington value, stems from the fact that free-free emission is a two-body process, which creates a nonlinear dependence on the density. Conversely, free-free cooling is relatively efficient in the X-ray luminous sources because they accrete closer to the Eddington rate. The combination of high temperature and low density in the gas around underfed black holes leads to the development of a vertically extended disc composed of collisionless plasma.

A number of previous studies have demonstrated that both viscid and inviscid accretion discs can display either shocked or shock-free (i.e., globally smooth) solutions (e.g., Chakrabarti 1989; Chakrabarti \& Molten 1993; Lu \& Yuan 1997; Das et al. 2001; Le \& Becker 2005, hereafter LB05; Becker et al. 2011, hereafter B11; Das et al. 2009, hereafter D09; Chattopadhyay \& Kumar 2016). Furthermore, it has been established that the acceleration of particles at a standing shock in the disc can be sufficient to power the observed strong outflows in radio-loud active galaxies containing supermassive black holes, such as M87, and also in the Galactic center source \sgr (LB05; B11). While the second law of thermodynamics would tend to prefer shocks due to the fact that they increase the entropy of the system, relative to a shock-free solution, this fact alone does not guarantee that shocks will occur. A number of fully relativistic simulations have suggested that shocks do occur in hot tenuous discs (e.g., Hawley et al. 1984a, 1984b; Chattopadhyay \& Kumar 2016), but despite this, the existence of shocks in discs is not a settled matter, with some models including shocks and some not. The conclusions regarding shocks have for the most part been reached in the context of rather idealized models that do not include the dynamical effect of the pressure contributed by the relativistic particles accelerated in the disc. Hence, one of our goals in this paper is to reexamine the question of whether a shock is a necessary component in an ADAF disc around a supermassive black hole when one includes the dynamical effect of the relativistic particle pressure.

\begin{figure}
\begin{subfigure}[b]{0.47\textwidth}
\centering
\includegraphics[width=1\columnwidth]{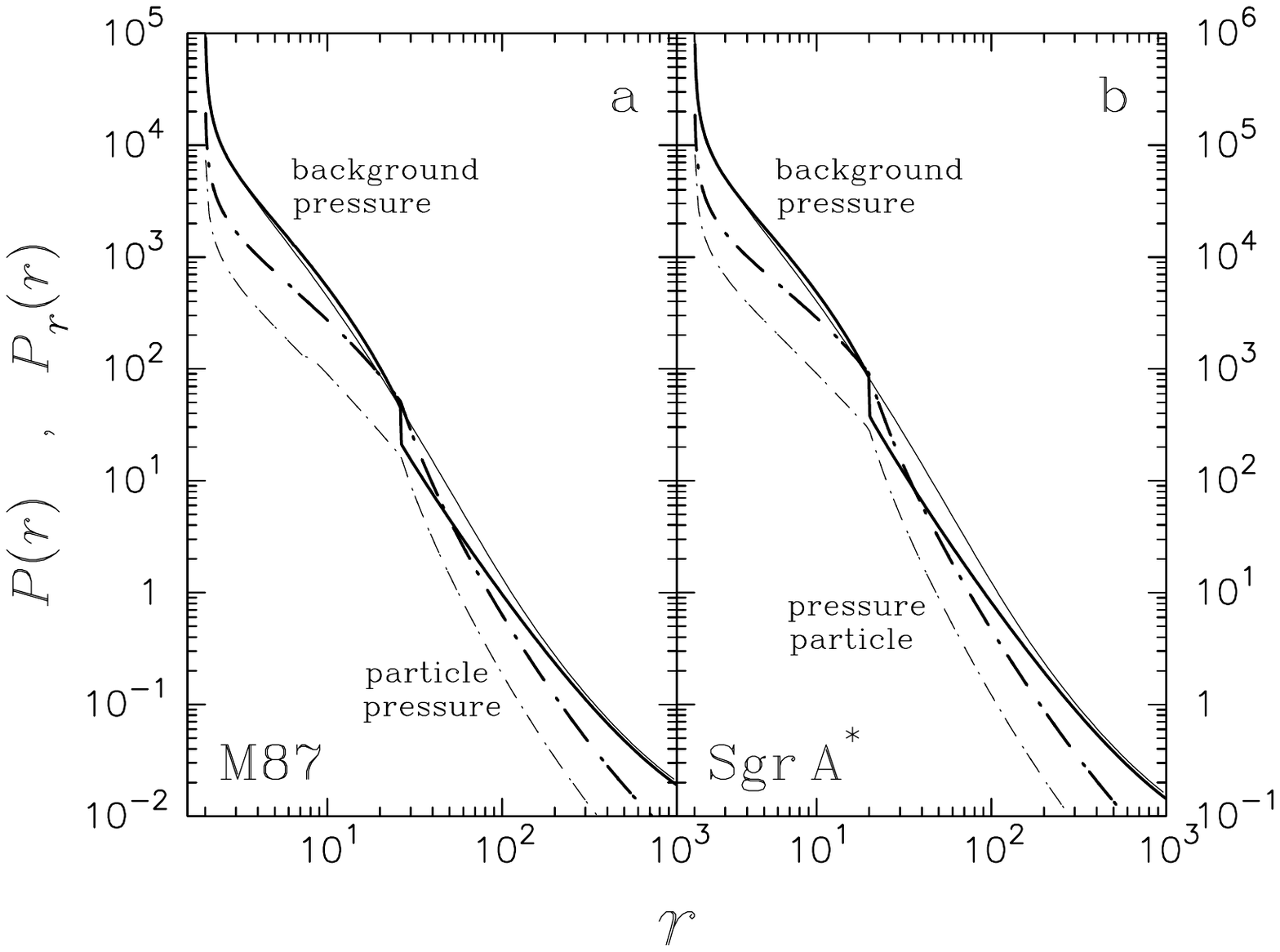}
\end{subfigure}
\qquad
\begin{subfigure}[b]{0.483\textwidth}
\hspace*{-2cm}\vspace{0.28cm}\includegraphics[width=0.92\columnwidth]{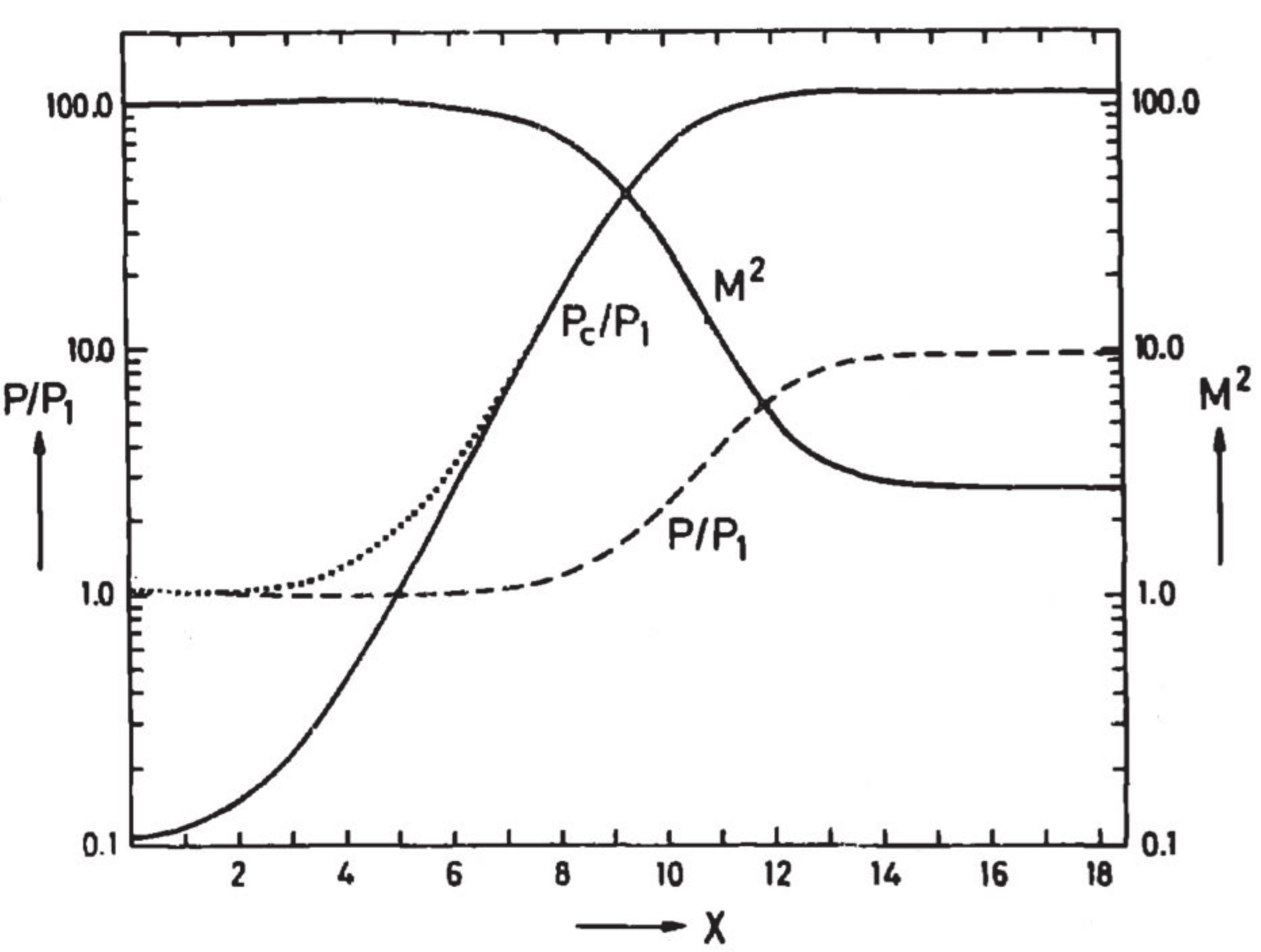}
\centering
\end{subfigure}
\caption{In the left-hand panel (taken from B11), it is seen that the relativistic particle pressure $P_{\rm rel}$ (dot-dashed lines) can exceed the thermal (background) pressure $P$ (solid lines) in the vicinity of the shock in a black hole accretion disc. In the right-hand panel (taken from Axford et al. 1977), it is likewise seen that the cosmic-ray pressure $P_c$ can exceed the thermal gas pressure $P$ in a strong plane-parallel shock driven by a supernova blast wave.}
\label{fig:fig1}
\end{figure}

The ADAF discs considered here are closely related to the viscous convection-dominated disc (CDAFs) with sub-Eddington accretion rates discussed by Quataert \& Gruzinov (2000), Narayan et. al (2000), Igumenshchev \& Abramowicz (2000). In CDAF discs accreting at a sub-Eddington rate, with relatively low viscosity parameter, $\alpha \lesssim 0.1$, convection accomplishes the outward transport of angular momentum, which facilitates the inward accretion of the matter. However, for $\alpha \gtrsim 0.1$, the energy transport is dominated by advection, and convection is unimportant for determining the disc structure. Since the viscosity parameter in real astrophysical accretion discs is likely to be significant, with $\alpha \gtrsim 0.1$, it follows that ADAF discs probably occur more frequently in nature than CDAF discs, at least in sources accreting at a sub-Eddington rate (Yuan \& Narayan 2014). Another possible disc structure is the magnetically-arrested (MAD) disc, discussed by Igumenshchev et al. (2003), and Igumenshchev (2008). In these discs, the vertical poloidal component of the magnetic field is strong enough to suppress accretion. Such structures are expected to form around rapidly-rotating black holes that are able to efficiently advect magnetized plasma, while avoiding reconnection, but it is not clear whether the advection efficiency is high enough in real discs to establish the MAD structure (Yuan \& Narayan 2014). Our focus here is on sources with relatively low accretion rates, such as M87 and \sgr, and therefore we will focus on the standard ADAF scenario, with details reviewed below.

\subsection{ADAF disc accretion}
\label{secADAFdiscaccretion}

The original self-similar ADAF model was originally introduced by Ichimaru (1977), and later standardized by Narayan \& Yi (1994, 1995a, 1995b), Abramowicz et al. (1995), Chen (1995), and Chen et al. (1995). In ADAF discs, the gas density is relatively low, the disc is optically thin to absorption, and the accretion rate $\dot M \ll\dot M_{\rm E}$, where $\dot M_{\rm E}$ is the Eddington accretion rate, which is related to the Eddington luminosity $L_{\rm E}$ and the radiative efficiency parameter $\beta \lesssim 0.1$ via
\begin{equation}
\dot M_{\rm E} \equiv c^{-2} \beta^{-1} L_{\rm E} \ .
\label{eqEddM}
\end{equation}
For pure, fully ionized hydrogen, the Eddington luminosity is given by
\begin{equation}
L_{\rm E} \equiv \frac{4\pi GMm_pc}{\sigma_{_{\rm T}}}
= 1.25 \times10^{38} \left(\frac{M}{M_\odot}\right)\,{\rm ergs \ s}^{-1} \ ,
\label{eqEddL}
\end{equation}
where $\sigma_{_{\rm T}}$, $M$, $m_p$, and $c$ denote the Thomson cross section, the black hole mass, the proton mass and the speed of light, respectively. The ADAF scenario is qualitatively similar to the accretion model developed by SLE, which likewise contains a two-temperature disc with the ion temperature greatly exceeding the electron temperature.

The earliest ADAF models utilized the standard Newtonian form for the gravitational potential, which is not applicable near the event horizon. The technical difficulties associated with fully implementing general relativity led to the development of the pseudo-Newtonian approximation, given by (Paczy\'nski \& Wiita 1980)
\begin{equation}
\Phi\left(r\right)\equiv \frac{-GM}{r-\rs} \ ,
\label{eqGravPot}
\end{equation}
where $\rs \equiv 2GM/c^2$ is the Schwarzschild radius for a black hole of mass $M$. This is a surprisingly accurate approximation that provides a convenient method for exploring the structure of the inner region of a sub-Keplerian disc. By adopting the pseudo-Newtonian, we are able to the treat the physical processes occurring within the accretion disc using a semiclassical methodology. Narayan et al. (1997) and Becker \& Subramanian (2005), amongst other authors, used this approach in developing their models for ADAF discs. The dynamical solutions obtained successfully describe the global structure of the accretion flow.

When a shock is present, interactions with magnetohydrodynamical (MHD) results in the first-order Fermi acceleration of charged particles. This process will either heat the gas, or, alternatively, it will lead the formation of a nonthermal distribution of relativistic particles. The distinction between these two possibilities depends on the relation between the disc half-thickness $H$ and the value of the mean free path for ion-ion Coulomb collisions, $\lambda_{ii}$, given by (e.g., Subramanian et al. 1996)
\begin{equation}
\lambda_{ii} = 1.8 \times 10^5\frac{T^2_i}{n_i\ln\Lambda} \ ,
\label{eqIonIon}
\end{equation}
where $n_i$ is the thermal ion number density and $\ln\Lambda$ is the Coulomb logarithm. For the parameter values typical of ADAF discs, we find that $\lambda_{ii} \gtrsim H$, and consequently the plasma is essentially collisionless. In this case, the energy gained by the particles as they cross the shock is not thermalized, and instead, a nonthermal relativistic particle distribution is generated. Although the particles do not interact with each other directly via Coulomb collisions, the two populations are still coupled through interactions with the MHD waves, which mediate the shock. In a series of previous investigations, it has been established that the particles accelerated in shocked ADAF discs can escape to power the outflows observed from radio-loud supermassive black holes (Le \& Becker 2004, hereafter LB04; LB05, Le \& Becker 2007; D09). These previous investigations have focused on the utilization of the pseudo-Newtonian approximation to general relativity in order to obtain semi-analytical results. However, in recent work, Chattopadhyay \& Kumar (2016) have also obtained shocked-disc solutions within the context of a fully relativistic simulation.

\subsection{Two-fluid model for cosmic-ray shocks}

The majority of the cosmic rays observed in our galaxy are thought to be accelerated by shock waves driven by supernova explosions (Axford et al. 1977). The exception is the population of ultra-high energy cosmic rays, whose origin is still not well understood, and which are probably created outside our galaxy. In the supernova-driven shock acceleration model, energetic charged particles scatter elastically with magnetic irregularities (MHD waves) convected with the background gas (Drury \& V\"olk 1981; Becker \& Kazanas 2001). The convergence of the MHD waves at the shock, combined with the effect of spatial diffusion, allows the cosmic rays to cross the shock multiple times, gaining energy continuously. Repeated shock crossings result in the characteristic power-law energy spectrum associated with first-order Fermi acceleration.

The early ``one-fluid'' models for the acceleration of cosmic rays in supernova-driven shocks neglected the dynamical effect of the particle pressure, so that the acceleration of the cosmic rays was treated using the ``test-particle'' approximation (e.g., Blandford \& Ostriker 1978). However, it was soon realized that the resulting cosmic ray pressure could exceed the pressure of the background thermal gas, and therefore the particle pressure should be included when analyzing the dynamical structure of the shock. In the next generation of models, this problem was remedied by treating the nonlinear coupling of the gas dynamics and the energization of the cosmic rays in a self-consistent manner. The resulting ``two-fluid'' model for diffusive shock acceleration has become an accepted paradigm for studying the self-consistent cosmic-ray shock acceleration problem (e.g., Drury \& V\"olk 1981). Although the divergence of the cosmic-ray pressure is removed in the two-fluid model, the cosmic-ray pressure is still comparable to the gas (thermal) pressure in the vicinity of the shock, as can be seen in the right panel of Fig. \ref{fig:fig1}, taken from Axford et al. (1977).

The two-fluid cosmic ray shock acceleration model can include both globally smooth solutions as well as solutions that contain discontinuous, gas-mediated sub-shocks (Ko, Chan, \& Webb 1997; Zank, Webb, \& Donohue 1993). In the case of a discontinuous shock, one observes a smooth deceleration precursor in the fluid just upstream from the shock. This precursor phenomenon is not observed in the classical case (in which the particle pressure is neglected), and is therefore a unique feature of the two-fluid shock model. We anticipate that this type of behaviour will also be observed in the analogous two-fluid model for disc accretion, once the pressure of the accelerated particles is included in the hydrodynamic equations.

\subsection{This paper}

The pressure of the accelerated particles was neglected in the early one-fluid studies of particle acceleration in shocked ADAF discs (LB04, LB05; D09). However, in a manner similar to the initial developments in the field of cosmic-ray acceleration, it was shown that in the one-fluid ADAF model, the pressure of the accelerated relativistic particles can actually exceed the pressure of the thermal background gas in the vicinity of the shock (see the left panel of Fig. \ref{fig:fig1}, taken from B11). This suggests the need to develop a new two-fluid model for the structure of ADAF discs that properly accounts for the dynamical effect of the relativistic particle pressure. Once the particle pressure is included in the disc model, we expect to see the appearance of a smooth deceleration precursor on the upstream side of the discontinuous shock, in analogy with the two-fluid model for cosmic-ray shock acceleration.

In this first study of the effect of particle pressure on the disc structure, we will focus on inviscid flows, deferring the study of vicious discs to later work. Our goal is to create a new, self-consistent, two-fluid disc accretion model that includes the dynamical effect of the pressure associated with both the relativistic particles and the background (thermal) gas. This represents a generalization of the work of LB04 and LB05. In this first paper in the series, we develop the formalism for computing the hydrodynamical structure of the disc, and we also derive a modified set of shock jump conditions that accounts for the dynamical effect of the relativistic particle pressure. The relativistic particle number and energy densities will be determined self-consistently along with the structure of the disc. Also, we shall explore the question of whether this new self-consistent ADAF disc model can admit both shocked and shock-free (i.e., globally smooth) solutions.

The organization of the paper is as follows. In Section \ref{sec:transflow} we discuss the general structure of the disc/shock model, and in Section \ref{sec:critpoint} we derive the associated critical conditions. The isothermal shock jump conditions are derived in Section \ref{sec:isoshock}, and in Section \ref{sec:asymp} we derive the asymptotic behaviours of the physical quantities at both large and small radii. In Section~\ref{sec:relmom} we review the transport formalism used to compute the relativistic number and energy densities, and in Section~\ref{sec:astrapp} we present detailed astrophysical applications using parameters appropriate for modeling the discs/outflows in M87 and \sgr. In Section \ref{sec:conclusion} we summarize our main conclusions and discuss the implications of our results.

\section{TRANSONIC FLOW STRUCTURE}
\label{sec:transflow}

\begin{figure}
\includegraphics[width=4in]{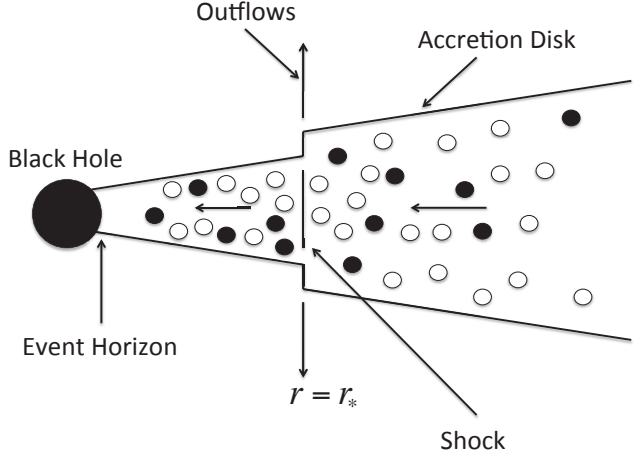}
\caption{Schematic diagram of our disc/shock/outflow model, including the relativistic particles (filled circles) injected at the standing shock location, and the MHD scattering centers (open circles) moving with the background gas through the disc. The compression of the scattering centers at the standing shock leads to efficient particle acceleration, which is analogous to the acceleration of cosmic rays in expanding, supernova-driven shock waves. The decrease in the disc thickness at the shock radius due to the loss of relativistic particle energy into the outflow is exaggerated here.}
\label{fig:fig2}
\end{figure}

The ambient gas fed into the outer region of an accretion flow onto a black hole is usually expected to be moving with a subsonic radial velocity. On the other hand, in the frame of reference of a stationary observer located just outside the event horizon, the radial inflow velocity approaches the speed of light, which exceeds any conceivable physical sound speed (Weinberg 1972). Taken together, these two facts imply that black hole accretion is a transonic phenomenon (Chakrabarti 1996). In the model considered here (depicted schematically in Fig. \ref{fig:fig2}), the gas is accelerated gravitationally toward the central mass, passes through an outer sonic point (where the radial velocity becomes supersonic), and then experiences a shock transition due to an obstruction near the event horizon, which is a consequence of the ``centrifugal barrier'' located between the inner and outer sonic points. The formation of the shock is described in simulations performed by Hawley et al. (1984a, 1984b) and Chattopadhyay \& Kumar (2016). Relativistic particles accelerated at the shock are transported throughout the disc until they either (1) escape via diffusion through the disc surface (forming the outflow from the upper/lower edges of the cylindrical standing shock), (2) advect with the flow through the event horizon, or (3) diffuse radially outward through the disc (see Section 3 of LB05 for further detail). In our model, it is assumed that the accelerated relativistic particles escape from the disc exclusively at the shock location. The basis for this assumption is twofold. First, we expect that the relativistic particle acceleration process will be concentrated at the shock. Second, the enhanced pressure of the relativistic particles in the vicinity of the shock can exceed the magnetic pressure, causing the magnetic field lines to transition to an open topology, giving rise to an outflow in a manner somewhat analogous to the formation of coronal holes in the solar atmosphere.

We employ the standard set of physical conservation equations discussed by Chakrabarti (1989) and Abramowicz \& Chakrabarti (1990) describing a vertically-averaged, one-dimensional, steady-state accretion disc that incorporates the effects of general relativity using the pseudo-Newtonian approximation for the gravitational potential (see equation~\ref{eqGravPot}). However, the conservation equations used by these authors will be generalized to include the effect of the relativistic particle pressure.

\subsection{Transport rates}

There are three conserved transport rates in viscous ADAF discs: the mass transport rate $\dot M$, the angular momentum transport rate $\dot J$, and the energy transport rate $\dot E$, which are all defined to be positive for inflow. The mass transport rate is given by
\begin{equation}
\dot M = 4\pi rH\rho \vel \ ,
\label{eqTranM}
\end{equation}
and the angular momentum transport rate is given by
\begin{equation}
\dot J = \dot M r^2\Omega-\cal{G} \ .
\label{eqTranJ}
\end{equation}
where $\rho$ is the mass density, $\vel$ is the radial velocity (defined to be positive for inflow), $\Omega$ is the angular velocity, ${\cal G}$ is the torque, and $H$ is the disc half-thickness.

The main goal of this paper is to explore the effect of the pressure of the relativistic particles accelerated in the disc on the dynamical structure of the accretion flow. Consequently, we need to implement a two-fluid model within the context of an ADAF accretion disc. In the two-fluid scenario adopted here, one component corresponds to the thermal gas, and the other to the relativistic particle population. Since the plasma in ADAF discs is collisionless, as demonstrated in equation~(\ref{eqIonIon}), it follows that there is no direct coupling between the relativistic and thermal ions via particle-particle collisions, and instead, they are indirectly coupled via collisions with MHD waves. Thus, the two components of the particle distribution are essentially independent, and each contributes separately to the total pressure. In our development of the required two-fluid model, we will follow the work of Becker \& Kazanas (2001) on the acceleration of cosmic rays at supernova-driven shock waves. To the best of our knowledge, this is the first time that such a two-fluid formalism has been applied in the context of an ADAF disc.

In the two-fluid approximation, the total radial energy transport rate, $\dot E$, is expressed using the linear combination
\begin{equation}
\dot E = \dot E_{\rm th} + \dot E_{\rm rel} \ ,
\label{eqEdot}
\end{equation}
where $\dot E_{\rm th}$ and $\dot E_{\rm rel}$ are the individual energy transport rates for the gas and the relativistic particles, respectively, given by
\begin{equation}
\dot E_{\rm th} = -{\cal G}\Omega+\dot M\left(\frac{1}{2}\vel^2_\phi + \frac{1}{2}\vel^2
+ \frac{P_{\rm th}+U_{\rm th}}{\rho} + \Phi\right) \ ,
\label{eqEdotG}
\end{equation}
and
\begin{equation}
\dot E_{\rm rel} = \dot M\left(\frac{P_{\rm rel}+U_{\rm rel}}{\rho} + \frac{\kappa}{\rho \vel}\frac{dU_{\rm rel}}{dr}\right) \ ,
\label{eqEdotR}
\end{equation}
with $\vel_\phi=r\Omega$ denoting the azimuthal velocity and $\kappa$ representing the spatial diffusion coefficient in the radial direction. We note that the sign convention adopted here for $\dot E_{\rm rel}$ is the opposite of the one used by LB05.

We can rewrite the total energy transport rate (equation \ref{eqEdot}) by combining it with the individual energy transport rates for the gas (equation \ref{eqEdotG}) and the relativistic particles (equation \ref{eqEdotR}), respectively, obtaining
\begin{equation}
\dot E = -{\cal G}\Omega + \dot M\left(\frac{1}{2}\vel^2_\phi
+ \frac{1}{2}\vel^2 + \frac{P+U}{\rho} + \Phi + \frac{\kappa}{\rho \vel}\frac{dU_{\rm rel}}{dr}\right) \ ,
\label{eqTranE}
\end{equation}
where the total internal energy density $U$ and pressure $P$ are given by
\begin{equation} 
U = U_{\rm th} + U_{\rm rel} \ , \qquad
P = P_{\rm th} + P_{\rm rel} \ .
\label{eqn:TotalPressure}
\end{equation}
Here, the gas and particle pressures are related to the respective internal energy densities via
\begin{equation}
P_{\rm th} = \left(\gamma_{\rm th}-1\right)U_{\rm th} \ ,\quad P_{\rm rel} = \left(\gamma_{\rm rel}-1\right)U_{\rm rel} \ ,
\label{eqPgPr}
\end{equation}
where $\gamma_{\rm th}$ and $\gamma_{\rm rel}$ denote the specific heat ratios for the gas and the relativistic particles, respectively. In the model considered here, we set $\gamma_{\rm rel} = 4/3$ for the relativistic particles and we assume that $\gamma_{\rm th} = 3/2$ to account for the pressure contribution from the expected equipartition magnetic field (e.g. Narayan et al. 1997). We note that previous single-fluid models incorporating the magnetrotational instability (MRI) exhibit sub-equipartition magnetic fields, which would correspond to setting $\gamma_{\rm th} = 5/3$ (see Yuan \& Narayan 2014, and references therein). However, as discussed by Blandford \& Begelman (1999), setting $\gamma_{\rm th} = 5/3$ within the context of an ADAF type model would create a singular mathematical structure, requiring the flow to be non-rotating. This would imply that for ADAF discs, the angular momentum would have to be dissipated at very large radii, potentially causing the thermal plasma to unbind from the disc. Hence we conclude that an ADAF model with $\gamma_{\rm th} = 5/3$ would be unphysical. Therefore, we assume in our model that the ion and magnetic energy densities are roughly comparable throughout the disc, and we set $\gamma_{\rm th} = 3/2$ accordingly.

Following Le \& Becker (2005 \& 2007), we describe the variation of the spatial diffusion coefficient using
\begin{equation}
\kappa(r) = \kappa_0 \, \vel \left(r\right)\rs\left(\frac{r}{\rs}-1\right)^2 \ ,
\label{eqKappa}
\end{equation}
where $\kappa_0$ is a dimensionless constant. Since the thermal gas and relativistic particles each contribute to the pressure support of the disc, the standard hydrostatic equilibrium relation for the disc half-thickness $H$ must now be generalized by writing
\begin{equation}
H^2(r) = \frac{\gamma_{\rm th}P}{\rho\,\OmegaK^2} \ ,
\label{eqHgen}
\end{equation}
where $\OmegaK$ is the Keplerian angular velocity of matter in a circular orbit at radius $r$ in the pseudo-Newtonian potential, given by (see equation~\ref{eqGravPot})
\begin{equation}
\OmegaK^2(r) \equiv \frac{GM}{r (r-\rs)^2} = \frac{1}{r}\frac{d\Phi}{dr} \ .
\label{eqOmegaK}
\end{equation}
We do not treat the vertical structure of the disc here, and instead we assume that each of the physical quantities represents an average over the vertical variation in the disc (see Appendix A in LB05). Our work is based fundamentally on the same set of equations employed by Narayan et al. (1997) in their analysis of the structure of ADAF discs, which are essentially the same equations utilized by Abramowicz et al. (1988) in their ``slim-disc'' model. As pointed out by Narayan et al. (1997), the main difference is that cooling is neglected in the ADAF model, whereas it is included in the slim-disc model. Hence, our utilization of the Narayan et al. (1997) equations is warranted in the context of interest here.

The adiabatic sound speeds $a_{\rm th}$ and $a_{\rm rel}$ for the gas and relativistic particles, respectively, are defined by
\begin{equation}
a_{\rm th}(r) \equiv \left(\frac{\gamma_{\rm th} P_{\rm th}}{\rho}\right)^{1/2} \ ,\qquad
a_{\rm rel}(r) \equiv \left(\frac{\gamma_{\rm rel} P_{\rm rel}}{\rho}\right)^{1/2} \ .
\label{eqAgAr}
\end{equation}
By combining equation~(\ref{eqHgen}) for the disk height with equation~(\ref{eqAgAr}) for the sound speeds, the disc half-thickness can be written as
\begin{equation}
H(r) = \frac{1}{\OmegaK}\left(\frac{\gamma_{\rm th}}{\gamma_{\rm rel}}a^2_{\rm rel}+a^2_{\rm th}\right)^{1/2} \ .
\label{eqH}
\end{equation}
In the limit $a_{\rm rel} \to 0$, we recover the standard hydrostatic relation used by Narayan et al. (1997) and Le \& Becker (2005, 2007), $H=a_{\rm th}/\OmegaK$.

The gradient of the angular velocity $\Omega$ is related to the torque $\cal{G}$ by (e.g., Frank et al. 2002)
\begin{equation}
{\cal G} = -4\pi r^3H\rho \nu \frac{d\Omega}{dr} \ ,
\label{eqG}
\end{equation}
where $\nu$ is the kinematic viscosity. The torque can be eliminated between equations~(\ref{eqTranJ}) for the angular momentum transport rate and equation~(\ref{eqTranE}) for the total energy transport rate, which can then be combined with equations~(\ref{eqn:TotalPressure}) and (\ref{eqAgAr}) for the pressures and sound speeds, respectively, to express the total energy transport rate per unit mass as
\begin{equation}
\epsilon \equiv \frac{\dot E}{\dot M} = \frac{1}{2}\vel^2 - \frac{1}{2}\frac{\ell^2}{r^2}
+ \frac{\ell_0 \ell}{r^2} +\frac{a^2_{\rm th}}{\gamma_{\rm th}-1} + \frac{a^2_{\rm rel}}{\gamma_{\rm rel}-1}
+ \Phi + \frac{\kappa}{\rho \vel}\frac{dU_{\rm rel}}{dr} \ ,
\label{eqEps1}
\end{equation}
where 
\begin{equation}
\ell(r) \equiv r^2\Omega(r)
\end{equation}
denotes the specific angular momentum at radius $r$, and
\begin{equation}
\ell_0 \equiv \frac{\dot J}{\dot M}
\end{equation}
represents the (constant) angular momentum transport rate per unit mass in the disc. We note that equation~(\ref{eqEps1}) for the total energy transport rate per unit mass, $\epsilon$, is applicable in both viscid and inviscid flows.

\subsection{Inviscid flow equations}

It was determined by D09 that the dynamical profiles in viscid and inviscid ADAF discs are very similar, provided the specific angular momentum $\ell_0$ in the supplied gas is low enough for the matter to reach the marginally stable orbital radius, where is pulled into the black hole regardless of the value of $\alpha$. It is therefore sufficient for our purposes to focus on inviscid ADAF discs, in which the pressure of the accelerated relativistic particles contributes significantly to the dynamical structure of the disc. We defer the consideration of particle pressure in viscous discs to future work.

With viscosity neglected, the torque $\cal G$ vanishes and the specific angular momentum conservation equation reduces to
\begin{equation}
\ell(r) = \ell_0 = {\rm constant}
\ ,
\label{eqConstL}
\end{equation}
so that $\ell$ is conserved throughout the disc. Likewise, equation~(\ref{eqEps1}) for the total energy transport rate $\epsilon$ now becomes
\begin{equation}
\epsilon = \frac{1}{2}\vel^2 + \frac{1}{2}\frac{\ell_0^2}{r^2} + \frac{a^2_{\rm th}}{\gamma_{\rm th}-1}
+ \Phi + \frac{a^2_{\rm rel}}{\gamma_{\rm rel}-1} + \frac{\kappa}{\rho \vel}\frac{dU_{\rm rel}}{dr} \ .
\label{eqEps2}
\end{equation}
Compared with past work (e.g., LB05; Le \& Becker 2007), the energy transport rate treated here includes two new contributions, represented by the final two terms in equation~(\ref{eqEps2}), which describe the transport of energy via the advection or diffusion of relativistic particles, respectively. The inclusion of these two new terms is a critical component in the development of a self-consistent theory for the structure of transonic ADAF discs, in analogy with the generalized energy equation employed in the two-fluid model of cosmic-ray shock acceleration (e.g., Axford et al. 1977; Drury \& V\"olk 1981; Becker \& Kazanas 2001).

Since radiative losses are negligible in ADAF discs, and we are neglecting viscous dissipation for the moment, the pressure of the thermal background gas varies adiabatically. In this case we can employ the standard adiabatic relation between the density $\rho$ and the pressure $P_{\rm th}$ of the thermal gas, given by
\begin{equation}
P_{\rm th} = D_0 \rho^{\gamma_{\rm th}} \ ,
\label{eqPgPrAd}
\end{equation}
where the parameter $D_0$ remains constant, except at the location of the isothermal shock if one exists in the flow.

Combining equations~(\ref{eqn:TotalPressure}) for the total pressure, (\ref{eqAgAr}) for the adiabatic sound speeds, and (\ref{eqPgPrAd}) for the thermal pressure yields an expression for the relativistic particle energy density $U_{\rm rel}$, which can be differentiated to obtain
\begin{equation}
\frac{dU_{\rm rel}}{dr} = \frac{\rho}{\gamma_{\rm rel}(\gamma_{\rm rel}-1)}\left[\frac{d a_{\rm rel}^2}{dr}
+ \frac{a_{\rm rel}^2}{a_{\rm th}^2(\gamma_{\rm th}-1)}\frac{d a_{\rm th}^2}{dr}\right] \ .
\label{eqDUrDr}
\end{equation}
This expression can be used to eliminate $dU_{\rm rel}/dr$ in equation~(\ref{eqEps2}) to rewrite the energy transport rate in terms of the sounds speeds, yielding
\begin{equation}
\epsilon = \frac{1}{2}\vel^2 + \frac{1}{2}\frac{\ell_0^2}{r^2} + \frac{a^2_{\rm th}}{\gamma_{\rm th}-1}
+ \frac{a^2_{\rm rel}}{\gamma_{\rm rel}-1} + \Phi + \frac{\kappa}{\vel\gamma_{\rm rel}
(\gamma_{\rm rel}-1)}\left[\frac{d a_{\rm rel}^2}{dr} + \frac{a_{\rm rel}^2}{a_{\rm th}^2
(\gamma_{\rm th}-1)}\frac{d a_{\rm th}^2}{dr}\right] \ .
\label{eqEps3}
\end{equation}
We note that in the special case of a non-diffusive disc ($\kappa=0$), equation~(\ref{eqEps3}) describes the advective transport propagation of the gas and relativistic particle internal energy density contributions.

In an adiabatic disc, the entropy of the thermal background gas is conserved. It is therefore convenient to define the gas entropy parameter, $K_{\rm th}$, which is related to the entropy per particle, $S_{\rm th}$, via (Becker \& Le 2003)
\begin{equation}
S_{\rm th} = k\ln K_{\rm th} + c_0 \ ,
\label{eqSg}
\end{equation}
where $k$ is the Boltzmann constant and $c_0$ is a constant that is independent of the state of the gas. By combining equations~(\ref{eqTranM}) for the mass transport rate, (\ref{eqAgAr}) for the adiabatic sound speeds, and (\ref{eqH}) for the disk half-thickness, we obtain
\begin{equation}
K_{\rm th} \equiv r^{3/2}(r-\rs) \vel a_{\rm th}^{2/(\gamma_{\rm th}-1)}\left(\frac{\gamma_{\rm th}}{\gamma_{\rm rel}} a^2_{\rm rel}
+ a^2_{\rm th}\right)^{1/2} \ .
\label{eqKg}
\end{equation}
The gas entropy parameter $K_{\rm th}$ is constant throughout an adiabatic disc, except at the location of an isothermal shock. It should be noted that equation~(\ref{eqKg}) is a generalization of the corresponding result obtained by Becker \& Le (2003) that now includes the relativistic particle sound speed $a_{\rm rel}$.

By analogy with $K_{\rm th}$, we also can define the entropy parameter for the relativistic particles, $K_{\rm rel}$, using
\begin{equation}
K_{\rm rel} \equiv r^{3/2}(r-\rs) \vel a_{\rm rel}^{2/(\gamma_{\rm rel}-1)}\left(\frac{\gamma_{\rm th}}{\gamma_{\rm rel}} a^2_{\rm rel}
+ a^2_{\rm th}\right)^{1/2} \ .
\label{eqKr}
\end{equation}
Near the event horizon, the flow velocity approaches $c$, and therefore it follows that diffusion is negligible as $r \to \rs$ (e.g., Weinberg 1972). Hence $K_{\rm rel}$ approaches a constant value near the horizon where the fluid becomes non-diffusive and purely adiabatic. However, at larger radii, $K_{\rm rel}$ will vary due to the spatial diffusion of the relativistic particles, which tends to increase the entropy of the particle distribution. The entropy ratio of the gas to the relativistic particles is obtained by dividing the gas entropy parameter (equation \ref{eqKg}) by the relativistic entropy parameter (equation \ref{eqKr}) to obtain
\begin{equation}
\frac{K_{\rm th}}{K_{\rm rel}} = \frac{a_{\rm th}^{2/(\gamma_{\rm th}-1)}}{a_{\rm rel}^{2/(\gamma_{\rm rel}-1)}} \ .
\label{eqKgKr}
\end{equation}
This ratio is one of the quantities that we will use to characterize the nature of the flow near the event horizon, as part of our determination of the global flow structure.

\subsection{Double-adiabatic wind equation ($\kappa=0$)}
\label{DAwindeq}

The wind equation is one of the fundamental differential equations that is used to study the transonic (critical) nature of accretion flows onto black holes. We can derive the wind equation by starting with the radial momentum equation, written as
\begin{equation}
\vel\frac{d\vel}{dr} = -\frac{1}{\rho}\frac{dP}{dr}-\frac{GM}{\left(r-\rs\right)^2} + \frac{\ell_0^2}{r^3} \ ,
\label{eqSSRM}
\end{equation}
where $P$ denotes the total pressure, including the thermal and relativistic components (equation~\ref{eqn:TotalPressure}). The total pressure can be expressed in terms of the sounds speeds $a_{\rm th}$ and $a_{\rm rel}$ by making use of equations~(\ref{eqn:TotalPressure}) and (\ref{eqAgAr}) to obtain the alternative form
\begin{equation}
\vel\frac{d\vel}{dr} = \frac{\ell_0^2}{r^3} - \frac{GM}{(r-\rs)^2}-\frac{1}{\gamma_{\rm rel}}\frac{d a_{\rm rel}^2}{dr}
- \frac{a_{\rm th}^2\gamma_{\rm rel} + a_{\rm rel}^2}{a_{\rm th}^2\gamma_{\rm rel} (\gamma_{\rm th}-1)}\frac{d a_{\rm th}^2}{dr} \ .
\label{eqDvDr1}
\end{equation}

Close to the event horizon, the situation simplifies substantially because the flow must become completely adiabatic in order to be consistent with general relativity, which requires that $\vel \to c$ as $r \to \rs$ (Weinberg 1972). In this limit, diffusion is negligible, and therefore both sound speeds $a_{\rm th}$ and $a_{\rm rel}$ must vary adiabatically. We refer to this special case, with $\kappa=0$, as the ``double-adiabatic'' model. In this situation, the particle pressure $P_{\rm rel}$ and the gas density $\rho$ are connected by the adiabatic relation
\begin{equation}
P_{\rm rel} = D_1\rho^{\gamma_{\rm rel}} \ ,
\label{eqPrAd}
\end{equation}
where $D_1$ is a constant, except at the location of a shock. Combining this relation with equation~(\ref{eqPgPrAd}) for the thermal pressure yields a symmetrical relationship between the two sound speeds in the double-adiabatic model,
\begin{equation}
\frac{d\arsq}{dr}\Bigg|_{\rm ad} = \frac{\arsq(\gamma_{\rm rel} - 1)}{\agsq(\gamma_{\rm th}-1)}
\frac{d\agsq}{dr} \ .
\label{eqDarDrMod2}
\end{equation}

We can derive a wind equation applicable in the double-adiabatic case by eliminating $d\vel/dr$ between equations~(\ref{eqDvDr1}) and (\ref{eqDarDrMod2}) and using the energy transport rate per unit mass (\ref{eqEps3}) to substitute for $da_{\rm rel}^2/dr$. After some algebra, the wind equation obtained for the double-adiabatic model is
\begin{equation}
\frac{d\agsq}{dr}\Bigg|_{\rm ad} = \frac{N_{\rm ad}}{D_{\rm ad}} \ ,
\label{eqDAgDrAdiabatic}
\end{equation}
where the numerator and denominator functions are defined by
\begin{equation}
\begin{split}
N_{\rm ad}& \equiv \frac{\ell_0^2}{r^3}-\frac{GM}{(r-\rs)^2} + \vel^2\frac{5r-3\rs}{2r(r-\rs)} \ ,\\
D_{\rm ad}& \equiv -\left[\frac{a_{\rm th}^2\gamma_{\rm rel}(\gamma_{\rm th}+1)+a_{\rm rel}^2\gamma_{\rm th}(\gamma_{\rm rel}+1)}
{2(a_{\rm rel}^2\gamma_{\rm th}+a_{\rm th}^2\gamma_{\rm rel})a_{\rm th}^2(\gamma_{\rm th}-1)}\right]
\left(\vel^2 - a^2_{\rm eff, ad}\right) \ ,
\label{eqNDMod2}
\end{split}
\end{equation}
and $a_{\rm eff, ad}$ denotes the effective sound speed for the double-adiabatic model,
\begin{equation}
a^2_{\rm eff, ad}(r) \equiv \frac{2(a_{\rm rel}^2\gamma_{\rm th}+a_{\rm th}^2\gamma_{\rm rel})(a_{\rm th}^2+a_{\rm rel}^2)}
{a_{\rm th}^2\gamma_{\rm rel}(\gamma_{\rm th}+1) + a_{\rm rel}^2\gamma_{\rm th}(\gamma_{\rm rel}+1)} \ .
\label{agEff1}
\end{equation}
These expressions will be combined with the double-adiabatic critical conditions derived in Section \ref{sec:DAcritpoint} to solve for the disc structure when both the thermal gas and the relativistic particles evolve adiabatically. It is interesting to note that in the limit $a_{\rm rel} \to 0$, equation~(\ref{agEff1}) reduces to
\begin{equation}
\lim_{a_{\rm rel} \to 0} a^2_{\rm eff, ad} \equiv \frac{2 a_{\rm th}^2}
{1+\gamma_{\rm th}} \ ,
\label{}
\end{equation}
which agrees with equation~(22) from LB05, who treated the adiabatic one-fluid case.

\subsection{Two-fluid wind equation with diffusion ($\kappa \ne 0$)}
\label{DIFFwindeq}

The double-adiabatic wind equation~(\ref{eqNDMod2}) describes the dynamical structure off the disc close to the event horizon, where diffusion of the relativistic particles is negligible compared with advection. However, at large radii, diffusion becomes dominant, and therefore it must be included in the set of dynamical equations in order to determine the disc structure. In order to treat this case, we need to employ the energy and entropy equations. We begin by deriving an expression for the velocity derivative $d\vel/dr$ by differentiating equation~(\ref{eqKg}) for the (constant) gas thermal entropy parameter, $K_{\rm th}$, which yields
\begin{equation}
- \frac{1}{\vel}\frac{d\vel}{dr} = \frac{3}{2r}+\frac{1}{r-\rs} + \frac{\gamma_{\rm th}}{2(a_{\rm rel}^2\gamma_{\rm th}
+ a_{\rm th}^2\gamma_{\rm rel})}\frac{d a_{\rm rel}^2}{dr}+\left[\frac{1}{a_{\rm th}^2(\gamma_{\rm th}-1)} + \frac{\gamma_{\rm rel}}{2(a_{\rm rel}^2\gamma_{\rm th}+ a_{\rm th}^2\gamma_{\rm rel})}\right]\frac{d a_{\rm th}^2}{dr} \ .
\label{eqDvDr2}
\end{equation}
Eliminating $d\vel/dr$ between equations~(\ref{eqDvDr1}) and (\ref{eqDvDr2}), and using the energy transport rate per unit mass (equation~\ref{eqEps3}) to substitute for $da_{\rm rel}^2/dr$, after some algebra we obtain the wind equation describing the diffusive case,
\begin{equation}
\frac{d\agsq}{dr}\Bigg|_{\rm diffusive} = \frac{N}{D} \ ,
\label{eqDAgDr}
\end{equation}
where the numerator and denominator functions, $N$ and $D$, are defined by
\begin{equation}
\begin{split}
N& \equiv \frac{\vel(\gamma_{\rm rel}-1)}{\kappa}\left[\vel^2\frac{\gamma_{\rm th}\gamma_{\rm rel}}{2(a_{\rm rel}^2\gamma_{\rm th}+a_{\rm th}^2\gamma_{\rm rel})}-1\right]
\left(\epsilon-\frac{1}{2}\vel^2-\frac{1}{2}\frac{\ell_0^2}{r^2}-\frac{a_{\rm th}^2}{\gamma_{\rm th}-1}-\frac{a_{\rm rel}^2}{\gamma_{\rm rel}-1}-\Phi\right) + \frac{\ell_0^2}{r^3} - \frac{GM}{(r-\rs)^2} + \vel^2\frac{5r-3\rs}{2r(r-\rs)} \ ,\\
D& \equiv -\left[\frac{a_{\rm rel}^2\gamma_{\rm th} + \gamma_{\rm rel} a_{\rm th}^2(\gamma_{\rm th}+1)}{2a_{\rm th}^2(\gamma_{\rm th}-1)(a_{\rm rel}^2\gamma_{\rm th}+a_{\rm th}^2\gamma_{\rm rel})}\right] \left(\vel^2-a^2_{{\rm eff},\kappa}\right) \ ,
\label{eqNDMod3}
\end{split}
\end{equation}
and $a_{{\rm eff}, \kappa}$ denotes the effective sound speed for the diffusive model,
\begin{equation}
a^2_{{\rm eff},\kappa}(r) \equiv \frac{2a_{\rm th}^2(a_{\rm rel}^2\gamma_{\rm th}+a_{\rm th}^2\gamma_{\rm rel})}
{a_{\rm rel}^2\gamma_{\rm th} + \gamma_{\rm rel}(\gamma_{\rm th}+1)a_{\rm th}^2} \ .
\label{eqAEffKappa}
\end{equation}

In our analysis of transonic disc flows, we will find it convenient to define the effective Mach number for the diffusive two-fluid model, $\mathscr{M}_{{\rm eff},\kappa}$, by writing
\begin{equation}
\mathscr{M}_{{\rm eff},\kappa} \equiv \frac{\vel}{a_{{\rm eff},\kappa}}
= \left[\frac{\mathscr{M}_{\rm rel}^{-2}\gamma_{\rm th} + \gamma_{\rm rel}(\gamma_{\rm th}+1)\mathscr{M}_{\rm th}^{-2}}
{2\mathscr{M}_{\rm th}^{-2}(\mathscr{M}_{\rm rel}^{-2}\gamma_{\rm th} + \mathscr{M}_{\rm th}^{-2}\gamma_{\rm rel})}\right]^{1/2} \ ,
\label{eqMeff}
\end{equation}
where $\mathscr{M}_{\rm th}$ and $\mathscr{M}_{\rm rel}$ denote the Mach numbers with respect to the gas and relativistic particle sound speeds, respectively, given by
\begin{equation}
\mathscr{M}_{\rm th} \equiv \frac{\vel}{a_{\rm th}} \ , \qquad
\mathscr{M}_{\rm rel} \equiv \frac{\vel}{a_{\rm rel}} \ .
\label{eqMachNumDefs}
\end{equation}

Critical points occur where the numerator and denominator functions $N$ and $D$ both vanish, so that $\vel = a_{{\rm eff},\kappa}$ and $\mathscr{M}_{{\rm eff},\kappa}=1$. Equation~(\ref{eqNDMod3}) is the two-fluid wind equation for the generalized case treated here, which incorporates the effects of the thermal and relativistic particle pressure, as well as the diffusion of the relativistic particle energy. In contrast with the adiabatic, one-fluid model treated by LB05, in the diffusive, two-fluid case, the dynamical structure of the disc cannot be determined using a root-finding procedure. Hence, we must numerically integrate the wind equation~(\ref{eqDAgDr}), supplemented by an additional differential equation for $a_{\rm rel}$, which is obtained by rearranging the energy equation~(\ref{eqEps3}) to obtain
\begin{equation}
\frac{d a_{\rm rel}^2}{dr} = \frac{\vel\gamma_{\rm rel} (\gamma_{\rm rel}-1)}{\kappa}
\left(\epsilon - \frac{1}{2}\vel^2 - \frac{1}{2}\frac{\ell_0^2}{r^2} - \frac{a^2_{\rm th}}{\gamma_{\rm th}-1}
- \frac{a^2_{\rm rel}}{\gamma_{\rm rel}-1} - \Phi \right) - \frac{a_{\rm rel}^2}{a_{\rm th}^2
(\gamma_{\rm th}-1)}\frac{d a_{\rm th}^2}{dr} \ .
\label{eqEps3alt}
\end{equation}
Finally, in order to close the system, we must also use the entropy equation~(\ref{eqKg}) to obtain an algebraic equation for $\vel$ in terms of $a_{\rm th}$ and $a_{\rm rel}$, given by
\begin{equation}
\vel = K_{\rm th} r^{-3/2}(r-\rs)^{-1} a_{\rm th}^{2/(1-\gamma_{\rm th})}
\left(\frac{\gamma_{\rm th}}{\gamma_{\rm rel}}a^2_{\rm rel}+ a^2_{\rm th}\right)^{-1/2} \ .
\label{eqVfromKg}
\end{equation}
The determination of the disc structure in the two-fluid model, with relativistic particle pressure included, requires the simultaneous solution of the differential wind equation~(\ref{eqDAgDr}), along with equation~(\ref{eqEps3alt}) for the relativistic sound speed derivative, supplemented by the algebraic velocity relation given by equation~(\ref{eqVfromKg}).

Once the profiles have been determined for the inflow velocity $\vel(r)$, the thermal sound speed $a_{\rm th}(r)$, and the relativistic particle sound speed $a_{\rm rel}(r)$, we can compute the disc half-thickness $H(r)$ using equation~(\ref{eqH}), and the radial distribution of the mass density can be evaluated using (see equation~\ref{eqTranM} for the mass transport rate)
\begin{equation}
\rho(r) = \frac{\dot M}{4 \pi H(r) \vel(r)} \ .
\label{eqDensity1}
\end{equation}
Based on these results, we can compute the thermal gas pressure and energy density using (see equations~\ref{eqAgAr} for the adiabatic sound speeds)
\begin{equation}
P_{\rm th}(r) = (\gamma_{\rm th}-1) U_{\rm th}(r) = \frac{\rho(r) a_{\rm th}^2(r)}{\gamma_{\rm th}} \ ,
\label{eqThermalP1}
\end{equation}
and likewise, the relativistic particle pressure and energy density can be computed using
\begin{equation}
P_{\rm rel}(r) = (\gamma_{\rm rel}-1) U_{\rm rel}(r) = \frac{\rho(r) a_{\rm rel}^2(r)}{\gamma_{\rm rel}} \ .
\label{eqRelP1}
\end{equation}

\section{CRITICAL POINT ANALYSIS}
\label{sec:critpoint}

In Sections \ref{DAwindeq} and \ref{DIFFwindeq} we derived the wind equations describing the double-adiabatic and diffusive two-fluid cases, with $\kappa=0$ and $\kappa\ne0$, respectively. Now we must understand the implications of the transonic (critical) nature of the accretion flow in both cases.

\subsection{Double-adiabatic critical conditions ($\kappa=0$)}
\label{sec:DAcritpoint}

In the double-adiabatic case ($\kappa=0$), the simultaneous vanishing of the functions $N_{\rm ad}$ and $D_{\rm ad}$ (equations~\ref{eqNDMod2}) yields the critical conditions
\begin{equation}
\frac{\ell_0^2}{\rc^3} - \frac{GM}{(\rc-\rs)^2} + \vcsq\frac{5\rc-3\rs}{2\rc(\rc-\rs)} = 0 \ ,
\label{eqNcMod2}
\end{equation}
\begin{equation}
\vcsq = a^2_{\rm eff, ad}(r_c) = \frac{2(\arcsq\gamma_{\rm th} + \agcsq\gamma_{\rm rel})(\agcsq + \arcsq)}
{\agcsq\gamma_{\rm rel}(\gamma_{\rm th}+1) + \arcsq\gamma_{\rm th}(\gamma_{\rm rel}+1)} \ ,
\label{eqDcMod2}
\end{equation}
where $a_{\rm eff, ad}$ is the effective sound speed in the double-adiabatic case, defined by equation~(\ref{agEff1}), and $\vel_c$, $a_{\rm th,c}$ and $a_{\rm rel,c}$ denote the values of the velocity and the thermal and relativistic sound speeds, respectively, at the critical radius, $r=r_c$. We note that these expressions reduce to the critical conditions derived by LB05 in the limit $a_{\rm rel,c} \to 0$, which is expected since the relativistic particle pressure is neglected in the one-fluid model treated by these authors.

The two-fluid, double-adiabatic model under consideration here is an extension of the adiabatic one-fluid model studied by LB05. In analogy with their investigation, we wish to develop a formalism that we can use to solve for the critical radius $r_c$, the critical velocity $\vel_c$, and the critical sound speeds $a_{\rm th,c}$ and $a_{\rm rel,c}$ for given input values of the energy transport rate $\epsilon$, the specific angular momentum $\ell_0$, and the entropy ratio $K_{\rm th}/K_{\rm rel}$ (equation~\ref{eqKgKr}). First, we use equation~(\ref{eqGravPot}) for the pseudo-Newtonian potential to rewrite the energy transport rate equation~(\ref{eqEps3}) at the critical point in the double-adiabatic case with $\kappa=0$ as
\begin{equation}
\epsilon = \frac{1}{2}\vel_c^2+\frac{1}{2}\frac{\ell_0^2}{r_c^2} + \frac{a^2_{\rm th,c}}
{\gamma_{\rm th}-1} + \frac{a^2_{\rm rel,c}}{\gamma_{\rm rel}-1} - \frac{GM}{r_c-\rs} \ .
\label{eqEps4}
\end{equation}
Next, we use equation~(\ref{eqNcMod2}) to eliminate $\vel_c^2$ in equation~(\ref{eqEps4}) and solve for $a_{\rm th,c}^2$ to obtain
\begin{equation}
\agcsq = (\gamma_{\rm th}-1) \left\{\epsilon-\frac{1}{2}\frac{\ell_0^2}{r_c^2} - \frac{\arcsq}{\gamma_{\rm rel}-1}
+ \frac{GM}{r_c-\rs} - \frac{r_c(r_c-\rs)}{5r_c-3\rs}
\left[\frac{GM}{(r_c-\rs)^2} - \frac{\ell_0^2}{r_c^3}\right] \right\} \ .
\label{eqAgcMod2}
\end{equation}
Since the entropy ratio $K_{\rm th}/K_{\rm rel}$ (equation~\ref{eqKgKr}) remains globally constant in the double-adiabatic model, its global value is equal to its value at the critical point. Setting $\gamma_{\rm th}=3/2$ and $\gamma_{\rm rel}=4/3$, we can therefore write the global value of the entropy ratio as
\begin{equation}
\left(\frac{K_{\rm th}}{K_{\rm rel}}\right)^{1/2} = \frac{a_{\rm th,c}^2}{a_{\rm rel,c}^3}
= \rm constant \ .
\label{eqKgcKrcMod2}
\end{equation}
We emphasize that this ratio is only constant in the double-adiabatic model under consideration here, and in the general case, with $\kappa \ne 0$, $K_{\rm th}/K_{\rm rel}$ will vary throughout the disc in response to the diffusion of the relativistic particles.

Eliminating $a_{\rm th,c}$ between equations~(\ref{eqAgcMod2}) and (\ref{eqKgcKrcMod2}) yields a cubic equation for $a_{\rm rel,c}$ given by
\begin{equation}
a_{\rm rel,c}^3 + {\cal N}\arcsq + {\cal P} = 0 \ ,
\label{eqArcMod2a}
\end{equation}
where
\begin{equation}
\begin{split}
{\cal N}& = \frac{\gamma_{\rm th}-1}{\gamma_{\rm rel}-1}\sqrt{\frac{K_{\rm rel}}{K_{\rm th}}} \ , \\
{\cal P}& = -(\gamma_{\rm th}-1)\sqrt{\frac{K_{\rm rel}}{K_{\rm th}}}\left\{\epsilon - \frac{1}{2}\frac{\ell_0^2}{r_c^2}
+ \frac{GM}{r_c-\rs} - \frac{r_c(r_c-\rs)}{5r_c-3\rs}
\left[\frac{GM}{(r_c-\rs)^2} - \frac{\ell_0^2}{r_c^3}\right] \right\} \ .
\label{eqArcMod2b}
\end{split}
\end{equation}
Of the three possible solutions to this cubic equation, only one is a physically acceptable real value, which can be computed in terms of the critical radius $r_c$ using
\begin{equation}
a_{\rm rel,c} = S + T - \frac{1}{3}{\cal N} \ ,
\label{eqArcMod2c}
\end{equation}
where 
\begin{equation}
S = \left(X+\sqrt{W^3+X^2}\right)^{1/3} \ , \quad T = \left(X-\sqrt{W^3+X^2}\right)^{1/3} \ ,
\end{equation}
and
\begin{equation}
W = - \frac{{\cal N}^2}{9} \ , \quad X = -\frac{1}{2}{\cal P} - \frac{1}{27}{\cal N}^3 \ .
\end{equation} 
Equation~(\ref{eqArcMod2c}) gives $a_{\rm rel,c}$ as an explicit algebraic function of the critical radius $r_c$.

We are now in a position to derive a single equation whose roots represent the possible values for the critical radius $r_c$. First we eliminate $\vel_c$ between equations~(\ref{eqNcMod2}) and (\ref{eqDcMod2}) and substitute for $a_{\rm th,c}$ using the entropy ratio (equation~\ref{eqKgcKrcMod2}) to obtain, after simplification,
\begin{equation}
\frac{\ell_0^2}{\rc^3}-\frac{GM}{(\rc-\rs)^2}
+ \frac{\arcsq\left[\gamma_{\rm th} + (K_{\rm th}/K_{\rm rel})^{1/2} a_{\rm rel,c}\gamma_{\rm rel}\right]
\left[(K_{\rm th}/K_{\rm rel})^{1/2} a_{\rm rel,c} + 1\right]}
{(K_{\rm th}/K_{\rm rel})^{1/2} a_{\rm rel,c}\gamma_{\rm rel}(\gamma_{\rm th}+1) + \gamma_{\rm th}(\gamma_{\rm rel}+1)}
\frac{5\rc-3\rs}{\rc(\rc-\rs)} = 0 \ ,
\label{eqNcMod2b}
\end{equation}
Using equation~(\ref{eqArcMod2c}) to substitute for $a_{\rm rel,c}$ in this expression yields a single nonlinear equation for the critical radius, $r_c$. This equation cannot be solved analytically, and therefore we must resort to a numerical root finding procedure to determine $r_c$ in terms of the fundamental parameters $\epsilon$, $\ell_0$, and $K_{\rm th}/K_{\rm rel}$. In general, the nonlinear equation admits three roots for $r_c$, which we refer to using the notation $r_{c1}$, $r_{c2}$, and $r_{c3}$ in order of decreasing radius.

Previous models have demonstrated that with the given parameters $\epsilon$, $\ell_0$, only one solution was viable for a shock or shock-free inviscid ADAF disc. In keeping with the earlier one-fluid adiabatic models, we find that in the double-adiabatic case studied here, multiple critical points are possible. Adhering to the categorization of the critical sonic points (e.g. LB05, Abramowicz \& Chakrabarti 1990), we focus on the innermost root $r_{c3}$ in our study, which is classified as an X-type critical point, and therefore a physically acceptable sonic point. It allows for a flow to exist that is transonic at $r_{c3}$, and then continues to be supersonic as it moves towards the event horizon. Once $r_{c3}$ is determined, we can substitute it into equation~(\ref{eqArcMod2c}) to compute $a_{\rm rel,c}$, and then we can use equation~(\ref{eqKgcKrcMod2}) to compute $a_{\rm th,c}$. Finally, application of equation~(\ref{eqKg}) yields the corresponding value for the conserved entropy parameter $K_{\rm th,c}$ at the critical point, given by
\begin{equation}
K_{\rm th,c}\equiv r_c^{3/2}(r_c-\rs)\vel_c a_{\rm th,c}^{2/\left(\gamma_{\rm th}-1\right)}
\left(\frac{\gamma_{\rm th}}{\gamma_{\rm rel}}\arcsq + \agcsq\right)^{1/2} \ .
\label{eqKgc}
\end{equation}
Since diffusion is negligible near the event horizon, the double-adiabatic inner critical radius $r_{c3}$ provides an accurate approximation of the location of the inner critical radius in the diffusive two-fluid case.

\subsection{Double-adiabatic flow solution ($\kappa=0$)}
\label{sec:DAflowsol}

Once the values of $a_{\rm rel,c}$ and $a_{\rm th,c}$ are known, we can use equations~(\ref{eqAgAr}) for the adiabatic sound speeds, along with equations~(\ref{eqPgPrAd}) and (\ref{eqPrAd}) for the adiabatic pressures, to express the density dependences of the gas and particle sound speeds as
\begin{equation}
a^2_{\rm th} = a^2_{\rm th,c} \left(\frac{\rho}{\rho_c}\right)^{\gamma_{\rm th}-1} \ ,
\qquad a^2_{\rm rel} = a^2_{\rm rel,c} \left(\frac{\rho}{\rho_c}\right)^{\gamma_{\rm rel}-1} \ .
\label{eqAgArAdRC}
\end{equation}
These expressions imply a symmetrical, adiabatic relation between the thermal and relativistic sounds speeds, which can be written as
\begin{equation}
\arsq = \arcsq\left(\frac{\agsq}{\agcsq}\right)^{(\gamma_{\rm rel} - 1)/(\gamma_{\rm th}-1)} \ .
\label{eqArAgfunMod2}
\end{equation}
In the double-adiabatic model, $\kappa=0$, and the energy equation~(\ref{eqEps3}) reduces to
\begin{equation}
\epsilon = \frac{1}{2}\vel^2+\frac{1}{2}\frac{\ell_0^2}{r^2}
+ \frac{a^2_{\rm th}}{\gamma_{\rm th}-1} + \frac{a^2_{\rm rel}}{\gamma_{\rm rel}-1} + \Phi \ .
\label{eqDoubAd}
\end{equation}
The flow velocity $\vel$ can be written in terms of the gas entropy $K_{\rm th,c}$ and the sound speeds $a_{\rm th}$ and $a_{\rm rel}$ using equation~(\ref{eqVfromKg}), which yields
\begin{equation}
\vel = K_{\rm th,c} r^{-3/2}\left(r-\rs\right)^{-1} a_{\rm th}^{2/\left(1-\gamma_{\rm th}\right)}\left(\frac{\gamma_{\rm th}}{\gamma_{\rm rel}}a^2_{\rm rel}+ a^2_{\rm th}\right)^{-1/2} \ .
\label{eqVfromKgc}
\end{equation}
By utilizing equations~(\ref{eqArAgfunMod2}) and (\ref{eqVfromKgc}), we can rewrite the energy equation~(\ref{eqDoubAd}) as an algebraic function of $a_{\rm th}$, obtaining
\begin{equation}
\epsilon = \frac{1}{2}\frac{\ell_0^2}{r^2} + \Phi + \frac{a_{\rm th}^2}{\gamma_{\rm th}-1} + \frac{a_{\rm rel,c}^2}{\gamma_{\rm rel}-1}\left(\frac{a_{\rm th}^2}{a_{\rm th,c}^2}\right)^{(\gamma_{\rm rel}-1)/(\gamma_{\rm th}-1)}
+ \frac{K_{\rm th,c}^2}{2\,r^3(r-\rs)^2 a_{\rm th}^{4/(\gamma_{\rm th}-1)}}
\left[\frac{\gamma_{\rm th}}{\gamma_{\rm rel}}a_{\rm rel,c}^2
\left(\frac{a_{\rm th}^2}{a_{\rm th,c}^2}\right)^{(\gamma_{\rm rel}-1)/(\gamma_{\rm th}-1)} + a_{\rm th}^2\right]^{-1} \ .
\label{eqEpsMod2Ag}
\end{equation}
Equation~(\ref{eqEpsMod2Ag}) can be solved using a simple root-finding procedure to determine the profile of $a_{\rm th}$ as a function of $r$, for any values of the parameters $\epsilon$, $\ell_0$, $a_{\rm th,c}$, $a_{\rm rel,c}$, and $K_{\rm th,c}$.

\subsection{Two-fluid critical conditions with diffusion ($\kappa\ne 0$)}
\label{sec:DIFFcritpoint}

By analogy with the double-adiabatic model, the diffusive two-fluid model obtained when $\kappa \ne 0$ also displays a critical behaviour. In the diffusive two-fluid model, the simultaneous vanishing of $N$ and $D$ (see equations~\ref{eqNDMod3} and \ref{eqAEffKappa}) yields the critical condition
\begin{equation}
\frac{\vel_c(\gamma_{\rm rel}-1)}{\kappa_c}\left[\frac{\gamma_{\rm th}\gamma_{\rm rel}\vcsq}{2(\arcsq\gamma_{\rm th}+\agcsq\gamma_{\rm rel})}-1\right]
\left(\epsilon-\frac{1}{2}\vcsq-\frac{1}{2}\frac{\ell_0^2}{\rc^2}-\frac{\agcsq}{\gamma_{\rm th}-1}-\frac{\arcsq}{\gamma_{\rm rel}-1}-\Phi_c\right)
+ \frac{\ell_0^2}{\rc^3}-\frac{GM}{(\rc-\rs)^2}+\frac{(5\rc-3\rs)\vcsq}{2\rc(\rc-\rs)} = 0 \ ,
\label{eqNcMod3}
\end{equation}
where $\kappa_c$ represents the diffusion coefficient at the critical radius $r=r_c$ (equation~\ref{eqKappa}), and the critical velocity $\vel_c$ is given by
\begin{equation}
\vcsq = a^2_{{\rm eff},\kappa}(r_c) = \frac{2 a_{\rm th,c}^2(\gamma_{\rm th} a_{\rm rel,c}^2 + \gamma_{\rm rel}a_{\rm th,c}^2)}
{\gamma_{\rm th} a_{\rm rel,c}^2 + \gamma_{\rm rel}(\gamma_{\rm th}+1)a_{\rm th,c}^2} \ ,
\label{eqVcMod3}
\end{equation}
where $a_{{\rm eff},\kappa}$ is the effective sound speed in the diffusive two-fluid case, defined by equation~(\ref{eqAEffKappa}).

Equations~(\ref{eqNcMod3}) and (\ref{eqVcMod3}) provide two constraints on the critical parameters $r_c$, $\vel_c$, $a_{\rm th,c}$, and $a_{\rm rel,c}$. Hence we need two more equations in order to close the system. We can also derive a useful relation that allows us to solve for $a_{\rm rel,c}$ as a function of $r_c$ and $a_{\rm th,c}$. By combining equation~(\ref{eqVcMod3}) for the critical velocity with equation~(\ref{eqKgc}) for the thermal entropy parameter, we obtain
\begin{equation}
\frac{2 r_c^3 (r_c-\rs)^2}{K^2_{\rm th,c} a_{\rm th,c}^{2(1+\gamma_{\rm th})/(1-\gamma_{\rm th})}}
\left(a_{\rm th,c}^2 + \frac{\gamma_{\rm th}}{\gamma_{\rm rel}}a_{\rm rel,c}^2\right)^2
- \left(a_{\rm th,c}^2 + \frac{\gamma_{\rm th}}{\gamma_{\rm rel}}a_{\rm rel,c}^2\right)^2
- \gamma_{\rm th} a_{\rm th,c}^2 = 0 \ .
\label{arcQuad}
\end{equation}
This quadratic equation can be solved to obtain an explicit solution for $a_{\rm rel,c}$ as a function of $a_{\rm th,c}$ and $r_c$.The result obtained is
\begin{equation}
a_{\rm rel,c}^2 = \frac{\gamma_{\rm rel}}{\gamma_{\rm th}}
\left\{-a_{\rm th,c}^2 + \frac{K_{\rm th,c}^2 a_{\rm th,c}^{4/(1-\gamma_{\rm th})}}{2 r_c^3 (r_c-\rs)^2}
\left[\frac{1}{2a_{\rm th,c}^2} + \sqrt{\frac{1}{4a_{\rm th,c}^4} + \frac{2\gamma_{\rm th} r_c^3 (r_c-\rs)^2}
{K_{\rm th,c}^2 a_{\rm th,c}^{4/(1-\gamma_{\rm th})}}}\right]\right\} \ ,
\label{arcQuad2}
\end{equation}
where the positive sign on the radical is selected in order to obtain a positive result for $a_{\rm rel,c}^2$ as required. Finally, we can use equation~(\ref{arcQuad2}) to substitute for $a_{\rm rel,c}$ in the critical condition (equation~\ref{eqNcMod3}), to obtain a single nonlinear equation whose roots give the possible values for $a_{\rm th,c}$ for a given value of $r_c$.

If the critical radius $r_c$ is known, then we can compute $a_{\rm th,c}$ using the combination of equations~(\ref{eqNcMod3}) and (\ref{arcQuad2}), and after that, we can compute $a_{\rm rel,c}$ and $\vel_c$ using equations~(\ref{arcQuad2}) and (\ref{eqVcMod3}), respectively. In the adiabatic, one-fluid model analyzed by LB05, one can derive an algebraic equation whose roots give the possible values of $r_c$. However, in the present case, the flow is not adiabatic with respect to the relativistic particles, and therefore we do not have enough information to derive an algebraic equation for $r_c$. In order to close the system and compute all of the critical quantities, we must therefore utilize an additional relation. There are two options available for closing the system, depending on whether we are treating the inner critical point at $r_{c3}$, or the outer critical point at $r_{c1}$. In the case of the inner critical point, we will utilize the double-adiabatic model to compute $r_{c3}$, since this model accurately describes the flow dynamics near the event horizon, where diffusion is negligible compared with advection. Conversely, in the case of the outer critical point, the flow is diffusive and therefore the double-adiabatic model does not apply there. In this case, we will utilize integration of the differential equations, combined with variation of the model parameters, until we observe that the numerator and denominator functions $N$ and $D$ (equations~\ref{eqNDMod3} and \ref{eqAEffKappa}) both vanish at the same location, which is interpreted as $r_{c1}$.

\section{ISOTHERMAL SHOCK MODEL}
\label{sec:isoshock}

LB05 studied the acceleration of relativistic particles due to the presence of a standing shock in an adiabatic one-fluid disc. Our goal here is to extend that study to treat diffusive discs in which the back-reaction of the pressure of the accelerated relativistic particles modifies the dynamics of the disc and the shock in a self-consistent way. Following LB05, we shall focus on isothermal shocks in order to understand how the structure of the disc responds to the presence of a shock. In keeping with the self-consistent approach taken here, we need to reconsider the isothermal shock jump conditions since they are expected to be influenced by the energy transport associated with the diffusion of the accelerated relativistic particles.

We designate $\epsilon_-$ and $\epsilon_+$ as the values of the energy transport parameter $\epsilon$ on the upstream and downstream sides of the isothermal shock, respectively. Physically, we require that $\epsilon_->\epsilon_+$ in order to account for the loss of energy through the upper and lower surfaces of the disc at the shock location, in response to the escape of relativistic particles. The shock jump conditions are determined by the conservation relations for mass, momentum, and energy, as employed by LB05. However, the results obtained here are more complex than those found by LB05 due to the incorporation of the diffusive energy transport associated with the relativistic particle population.

\subsection{Generalized isothermal shock jump conditions}
\label{GenIsothermalShockConditions}

Adopting the premise that the escape of the relativistic particles from the disc results in negligible mass loss, we assume that the mass accretion rate $\dot M$ is conserved throughout the disc, including at the shock location, $r=r_*$. We will revisit this assumption in Section~\ref{sec:astrapp}. Hence we can write the mass conservation condition at the shock as
\begin{equation}
\Delta\dot M = 0 \ ,
\label{eqDeltaM}
\end{equation}
where the operator $\Delta$ is defined by
\begin{equation}
\Delta f \equiv \lim_{\delta\to 0} f(r_*-\delta) - f(r_*+\delta) = f_+ - f_- \ ,
\label{eqDeltaM2}
\end{equation}
which denotes the difference between post-shock (``+'') and pre-shock (``-") values for any physical quantity. We assume that the outflow produces no torque on the disc, and therefore the specific angular momentum $\dot J$ defined in equation~(\ref{eqTranJ}) is conserved across the shock. Hence we find that
\begin{equation}
\Delta\dot J = 0 \ .
\label{eqDeltaJ}
\end{equation}
Likewise, the radial momentum transport rate, $\dot I$, defined by
\begin{equation}
\dot I \equiv 4\pi rH(P_{\rm th}+P_{\rm rel}+\rho \vel^2) \ ,
\label{eqTranI}
\end{equation}
is also conserved across the shock, and therefore
\begin{equation}
\Delta\dot I = 0 \ .
\label{eqDeltaI}
\end{equation}

Based on equation~(\ref{eqTranM}) for the mass accretion rate and equation~(\ref{eqH}) for the disc half-thickness, we find that the continuity of $\dot M$ across the shock implies that
\begin{equation}
\left(\frac{\gamma_{\rm th}}{\gamma_{\rm rel}}a^2_{\rm rel+} + a^2_{\rm th+}\right)^{1/2}\rho_+ \vel_+
= \left(\frac{\gamma_{\rm th}}{\gamma_{\rm rel}}a^2_{\rm rel-} + a^2_{\rm th-}\right)^{1/2}\rho_- \vel_- \ ,
\label{eqJumpTemp1}
\end{equation}
where the subscripts ``-'' and ``+'' denote quantities measured just upstream and downstream from the shock, respectively. Next, the continuity of the ratio $\dot I/\dot M$ can be combined with equation~(\ref{eqTranM}) for the accretion rate, equation~(\ref{eqAgAr}) for the adiabatic sound speeds, and equation~(\ref{eqTranI}) for the radial momentum transport rate to conclude that
\begin{equation}
\frac{1}{\vel_+}\left(\frac{\gamma_{\rm th}}{\gamma_{\rm rel}} a_{\rm rel+}^2 + a_{\rm th+}^2\right)
+ \gamma_{\rm th} \vel_+
= \frac{1}{\vel_-}\left(\frac{\gamma_{\rm th}}{\gamma_{\rm rel}} a_{\rm rel-}^2 + a_{\rm th-}^2\right) + \gamma_{\rm th} \vel_- \ ,
\label{eqJumpTemp2}
\end{equation}
In the case of an isothermal shock, which is our focus here, we also have the additional relation
\begin{equation}
a_{\rm th+}=a_{\rm th-} \ ,
\label{eqIsoAg}
\end{equation}
which we can use to eliminate $a_{\rm th+}$ in equations~(\ref{eqJumpTemp1}) and (\ref{eqJumpTemp2}), so that they reduce to
\begin{equation}
\left(\frac{\gamma_{\rm th}}{\gamma_{\rm rel}}a^2_{\rm rel+}+a^2_{\rm th-}\right)^{1/2}\rho_+\vel_+
= \left(\frac{\gamma_{\rm th}}{\gamma_{\rm rel}}a^2_{\rm rel-}+a^2_{\rm th-}\right)^{1/2}\rho_-\vel_- \ ,
\label{eqJump1Red}
\end{equation}
and
\begin{equation}
\frac{1}{\vel_+}\left(\frac{\gamma_{\rm th}}{\gamma_{\rm rel}} a_{\rm rel+}^2 + a_{\rm th-}^2\right) + \gamma_{\rm th} \vel_+
= \frac{1}{\vel_-}\left(\frac{\gamma_{\rm th}}{\gamma_{\rm rel}} a_{\rm rel-}^2 + a_{\rm th-}^2\right) + \gamma_{\rm th} \vel_- \ ,
\label{eqJump2Red}
\end{equation}
respectively.

From equation~(\ref{eqJump1Red}), we can determine the shock compression ratio for this new model, denoted by $R_*$. The result obtained is
\begin{equation}
R_* \equiv \frac{\rho_+}{\rho_-} = \frac{\vel_-}{\vel_+}
\frac{\left(\frac{\gamma_{\rm th}}{\gamma_{\rm rel}}a^2_{\rm rel-} + a^2_{\rm th-}\right)^{1/2}}
{\left(\frac{\gamma_{\rm th}}{\gamma_{\rm rel}}a^2_{\rm rel+} + a^2_{\rm th-}\right)^{1/2}} \ \ > \ 1 \ ,
\label{eqCompressionRatio}
\end{equation}
so that the gas density increases across the shock as expected. Likewise, we can also determine the gas entropy jump at the isothermal shock by combining equation~(\ref{eqKg}) for the thermal entropy parameter with equation~(\ref{eqIsoAg}) for the isothermal condition to obtain for the thermal entropy jump ratio
\begin{equation}
\frac{K_{\rm th+}}{K_{\rm th-}} = \frac{\vel_+}{\vel_-} \frac{\left(\frac{\gamma_{\rm th}}{\gamma_{\rm rel}} a^2_{\rm rel+}
+ a^2_{\rm th-}\right)^{1/2}}
{\left(\frac{\gamma_{\rm th}}{\gamma_{\rm rel}}a^2_{\rm rel-} + a^2_{\rm th-}\right)^{1/2}} \ \ < \ 1 \ .
\label{eqJumpEnt1}
\end{equation}
Note that the gas entropy decreases across the shock in response to the loss of entropy from the disc into the outflow at the shock location.

The relativistic energy density $U_{\rm rel}(r)$ is a continuous function of radius $r$ throughout the disc, so that $\Delta U_{\rm rel}=0$ across the shock. This implies that the particle pressure is also conserved, and hence $\Delta P_{\rm rel}=0$. The conservation of $U_{\rm rel}$ at the shock is required in order to avoid the generation of an infinite diffusive energy flux at the shock (see Appendix~A in Becker \& Kazanas 2001). According to equation~(\ref{eqAgAr}), the constancy of $P_{\rm rel}$ across the shock implies that the upstream and downstream relativistic sound speeds are related via
\begin{equation}
\frac{a_{\rm rel+}}{a_{\rm rel-}} = \left(\frac{\rho_+}{\rho_-}\right)^{-1/2} \ .
\label{eqDenJump}
\end{equation}
By combining equation~(\ref{eqDenJump}) with equation~(\ref{eqCompressionRatio}) for the compression ratio and equation~(\ref{eqJumpEnt1}) for the thermal entropy jump, we can derive another expression for the thermal entropy jump, given by
\begin{equation}
\frac{K_{\rm th+}}{K_{\rm th-}} = \frac{a^2_{\rm rel+}}{a^2_{\rm rel-}} \ .
\label{eqJumpEnt2}
\end{equation}

\subsection{Velocity jump condition}
\label{VelJump}

The relations derived above can be combined to obtain a single nonlinear equation whose roots express the possible values for the velocity jump ratio at the shock, $Q$, defined by
\begin{equation}
Q \equiv \frac{\vel_+}{\vel_-} \ .
\label{eqJumpRat}
\end{equation}
First we solve equation~(\ref{eqJump2Red}) to obtain an expression for the downstream relativistic particle sound speed, $a_{\rm rel+}$, given by
\begin{equation}
a_{\rm rel+}^2 = \frac{\gamma_{\rm rel}}{\gamma_{\rm th}}\left(\frac{\vel_+}{\vel_-}-1\right)a_{\rm th-}^2
+ \frac{\vel_+}{\vel_-}a_{\rm rel-}^2 + \gamma_{\rm rel}\vel_+\left(\vel_- - \vel_+\right) \ .
\label{eqJumpArplus}
\end{equation}
After some algebra, we can rewrite this in the equivalent form
\begin{equation}
\mathscr{M}_{\rm rel+}^{-2} Q^2= \frac{\gamma_{\rm rel}}{\gamma_{\rm th}}\left(Q-1\right) \mathscr{M}_{\rm th-}^{-2}
+ Q \, \mathscr{M}_{\rm rel-}^{-2} + \gamma_{\rm rel} \, Q \left(1-Q\right) \ ,
\label{eqJumpArplus1}
\end{equation}
where
\begin{equation}
\mathscr{M}_{\rm th-} \equiv \frac{\vel_-}{a_{\rm th-}} \ , \qquad
\mathscr{M}_{\rm rel-} \equiv \frac{\vel_-}{a_{\rm rel-}} \ ,
\label{eqMachGR}
\end{equation}
denote the upstream Mach numbers associated with the thermal gas and relativistic particle sound speeds, respectively. We can also combine our two expressions for the thermal entropy jump ratio (equations~\ref{eqJumpEnt1} and \ref{eqJumpEnt2}) to obtain another relation between the upstream and downstream relativistic particle sound speeds, given by
\begin{equation}
\frac{a^4_{\rm rel+}}{a^4_{\rm rel-}} = \frac{\vel^2_+}{\vel^2_-} \frac{\frac{\gamma_{\rm th}}{\gamma_{\rm rel}}a^2_{\rm rel+}+a^2_{\rm th-}}
{\frac{\gamma_{\rm th}}{\gamma_{\rm rel}}a^2_{\rm rel-}+a^2_{\rm th-}} \ ,
\label{eqShockJump1}
\end{equation}
which can be rewritten in terms of the Mach numbers as
\begin{equation}
\mathscr{M}^{-4}_{\rm rel+} Q^4 = \mathscr{M}^{-4}_{\rm rel-} Q^2 \left(\frac{Q^2 \frac{\gamma_{\rm th}}{\gamma_{\rm rel}}
\mathscr{M}^{-2}_{\rm rel+} + \mathscr{M}^{-2}_{\rm th-}}
{\frac{\gamma_{\rm th}}{\gamma_{\rm rel}} \mathscr{M}^{-2}_{\rm rel-} + \mathscr{M}^{-2}_{\rm th-}}\right) \ .
\label{eqShockJump2}
\end{equation}

Eliminating $\mathscr{M}_{\rm rel+}$ between equations~(\ref{eqJumpArplus1}) and (\ref{eqShockJump2}), one obtains, after some algebra, a quartic equation for the isothermal shock velocity jump ratio, $Q=\vel_+/\vel_-$, in terms of the upstream Mach numbers, $\mathscr{M}_{\rm th-}$ and $\mathscr{M}_{\rm rel-}$. One root of the quartic equation is the trivial upstream root, $Q=1$. We can therefore divide the quartic equation by the factor $(Q-1)$ to obtain the reduced cubic equation, 
\begin{equation}
Q^3{\cal F} + Q^2{\cal H} + Q{\cal I} + {\cal J} = 0 \ ,
\label{eqQJumpCubic}
\end{equation}
where
\begin{equation}
\begin{split}
{\cal F} & =1+\gamma_{\rm rel}^{-1}\mathscr{M}_{\rm rel-}^{-2}-\left(\gamma_{\rm rel}\mathscr{M}_{\rm rel-}^2
+ \gamma_{\rm th}\mathscr{M}_{\rm th-}^2\right)^{-1} \ ,\\
{\cal H} & =-2\gamma_{\rm th}^{-1}\mathscr{M}_{\rm th-}^{-2}-\gamma_{\rm rel}^{-2}\mathscr{M}_{\rm rel-}^{-4}
\left(1+\gamma_{\rm rel}\mathscr{M}_{\rm rel-}^2\right)^2 \ ,\\
{\cal I} & =\gamma_{\rm th}^{-2}\mathscr{M}_{\rm th-}^{-4}\left[2\gamma_{\rm th}\mathscr{M}_{\rm th-}^2
\left(1+\gamma_{\rm rel}^{-1}\mathscr{M}_{\rm rel-}^{-2}\right) +1\right] \ ,\\
{\cal J} & =-\gamma_{\rm th}^{-2}\mathscr{M}_{\rm th-}^{-4} \ .
\label{eqQjumpCubic1}
\end{split}
\end{equation}
Only one of the three solutions is physically valid. 

\begin{figure}
\includegraphics[width=4in]{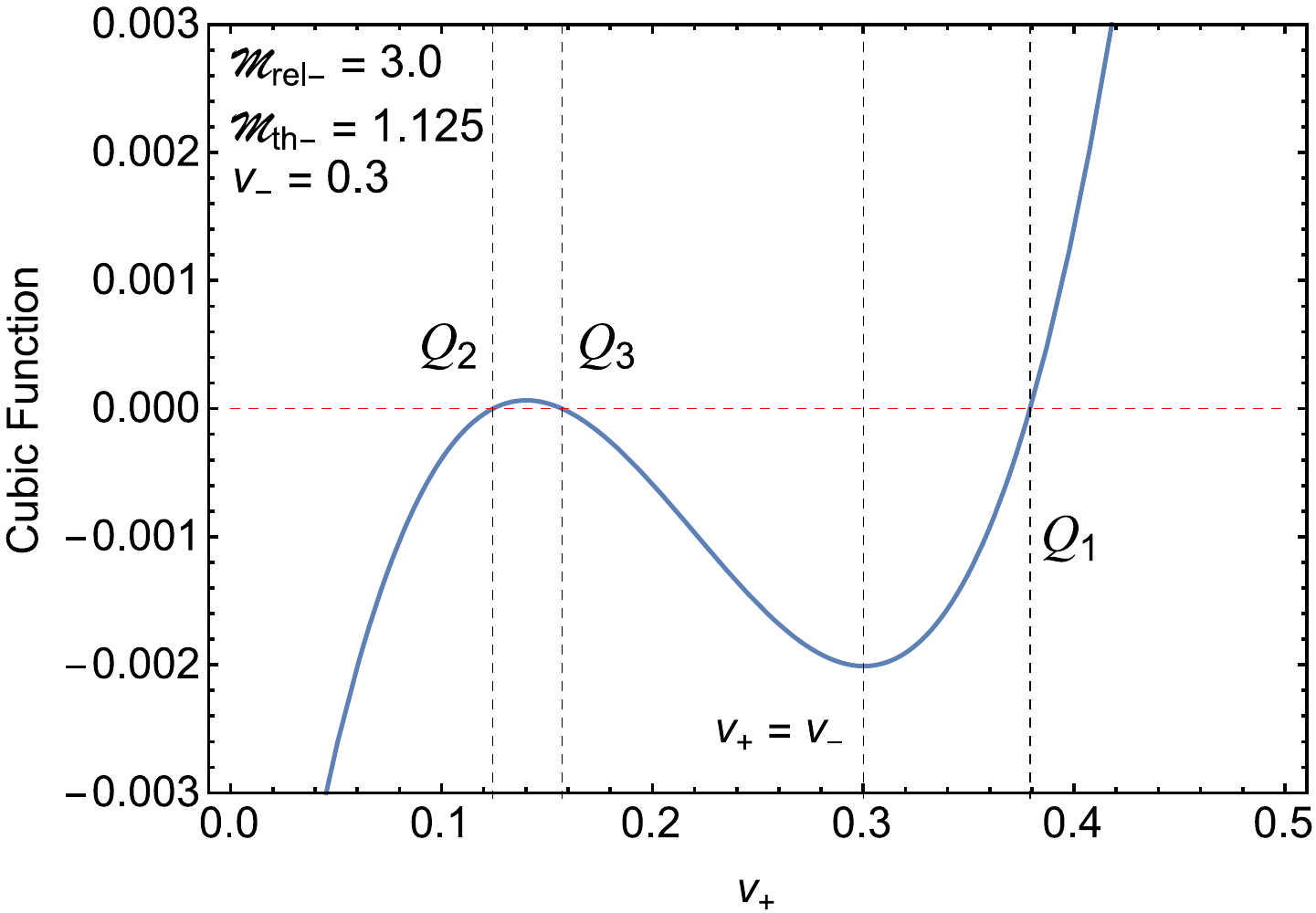}
\caption{A depiction of the cubic function (equation~\ref{eqQJumpCubic}, solid line), the roots of which determine the shock velocity jump $Q=\vel_+/\vel_-$, for a typical set of the parameters $\mathscr{M}_{\rm th-}$, $\mathscr{M}_{\rm rel-}$, and $\vel_-$. Only the root $Q_3$ is physically acceptable; see the discussion in the text.}
\label{fig:fig3}
\end{figure}

The three solutions to the cubic equation are given by,
\begin{equation}
\begin{split}
Q_1& =S+T-\frac{1}{3}\frac{\cal H}{\cal F} \ ,\\
Q_2& =-\frac{1}{2}\left(S+T\right)-\frac{1}{3}\frac{\cal H}{\cal F}+\frac{1}{2}i\sqrt{3}\left(S-T\right) \ ,\\
Q_3& =-\frac{1}{2}\left(S+T\right)-\frac{1}{3}\frac{\cal H}{\cal F}-\frac{1}{2}i\sqrt{3}\left(S-T\right) \ ,
\label{eqQ1}
\end{split}
\end{equation}
where 
\begin{equation}
\begin{split}
S& = \left(X+\sqrt{W^3+X^2}\right)^{1/3} \ , \\
T& = \left(X-\sqrt{W^3+X^2}\right)^{1/3} \ ,
\end{split}
\end{equation}
and
\begin{equation}
\begin{split}
W &= \frac{1}{9}\left(\frac{3\cal I}{\cal F} - \frac{{\cal H}^2}{{\cal F}^2}\right) \ ,\\
X &=\frac{1}{54}\left(\frac{9{\cal H}{\cal I}}{{\cal F}^2} - \frac{27\cal J}{\cal F} - \frac{2{\cal H}^3}{{\cal F}^3}\right) \ .
\end{split}
\end{equation}
For an arbitrary set of parameters (see e.g. Fig. \ref{fig:fig3}), numerical evaluation of the three roots shows that $Q_1$ is unphysical because $Q_1 > 1$, which implies the existence of an ``anti-shock'' with $\vel_+ > \vel_-$, which is impossible because it would violate the second law of thermodynamics. The remaining two roots, $Q_2$ and $Q_3$, are both less than unity as required. However, when these two roots are substituted into equation~(\ref{eqJumpArplus}) for the downstream relativistic particle sound speed, one obtains $a^2_{\rm rel+} < 0$ for $Q_2$ and $a^2_{\rm rel+} > 0$ for $Q_3$. Since we must have a positive value for $a^2_{\rm rel+}$, it follows that the only physically valid root for the shock velocity jump ratio is $Q=Q_3$, evaluated using the third relation in equations~(\ref{eqQ1}). Despite the appearance of the imaginary number $i$ in equation~(\ref{eqQ1}), it is worth noting that the value of $Q_3$ is real, because $S$ and $T$ are complex conjugates.

\subsection{Energy transport}
\label{sec:entrans}

In the dynamical model considered by LB05, all of the terms in the energy transport rate $\dot E$ reflect contributions due the thermal background gas. However, in the situation considered here, both the relativistic particles and the gas contribute to $\dot E$. Hence we need to develop separate energy jump conditions for each of these two populations. According to equation~(\ref{eqEdot}), the total energy transport rate is given by the sum of the gas and particle components, $\dot E = \dot E_{\rm th} + \dot E_{\rm rel}$. In the inviscid case treated here, the energy transport rate for the gas, $\dot E_{\rm th}$, is given by (cf. equations~\ref{eqEdotG} and \ref{eqEdotR})
\begin{equation}
\dot E_{\rm th} = \dot M\left(\frac{1}{2} \vel^2 + \frac{1}{2}\frac{\ell_0^2}{r^2} + \Phi
+ \frac{\agsq}{\gamma_{\rm th}-1}\right) \ ,
\label{}
\end{equation}
and the energy transport rate for the relativistic particles, $\dot E_{\rm rel}$, can be written as
\begin{equation}
\dot E_{\rm rel} = \dot M\left(\frac{\arsq}{\gamma_{\rm rel}-1} + \frac{\kappa}{\rho \vel}\frac{dU_{\rm rel}}{dr}\right) \ .
\label{eqErA}
\end{equation}
The jump in the total energy transport rate at the shock, $\dot E$, can be broken in gas and particle components by writing
\begin{equation}
\Delta\dot E = \Delta\dot E_{\rm th} + \Delta\dot E_{\rm rel} \ ,
\label{eqDeltaEdot}
\end{equation}
where 
\begin{equation}
\Delta\dot E_{\rm th} = \dot M\left(\frac{1}{2}\Delta \vel^2 + \frac{\Delta\agsq}{\gamma_{\rm th}-1}\right) \ ,
\label{eqDeltaEG}
\end{equation}
and
\begin{equation}
\Delta\dot E_{\rm rel} = \dot M\left[\frac{\Delta\arsq}{\gamma_{\rm rel}-1}
+ \Delta\left(\frac{\kappa}{\rho \vel}\frac{dU_{\rm rel}}{dr}\right)\right] \ .
\label{eqDeltaEr}
\end{equation}

In order to properly conserve energy in our model, the jump in $\dot E$ at the shock must be equal to the energy that is fed into the jet outflow due to the escape of accelerated relativistic particles from the disc at the shock location. This energy conservation principle is expressed by the statement
\begin{equation}
L_{\rm jet} = - \Delta\dot E \ \propto \ {\rm ergs \ s}^{-1} \ ,
\label{eqDeltaEnew}
\end{equation}
where $L_{\rm jet}$ is the kinetic luminosity of the jet. Next we must ask how the total energy jump $\Delta\dot E$ is distributed between the relativistic particles and the thermal gas. The relativistic particle energy transport rate, $\Delta E_{\rm rel}$, must be continuous across the shock, as shown by Becker \& Kazanas (2001). If this were not the case, then it would imply the presence of a discontinuity in $U_{\rm rel}$, which would create an infinite (and therefore patently unphysical) energy flux at the shock. Hence we must have
\begin{equation}
\Delta\dot E_{\rm rel} = 0 \ .
\label{eqDUrDrJump2}
\end{equation}
This relation indicates that there is a balance between the energy injection rate from the background thermal flow into the relativistic particle population and the escape of particle energy from the disc at the shock location.

Proceeding, we recall that in an isothermal shock, $\Delta a_{\rm th}^2=0$, and therefore equation~(\ref{eqDeltaEG}) reduces to
\begin{equation}
\Delta\dot E_{\rm th} = \frac{1}{2}\dot M \Delta \vel^2 \ .
\label{eqDeltaEG2}
\end{equation}
By combining equations~(\ref{eqDeltaEdot}), (\ref{eqDeltaEnew}), (\ref{eqDUrDrJump2}), and (\ref{eqDeltaEG2}), we are led to the conclusion that
\begin{equation}
L_{\rm jet} = - \Delta\dot E_{\rm th} = - \frac{1}{2}\dot M \Delta \vel^2 \ .
\label{eqLshock}
\end{equation}
In terms of the dimensionless energy transport rate, $\epsilon \equiv \dot E/\dot M$, we obtain the equivalent result
\begin{equation}
L_{\rm jet} = - \dot M \Delta\epsilon > 0 \ ,
\label{eqLshock2}
\end{equation}
where
\begin{equation}
\Delta\epsilon \equiv \epsilon_+ - \epsilon_- = \frac{\vel_+^2 - \vel_-^2}{2} < 0 \ .
\label{eqJump4Red}
\end{equation}
Equation~(\ref{eqJump4Red}) allows us to compute the jump in the dimensionless energy transport rate $\epsilon$ in terms of the velocity jump, which is evaluated using the final relation in equations~(\ref{eqQ1}) for the velocity jump ratio $Q_3$.

\subsection{Flow structure and shock location}

We know from the work of Chakrabarti (1989), Abramowicz \& Chakrabarti (1990), Das et al. (2001), and LB05 that for a shock to exist in the flow, it must be located between two critical points, and it must satisfy the jump conditions given by equations~(\ref{eqJumpArplus}) for the downstream relativistic particle sound speed, equation~(\ref{eqQ1}) for the velocity jump ratio $Q_3$, and equation~(\ref{eqJump4Red}) for energy transport rate jump. In the original one-fluid, adiabatic model studied by these authors, the determination of the shock location in the flow was carried out by finding the root of an algebraic equation. The situation is not so simple once we have implemented the diffusive, two-fluid model being developed here. Instead, we must employ an iterative procedure involving the initialization of the flow variables at an inner boundary near the event horizon, followed by numerical integration in the outward direction of the coupled equations~(\ref{eqDAgDr}) and (\ref{eqEps3alt}) for the first-order derivatives of the thermal and relativistic particle sound speeds, respectively, supplemented by the algebraic relation given by equation~(\ref{eqVfromKg}) for the inflow velocity. For a given flow structure, we must then determine if a shock can be placed in the flow at any candidate shock radius, denoted by $r_*$, that lies beyond the inner sonic point, located at $r_{c3}$. Once the candidate shock location is selected, the integration is continued in the outward direction, starting on the upstream side of the shock. In the outer region, the flow must pass through another critical point at radius $r_{c1}$, beyond which the flow is subsonic out to infinity. At a very large distance, the flow should approach zero accretion velocity, and the sound speeds should approach constant values, indicative of conditions in the surrounding medium. We summarize the various steps in the simulation process below.

The procedure begins by selecting the values for the fundamental parameters $\epsilon_+$, $\ell_0$, $\kappa_0$, and $K_{\rm th}/K_{\rm rel}$. We set the adiabatic indices for the thermal gas and the relativistic particles using $\gamma_{\rm th}=3/2$ and $\gamma_{\rm rel}=4/3$, respectively. The next step is to utilize the double-adiabatic model (see Section \ref{sec:DAcritpoint}) in the inner region to establish the flow structure near the horizon, starting at radius $r=2.1\,GM/c^2$. The double-adiabatic model provides an accurate approximation to the diffusive model close to the event horizon since diffusion becomes negligible compared to advection as $r \to 2\,GM/c^2$ (e.g., Weinberg 1972). The location of the inner critical point at radius $r_{c3}$ is computed using the double-adiabatic model, as outlined in Section \ref{sec:DAcritpoint}. Once $r_{c3}$ is known, we can compute the associated values for the critical velocity, $\vel_{c3}$, the critical gas sound speed, $a_{\rm th,c3}$, the critical relativistic particle sound speed, $a_{\rm rel,c3}$, and the critical gas entropy parameter, $K_{\rm th,c3}$. As explained in Section \ref{sec:DAflowsol}, the radial profile for the gas sound speed $a_{\rm th}(r)$ in the supersonic region $2.1\,GM/c^2 < r < r_{c3}$ is computed using a root-finding procedure based on the double-adiabatic energy equation~(\ref{eqEpsMod2Ag}), and the relativistic particle sound speed $a_{\rm rel}(r)$ is then calculated using equation~(\ref{eqArAgfunMod2}).

The double-adiabatic model accurately describes the structure of the accretion disc from the event horizon out to the inner critical radius at $r=r_{c3}$, including a determination of the gas entropy parameter, $K_{\rm th,c3}$. In the region beyond the inner critical radius, the effects of diffusion become important, and therefore the determination of the flow structure for $r > r_{c3}$ requires integration of the diffusive wind equation~(\ref{eqDAgDr}), coupled with equations~(\ref{eqEps3alt}) and (\ref{eqVfromKg}), as discussed in Section \ref{DIFFwindeq}. The determination of the value of $K_{\rm th,c3}$ using the double-adiabatic model is important because the gas entropy parameter $K_{\rm th}$ is a global constant throughout the disc (except at the shock location), and therefore this value carries over into the integration of the two-fluid conservation equations in the region $r > r_{c3}$. In order to transition to the diffusive two-fluid model, we need to recompute the critical quantities $\vel_{c3}$, $a_{\rm th,c3}$, and $a_{\rm rel,c3}$ in a manner consistent with the diffusive critical conditions. We can accomplish this by solving equations~(\ref{eqNcMod3}) for the critical conditions, (\ref{eqVcMod3}) for the critical velocity, and (\ref{arcQuad2}) for the relativistic sound speed at the critical point, based on the assumption that the values of $r_{c3}$ and $K_{\rm th,c3}$ can be adopted directly from the double-adiabatic model. The diffusive and double-adiabatic models are expected to agree in this region of the disc, where diffusion is supposed to be negligible, and therefore we should expect to find little change in the critical quantities when we move from the double-adiabatic critical conditions to the diffusive conditions. A comparison between the two sets of critical quantities is therefore an interesting way to test of the integrity of the model. In our numerical applications, we find that the two sets of critical quantities are extremely close to each other, which helps to validate our model assumptions.

Once the flow structure has been established in the inner region, $2.1\,GM/c^2 \le r \le r_{c3}$, the integration of the coupled equations~(\ref{eqDAgDr}), (\ref{eqEps3alt}) and (\ref{eqVfromKg}) must be continued in the subsonic region, $r > r_{c3}$. This integration begins at a radius just outside $r_{c3}$, which is slightly offset from the precise critical point, at $r=r_{c3}$, because that is a singular point of the differential equations. In order to offset the starting location from the critical point, we need to employ linear extrapolation to compute corrected values for the flow variables, which requires knowledge of the derivatives of the flow variables at $r_{c3}$. The required derivatives are computed using L'H\^opital's rule, following essentially the same procedure employed by LB05. Once the integration of equation~(\ref{eqDAgDr}), (\ref{eqEps3alt}) and (\ref{eqVfromKg}) has been completed in the region $r > r_{c3}$, the next step is to adopt a candidate (provisional) value for the shock radius, $r_*$, which provides associated values for the post-shock quantities $\vel_+$, $a_{\rm th+}$, and $a_{\rm rel+}$. Since the thermal gas behaves adiabatically in the post-shock region, it follows that
\begin{equation}
K_{\rm th+} = K_{\rm th,c3} \ ,
\label{eqKcon}
\end{equation}
where $K_{\rm th,c3}$ is computed using the double-adiabatic model, as discussed above. It is important to emphasize that the integration to determine the disc structure proceeds in the outward direction, and therefore we need to employ ``reverse'' jump conditions in order to cross over the shock from the downstream side, with quantities $\vel_+$, $a_{\rm th+}$, $a_{\rm rel+}$, $K_{\rm th+}$, and $\epsilon_+$ to the upstream side, with quantities $\vel_-$, $a_{\rm th-}$, $a_{\rm rel-}$, $K_{\rm th-}$, and $\epsilon_-$. The necessary reverse-jump conditions are derived in Appendix~\ref{AppendixReverseJump} by exploiting the symmetry of the momentum and mass conservations relations at the shock. The value of the upstream velocity $\vel_-$ is obtained using the third relation in equations~(\ref{eqQ1b}) for the inverse velocity jump ratio, $Q_* \equiv \vel_-/\vel_+$ (see equation~\ref{eqJumpRatReverse}), and then $K_{\rm th-}$, $a_{\rm rel-}$, and $\epsilon_-$ are computed using equations~(\ref{eqJumpEnt2}) for the thermal entropy jump, (\ref{eqShockJump1}) for the relativistic particle sound speed jump, and (\ref{eqJump4Red}) for the energy jump, respectively. Since the shock is assumed to be isothermal, there is no jump in $a_{\rm th}$, and therefore $a_{\rm th-}=a_{\rm th+}$ (see equation~\ref{eqIsoAg}).

Beyond the shock radius, in the region $r > r_*$, the integration is continued until an outer critical point $r_{c1}$ is determined, which is defined as the location where both $N$ and $D$ vanish (equations~\ref{eqNDMod3}). Through an iterative process, the value of the shock radius $r_*$ is varied until $N$ and $D$ vanish at the same location, which is then identified as the outer critical radius, $r_{c1}$. Outside the outer critical point, in the region $r > r_{c1}$, the integration is continued using the same linear extrapolation method based on L'H\^opital's rule that was applied at the inner sonic point, $r_{c3}$. The analysis of the shock location discussed above allows us to compute the structure of shocked disc solutions for a given set of parameters $\epsilon_+$, $\ell_0$, $\kappa_0$, and $K_{\rm th}/K_{\rm rel}$. The dynamical results derived using this iterative procedure are used in Section \ref{sec:astrapp} to model the outflows observed in M87 and \sgr.

\section{ASYMPTOTIC BEHAVIOURS}
\label{sec:asymp}

In the diffusive, two-fluid model considered here, the structure of the accretion disc is determined by solving numerically a system of hydrodynamical conservation equations, which includes two differential equations and one algebraic relation. The solution of the set of equations is complicated by the fact that the event horizon at radius $r=\rs$ is also a singular point of the equations. Hence the starting point for the outward integration cannot be the horizon itself, but instead it must be offset slightly from the horizon. In our astrophysical applications, the starting point for the outward integration is $2.1\,GM/c^2$. In order to integrate the system of equations, we must specify values for the physical variables at the starting radius, and this in turn requires the development and utilization of a set of asymptotic relations that describe conditions near the horizon. Likewise, we will also need to analyze the asymptotic behaviours of the physical variables at a large distance from the black hole, in order to ensure that the results obtained using the outward integration are physically reasonable. We discuss the required asymptotic relations here, with further details provided in Appendix~\ref{AppendixAsymptoticRelations}.

\subsection{Asymptotic behaviour near the horizon}
\label{sec:asympHor}

Near the event horizon, the radial velocity $\vel$ approaches the free-fall velocity $\vel^2_{\rm ff}(r) \equiv 2GM/(r-\rs)$, so that (Becker \& Le 2003)
\begin{equation}
\vel^2(r) \propto (r-\rs)^{-1} \ ,\qquad r\to\rs \ .
\label{eqAsyRS1a}
\end{equation}
It should be noted that since the velocity $\vel$ diverges as $r \to \rs$ it is more correctly interpreted as the radial component of the four-velocity (Becker \& Le 2003; Becker \& Subramanian 2005). We show in Appendix~\ref{AppendixAsymptoticRelations} that diffusion is negligible near the event horizon, and therefore this region is accurately described by the double-adiabatic model discussed in Section~\ref{DAwindeq}. It follows that near the horizon, equation~(\ref{eqAsyRS1a}) can be combined with equations~(\ref{eqKg}) for the gas entropy parameter and equation~(\ref{eqArAgfunMod2}) expressing the adiabatic relation between the thermal and relativistic sounds speeds to conclude that the asymptotic behaviour of the thermal sound speed, $a_{\rm th}$, near the horizon is given by
\begin{equation}
a_{\rm th}^2(r) \propto (r-\rs)^{\left(1-\gamma_{\rm th}\right)/\left(1+\gamma_{\rm th}\right)} \ ,\qquad r\to\rs \ .
\label{eqAsyRS1b}
\end{equation}
The corresponding asymptotic variations of the disc half-thickness $H$ (equation~\ref{eqH}) and the gas density $\rho$ (equation~\ref{eqTranM}), respectively, can be written as
\begin{equation}
H(r) \propto (r-\rs)^{\left(\gamma_{\rm th}+3\right)/\left(2\gamma_{\rm th}+2\right)} \ , \qquad
\rho(r) \propto (r-\rs)^{-1/\left(\gamma_{\rm th}+1\right)} \ ,\qquad r\to\rs \ .
\label{eqAsyRS2}
\end{equation}
Close to the event horizon, the particle transport is dominated by advection rather than diffusion, and therefore the relativistic particle distribution evolves adiabatically in this region. The corresponding asymptotic forms for the relativistic particle number and energy densities are given by (see Appendix~\ref{AppendixAsymptoticRelations} for further details)
\begin{equation}
n_{\rm rel}(r) \propto \left(r-\rs\right)^{-1/\left(\gamma_{\rm th}+1\right)} \ ,\qquad r\to\rs \ ,
\label{eqNasympIn}
\end{equation}
and for the energy density
\begin{equation}
U_{\rm rel}(r) \propto \left(r-\rs\right)^{-4/\left(3\gamma_{\rm th}+3\right)} \ ,\qquad r\to\rs \ ,
\label{eqUasympIn}
\end{equation}
respectively.

\subsection{Asymptotic behaviour at infinity}

In the limit $r\to\infty$, the gas and relativistic particle sound speeds, $a_{\rm th}$ and $a_{\rm rel}$, respectively, are expected to approach constants indicative of the surrounding medium. In this case, we show in Appendix~\ref{AppendixAsymptoticRelations} that the constancy of the gas entropy $K_{\rm th}$ (equation~\ref{eqKg}) implies that the accretion velocity $\vel$ varies as
\begin{equation}
\vel(r) \propto r^{-5/2} \ , \qquad r \to \infty \ .
\label{eqAsyInf2}
\end{equation}
We also conclude that the disc half-thickness $H$ (equation~\ref{eqH}) and the gas density $\rho$ (equation~\ref{eqTranM}) vary as
\begin{equation}
H(r) \propto r^{3/2} \ , \qquad \rho \to {\rm constant} \ , \qquad r \to \infty \ .
\label{eqAsyInf3}
\end{equation}
Far from the black hole, the accretion velocity diminishes to zero according to equation~(\ref{eqAsyInf2}), and therefore one expects that the particle transport in the disc is dominated by outward-bound diffusion. In Appendix~\ref{AppendixAsymptoticRelations}, we use asymptotic analysis of the transport equation at a large distance from the black hole to demonstrate that the variations of the relativistic particle number density $n_{\rm rel}$ and energy density $U_{\rm rel}$ are given by
\begin{equation}
n_{\rm rel} \approx n_{\rm rel,\infty} \left(\frac{C_1}{r} + 1\right) \ , \qquad r \to \infty \ ,
\label{eqNasympOut}
\end{equation}
and
\begin{equation}
U_{\rm rel} \approx U_{\rm rel,\infty} \left(\frac{C_1}{r} + 1\right) \ , \qquad r \to \infty \ ,
\label{eqUasympOut}
\end{equation}
where $n_{\rm rel,\infty}$ and $U_{\rm rel,\infty}$ denote values measured at infinity, and the constant $C_1$ is set by requiring that the solution for $U_{\rm rel}$ agree with the dynamical solution for the relativistic particle energy density.

\section{RELATIVISTIC PARTICLE TRANSPORT}
\label{sec:relmom}

Our primary goal in this paper is to analyze the transport and acceleration of relativistic ions in an advection-dominated accretion disc, and to understand the effects of particle pressure and particle diffusion on the dynamical structure of the disc. As such, our focus up to this point has been on the dynamical equations that describe the disc structure. By solving these equations via numerical integration, we are able to obtain the profiles of the physical variables $\vel(r)$, $a_{\rm th}(r)$, and $a_{\rm rel}(r)$. In particular, the profile of the relativistic particle sound speed $a_{\rm rel}(r)$ can be used to compute the particle pressure using equation~(\ref{eqRelP1}). However, the particle pressure alone does not provide a complete picture of the particle propagation in the disc, or the energetics of the associated outflows. In order to obtain a complete understanding, we need to consider the particle transport equation governing the distribution function for the relativistic ions. In the model considered here, the particle transport equation includes terms describing spatial diffusion, Fermi energization, advection, and particle escape. This is essentially the same formalism considered by LB05, except they applied it in the context of a one-fluid dynamical model, whereas we will apply it in the context of the two-fluid model considered here.

A full analysis of the transport equation, including a solution for the relativistic particle Green's function, is deferred to Paper~II. However, we can gain some useful insight by examining the ``moment'' equations that are obtained by integrating the full transport equation with respect to the particle energy, $E$. By analyzing the resulting ordinary differential equations, we can obtain solutions for the profiles of the relativistic particle number and energy densities, $n_{\rm rel}(r)$ and $U_{\rm rel}(r)$, respectively. The number density profile provides us with the additional information we need to understand the energetics of the particle population in the disc and the outflow, which is assumed to escape from the disc at the shock radius, $r_*$. The solution obtained for the particle energy density, $U_{\rm rel}$, is also important because it provides us with a self-consistency check on the entire model, since the associated particle pressure, $P_{\rm rel}=(\gamma_{\rm rel}-1) U_{\rm rel}$, must equal the particle pressure computed using the dynamical conservation equations (see equation~\ref{eqRelP1}).

\subsection{Relativistic particle number density}

The governing transport equation for the particle number density, $n_{\rm rel}(r)$, is given by LB05 as
\begin{equation}
\frac{d\dot N_{\rm rel}}{dr} = \dot N_0\delta(r-r_*) - 4 \pi r_* H_* A_0 c n_{\rm rel}(r-r_*) \ ,
\label{eqTransNr}
\end{equation}
where $A_0$ is the dimensionless escape parameter (see Appendix~\ref{AppenShockWidth}), and $\dot N_{\rm rel}(r)$ denotes the relativistic particle transport rate, which is related to $n_{\rm rel}$ via
\begin{equation}
\dot N_{\rm rel}(r) \equiv - 4\pi rH\left(\vel n_{\rm rel} + \kappa\frac{dn_{\rm rel}}{dr}\right) \ .
\label{eqNr}
\end{equation}
Note that $\dot N_{\rm rel} > 0$ for outwardly-directed transport, and $\dot N$ is discontinuous at the source/shock radius $r_*$ due to the combined effect of particle injection and escape. The first and second terms on the right-hand side of equation~(\ref{eqNr}) represent particle advection and diffusion, respectively, and we remind the reader that $\vel > 0$ for inflow. The particle transport has two spatial regions in the calculations, designated domain I ($r > r_*$) and domain II ($r < r_*$), and the global solution is written in the form
\begin{equation}
\dot N_{\rm rel}(r) = \begin{cases}
\dot N_{\rm I} \ , & r>r_* \ ,\\ 
\dot N_{\rm II} \ , & r<r_* \ ,\\ 
\end{cases}
\label{}
\end{equation}
where $\dot N_{\rm I} > 0$ and $\dot N_{\rm II} < 0$ denote the rates at which particles are radially transported outward along the disc and inward toward the event horizon, respectively, from the source location. Integrating equation~(\ref{eqTransNr}) in a very small region around $r=r_*$ gives the magnitude of the jump in the particle transport rate,
\begin{equation}
\dot N_{\rm I} - \dot N_{\rm II} = \dot N_0 - \dot N_{\rm esc} \ ,
\qquad \dot N_{\rm esc} \equiv 4\pi r_*H_*A_0 c n_* \ ,
\label{eqNesc}
\end{equation}
where $n_*\equiv n_{\rm rel}(r_*)$, and $\dot N_{\rm esc}$ is the positive rate at which particles escape the disc at the shock location in order to form the jet outflow. In Appendix~\ref{AppendixEnergyMoments}, we demonstrate that the vertically-averaged transport equation for the total relativistic number density (e.g. equation~\ref{eqVertIntTransNr}) is given as,
\begin{equation}
H \vel_r \frac{dn_{\rm rel}}{dr} = -\frac{n_{\rm rel}}{r}\frac{d}{dr}\left(rH \vel_r\right) + \frac{1}{r}\frac{d}{dr}\left(rH\kappa\frac{dn_{\rm rel}}{dr}\right) + \frac{\dot N_0 \delta(r-r_*)}{4\pi r_*} - A_0 c H_* n_* \delta(r-r_*) \ ,
\label{eqVertIntTransNr2}
\end{equation}
where $\vel_r \equiv - \vel < 0$. Equation~(\ref{eqVertIntTransNr2}) can also be obtained by combining equations~(\ref{eqTransNr}) and (\ref{eqNr}). It should be noted that the discontinuity in $\dot N_{\rm rel}$ at the shock location produces a corresponding jump in the derivative $dn_{\rm rel}/dr$ by virtue of equation~(\ref{eqNr}).

The global solution for the particle number density $n_{\rm rel}=I_2$ can be expressed as
\begin{equation}
n_{\rm rel}(r) = \begin{cases}
A Q_{\rm I}(r) \ , & r \ge r_* \ , \\
B Q_{\rm II}(r) \ , & r \le r_* \ ,
\end{cases}
\label{eqGlobalNr}
\end{equation}
where $A$ and $B$ are normalization constants, and the functions $Q_{\rm I}(r)$ and $Q_{\rm II}(r)$ satisfy the homogeneous differential equation~(cf. equation~\ref{eqVertIntTransNr2}),
\begin{equation}
H \vel_r \frac{dQ}{dr} = -\frac{Q}{r}\frac{d}{dr}\left(rH \vel_r\right)
+ \frac{1}{r}\frac{d}{dr}\left(rH\kappa\frac{dQ}{dr}\right) \ ,
\label{eqHomoDifEqNr}
\end{equation}
coupled with the boundary conditions (see equations~\ref{eqNasympIn} and \ref{eqNasympOut})
\begin{equation}
Q_{\rm I}(r_{\rm out}) = \frac{C_1}{r_{\rm out}} + 1 \ , \qquad
Q_{\rm II}(r_{\rm in}) = \left(\frac{r_{\rm in}}{\rs}-1\right)^{-1/(\gamma_{\rm th}+1)} \ ,
\label{eqBoundaryQNr}
\end{equation} 
where $C_1$ is a constant and $r_{\rm in}$ and $r_{\rm out}$ denote the radii at which the inner and outer boundary conditions are applied, respectively.

The constants $A$ and $B$ are determined by setting $n=2$ in equations~(\ref{eqAIn}) and (\ref{eqBIn}), which yields
\begin{equation}
A = B \frac{Q_{\rm II}} {Q_{\rm I}} \Bigg|_{r=r_*} \ ,
\label{}
\end{equation}
\begin{equation}
B = \frac{\dot N_0}{4\pi r_*} Q_{\rm I}\left[(H_+ \vel_+ - H_- \vel_-) Q_{\rm I} Q_{\rm II}
- H_- \kappa_- Q_{\rm II}Q'_{\rm I} + H_+ \kappa_+ Q_{\rm I}Q'_{\rm II} + H_* A_0 c
Q_{\rm II} Q_{\rm I}\right]^{-1} \Bigg|_{r=r_*} \ ,
\label{}
\end{equation}
where the primes denote differentiation with respect to radius. The solutions for the functions $Q_{\rm I}(r)$ and $Q_{\rm II}(r)$ are obtained by numerically integrating equation~(\ref{eqHomoDifEqNr}), subject to the boundary conditions given by equations~(\ref{eqBoundaryQNr}). Once the constants $A$ and $B$ are computed, the global solution for $n_{\rm rel}(r)$ is evaluated using equation~(\ref{eqGlobalNr}). This completes the solution procedure for the relativistic particle number density $n_{\rm rel}(r)$.

\subsection{Relativistic particle energy density}

The differential equation that is satisfied by the relativistic particle energy density $U_{\rm rel}$ (cf. equation~\ref{eqVertIntTransUr}) is given by
\begin{equation}
H \vel_r \frac{dU_{\rm rel}}{dr} = -\gamma_{\rm rel}\frac{U_{\rm rel}}{r}\left(rH \vel_r\right) + \frac{1}{r}\frac{d}{dr}\left(rH\kappa\frac{dU_{\rm rel}}{dr}\right) + \frac{\dot N_0 E_0\delta(r-r_*)}{4\pi r_*} - A_0 c H_* U_{\rm rel}\delta(r-r_*) \ ,
\label{eqVertIntTransUrV2}
\end{equation}
where $\vel_r \equiv -\vel$. This expression can be rewritten in flux-conservation form as
\begin{equation}
\frac{d\dot E_{\rm rel}}{dr} = 4\pi rH\left[\frac{\vel}{3}\frac{dU_{\rm rel}}{dr}
- \frac{\dot N_0 E_0 \delta(r-r_*)}{4\pi r_*H_*} + A_0 c U_{\rm rel}\delta(r-r_*)\right] \ ,
\label{eqFluxR}
\end{equation}
where
\begin{equation}
\dot E_{\rm rel}(r) \equiv 4\pi rH\left(\frac{4}{3} \vel U_{\rm rel} + \kappa\frac{dU_{\rm rel}}{dr}\right)
\label{eqEr}
\end{equation}
represents the radial energy transport rate for the relativistic particles, and $\dot E_{\rm rel} > 0$ for outwardly directed transport. We note that equation~(\ref{eqEr}) is equivalent to equation~(\ref{eqErA}).

In Section~\ref{sec:entrans}, we demonstrated that the relativistic particle energy transport rate, $\dot E_{\rm rel}$, is continuous at the shock radius, $r=r_*$, because the energy injected into the particle distribution from the source is balanced by the escape of particle energy into the outflow (see equation~\ref{eqDUrDrJump2}). We can therefore integrate equation~(\ref{eqFluxR}) in a very small region around $r=r_*$ to obtain
\begin{equation}
\Delta\dot E_{\rm rel} = -\lim_{\delta r\to 0}\int^{r_*+\delta r}_{r_*-\delta r} \frac{d\dot E_{\rm rel}}{dr}dr
= \dot N_0E_0 - L_{\rm esc} = 0 \ ,
\label{eqDeltaER2}
\end{equation}
where the negative sign appears because $\Delta$ represents the difference between the post-shock (``+'') and pre-shock (``-'') values, and
\begin{equation}
L_{\rm esc} \equiv 4\pi r_*H_*A_0cU_{\rm rel}(r_*) \ \propto \ {\rm ergs \ s}^{-1} \ .
\label{eqDeltaER3}
\end{equation}
In order to obtain Equation~(\ref{eqDeltaER2}), we also had to assume that $dU_{\rm rel}/dr$ is not singular at the shock location, so that the spatial diffusion flux is finite there. This condition implies that
\begin{equation}
\lim_{\delta r\to 0}\int^{r_*+\delta r}_{r_*-\delta r} \frac{\vel}{3}\frac{d U_{\rm rel}}{dr}dr = 0 \ .
\end{equation}
The power in the injected relativistic seed particles, $\dot N_0 E_0$, must come from the thermal disc population, and therefore we can also write (see equation~\ref{eqLshock2})
\begin{equation}
L_{\rm jet} = - \dot M \Delta\epsilon = \dot N_0E_0 = L_{\rm esc} \ .
\label{eqEnergyCons1}
\end{equation}
Equation~(\ref{eqEnergyCons1}) expresses the global conservation of energy in our model.

The global solution for $U_{\rm rel}(r)$, obtained by numerically solving equation~(\ref{eqVertIntTransUrV2}), can be written as
\begin{equation}
U_{\rm rel}(r) = \begin{cases}
A Q_{\rm I}(r) \ , & r \ge r_* \ , \\
B Q_{\rm II}(r) \ , & r \le r_* \ ,
\end{cases}
\label{eqGlobalUr}
\end{equation}
where $A$ and $B$ are normalization constants. The functions $Q_{\rm I}(r)$ and $Q_{\rm II}(r)$ satisfy the homogeneous differential equation (cf. equation~\ref{eqVertIntTransUrV2})
\begin{equation}
H \vel_r \frac{dQ}{dr} = -\gamma_{\rm rel}\frac{Q}{r}\left(rH\vel_r\right)
+ \frac{1}{r}\frac{d}{dr}\left(rH\kappa\frac{dQ}{dr}\right) \ .
\label{eqVertIntTransUrHomo}
\end{equation}
The associated boundary conditions for $Q_{\rm I}(r)$ and $Q_{\rm II}(r)$ are (see equations~\ref{eqUasympIn} and \ref{eqUasympOut})
\begin{equation}
Q_{\rm I}(r_{\rm out}) = \frac{C_1}{r_{\rm out}} + 1 \ , \qquad
Q_{\rm II}(r_{\rm in}) = \left(\frac{r_{\rm in}}{\rs}-1\right)^{-4/(3\gamma_{\rm th}+3)} \ ,
\label{eqBoundaryQUr}
\end{equation}
where $C_1$ is a constant and $r_{\rm in}$ and $r_{\rm out}$ denote the inner and outer boundaries for the integration, respectively. The value of $C_1$ is chosen to agree with the dynamical solution for $U_{\rm rel}$ far from the black hole.

The requirement that the diffusive flux must remain finite across the shock implies that $\Delta\dot E_{\rm rel}=0$ (see equation~\ref{eqDUrDrJump2}), which can be combined with equation~(\ref{eqDeltaEr}) for $\Delta\dot E_{\rm rel}$ to conclude that
\begin{equation}
\Delta\left(\frac{\kappa}{\rho \vel}\frac{dU_{\rm rel}}{dr}\right) = \frac{-\Delta\arsq}{\gamma_{\rm rel}-1} \ .
\label{eqDeltaErNEW}
\end{equation}
The constants $A$ and $B$ are determined by ensuring the derivative $dU_{\rm rel}/dr$ satisfies this jump condition, along with the condition that $U_{\rm rel}$ is continuous at $r=r_*$. The detailed derivation is carried out in Appendix~\ref{AppendixEnergyMoments}, and the results obtained for the case with $n=3$ are (see equations~\ref{eqAIn} and \ref{eqBIn})
\begin{equation}
A = B\frac{Q_{\rm II}}{Q_{\rm I}} \Bigg|_{r=r_*} \ ,
\label{eqUrAconst}
\end{equation}
\begin{equation}
B = \frac{\dot N_0 E_0}{4\pi r_*}{Q_{\rm I}}\left[\frac{4}{3} \left(H_+ \vel_+ - H_- \vel_-\right)
Q_{\rm II} Q_{\rm I} + H_+ \kappa_+ Q_{\rm I}Q'_{\rm II} - H_- \kappa_- Q_{\rm II} Q'_{\rm I}
+ A_0 H_* c Q_{\rm I} Q_{\rm II}\right]^{-1}\Bigg|_{r=r_*} \ ,
\label{eqUrBconst}
\end{equation}
where the primes denote differentiation with respect to radius. By substituting equations~(\ref{eqUrAconst}) and (\ref{eqUrBconst}) into equation~(\ref{eqGlobalUr}), we obtain the global solution for the relativistic particle energy density, $U_{\rm rel}(r)$. We can use the solution for $U_{\rm rel}(r)$ to check the self-consistency of our combined model for the acceleration and transport of relativistic particles in the disc, since the results obtained using equation~(\ref{eqGlobalUr}) must agree with the dynamical solution for $U_{\rm rel}(r)$ obtained from the hydrodynamical solution, given by equation~(\ref{eqRelP1}).

\section{ASTROPHYSICAL APPLICATIONS}
\label{sec:astrapp}

Our goal is to determine the properties of the integrated disc/shock/outflow model for a given source, based on observationally constrained values for the black hole mass mass $M$ and the jet kinetic power $L_{\rm jet}$. In all of our calculations, we set the specific heat ratios using $\gamma_{\rm th}=3/2$ and $\gamma_{\rm rel}=4/3$. We then vary the remaining model free parameters $\epsilon_+$, $\ell_0$, $\kappa_0$, and $K_{\rm th}/K_{\rm rel}$ in order to obtain models that comply with the observationally estimates for the source. The sound speed profiles $a_{\rm th}(r)$ and $a_{\rm rel}(r)$ are computed by numerically integrating equation~(\ref{eqDAgDr}) for the thermal sound speed derivative and equation~(\ref{eqEps3alt}) for the relativistic sound speed derivative, supplemented by equation~(\ref{eqVfromKg}) for the inflow velocity, which is used to compute the inflow velocity $\vel$ in terms of the sound speeds $a_{\rm th}$ and $a_{\rm rel}$. Once the profiles for $\vel(r)$, $a_{\rm th}(r)$, and $a_{\rm rel}(r)$ have been obtained, the corresponding pressure and energy density profiles for the gas and relativistic particles can be computed using equations~(\ref{eqThermalP1}) and (\ref{eqRelP1}). The computational procedure and the results obtained are discussed in further detail below.

\subsection{Model parameters}

Four different accretion/shock scenarios are explored in detail here. All of the model profiles are stated in terms of dimensional variables, and therefore they can be scaled to any black hole mass $M$. The simulations of the disc structure in M87 and \sgr are based on the published observational estimates for $M$, $\dot M$, and $L_{\rm jet}$ discussed below. In the case of M87, we set $M = 3 \times 10^9\,\msun$ (e.g., Ford et al. 1994), and for \sgr, we set $M = 2.6 \times 10^6\,\msun$ (e.g., Sch\"odel et al. 2002). For the kinetic luminosity of the outflow in M87, we adopt the value $L_{\rm jet} = 5.5 \times 10^{43}\,{\rm ergs \ s}^{-1}$ (Reynolds et al. 1996; Bicknell \& Begelman 1996; Owen et al. 2000). The kinetic luminosity of the outflow in \sgr is rather uncertain, and the published values encompass a wide range (e.g., Yuan 2000; Yuan et al. 2002). Here, we refer to the results of Falcke \& Biermann (1999), who obtained $L_{\rm jet} = 5 \times 10^{38}\,{\rm ergs \ s}^{-1}$. The values used for the theory parameters $\ell_0$, $\kappa_0$, $K_{\rm th}/K_{\rm rel}$, $\epsilon_+$, $\epsilon_-$, $r_{c1}$, $r_{c3}$, $r_*$, $H_*$, $R_*$ and $T_*$ in our four models are reported in Table \ref{table:table1}. Here, $T_*$ is the ion temperature at the shock location, which is related to the thermal pressure via $P_* = n_* kT_*$, where $k$ is Boltzmann constant, and the ion number density at the shock radius, $n_*$, is related to the mass density via $n_* = \rho(r_*)/m_p$, where $m_p$ is the proton mass. The parameters associated with the shock jump conditions, transport equation, and the specific sources, are reported in Tables \ref{table:table2}, \ref{table:table3}, and \ref{table:table4}, respectively.

As a point of departure for the new two-fluid model developed here, we will focus on the $\ell_0$ and $\kappa_0$ values that LB05 used to analyze M87 (their model 2) and \sgr (their model 5), which correspond to our Models A and B, respectively. With the values of $\ell_0$ and $\kappa_0$ thus determined, we proceed to vary the remaining parameters in our two-fluid model until we obtain the {\it maximum possible value} for the asymptotic (terminal) Lorentz factor of the escaping particles, given by
\begin{equation}
\Gamma_\infty = \frac{E_{\rm esc}}{m_p c^2} \ ,
\label{eqGammaInf}
\end{equation}
where the mean energy of the particles escaping at the shock location, $r=r_*$, is computed using
\begin{equation}
E_{\rm esc} \equiv \frac{U_{\rm rel}(r_*)}{n_{\rm rel}(r_*)} \ .
\label{eqEesc}
\end{equation}
The results obtained for $\Gamma_\infty$ are listed in Table~\ref{table:table3}. In the case of M87, our values for $\Gamma_\infty$ are in good agreement with Abdo et al. (2009), who estimated $\Gamma_\infty \sim 2.3$. In the case of \sgr, our results for $\Gamma_\infty$ agree fairly well with the findings of Yusef-Zadeh et al. (2012), who estimated that $\Gamma_\infty \sim 3$. In addition to Models~A and B, we also consider Models~C and D, in which we maintain the values for $\ell_0$ used in Models A and B, but we allow the value of $\kappa_0$ to vary in order to obtain the value $\Gamma_\infty \approx 2.3$ quoted by Abdo et al. (2009) for M87. For illustrative purposes, in the discussion below we mainly focus on the details of the disc structure and particle transport obtained in Models A and B.

The computational domain for the disc-shock structure simulations ranges from the inner radius at $r_{\rm in} = 2.1\,GM/c^2$ to the outer radius at $r_{\rm out} = 5,000\,GM/c^2$. Our numerical examples use natural gravitational units ($GM=c=1$ and $\rs=2$). Global energy conservation in our model requires that $L_{\rm esc} = L_{\rm jet}$ (see equation~\ref{eqEnergyCons1}), and therefore the accretion rate $\dot M$ is dependent on $\Delta\epsilon$ via equation~(\ref{eqLshock}). The model values for $\dot M$ and $L_{\rm jet}$ are indicated in Table \ref{table:table4}, and the ratio of the jet mass outflow rate, $\dot M_{\rm esc}$, compared with the disc accretion rate $\dot M$ for each model is included in Table \ref{table:table3}. The low values obtained for the ratio $\dot M_{\rm esc}/\dot M$ justify our assumption of a constant mass accretion rate across the shock (see equation~\ref{eqDeltaM}).

\subsection{Disc structure and particle transport}

\begin{table}
\centering
\caption{Disc structure parameters. All quantities are expressed in gravitational units $\left(GM=c=1\right)$, except $T_*$, which is written in units of $10^{11}$~K.}
\begin{tabular}[width=1\columnwidth]{lccrcccccccc}
\hline\hline
Model & $\ell_0$ & $\kappa_0$ & $K_{\rm th}/K_{\rm rel}$ & $\epsilon_+$ & $\epsilon_-$ & $r_{c1}$ & $r_{c3}$ & $r_*$ & $H_*$ & $R_*$ & $T_*$\\
\hline
A & 3.1340 & 0.02044 & 7,400 & -0.006100 & -0.000429 & 110.29 & 5.964 & 12.565 & 6.20 & 1.61 & 1.50\\
B & 3.1524 & 0.02819 & 7,700 & -0.007500 & -0.001502 & 123.52 & 5.937 & 11.478 & 5.46 & 1.61 & 1.59\\
C & 3.1340 & 0.03000 & 65,000 & -0.007500 & -0.001073 & 131.75 & 5.898 & 14.780 & 7.49 & 1.69 & 1.41\\
D & 3.1524 & 0.05500 & 260,000 & -0.009900 & -0.003784 & 61.110 & 5.886 & 14.156 & 6.91 & 1.61 & 1.45\\
\hline
\end{tabular}
\label{table:table1} 
\end{table}
\begin{table}
\centering
\caption{Shock jump conditions.}
\begin{tabular}{lcccccccccc}
\hline\hline
Model & $\vel_+/\vel_-$ & $\rho_+/\rho_-$ & $a_{\rm th+}=a_{\rm th-}$ & $a_{\rm rel+}$ & $a_{\rm rel-}$ & $H_+/H_-$ & $\mathscr{M}_{{\rm eff},\kappa}$\\
\hline
A & 0.659 & 1.61 & 0.144 & 0.0676 & 0.0857 & 0.945 & 1.0017\\
B & 0.659 & 1.61 & 0.148 & 0.0694 & 0.0880 & 0.945 & 1.0018\\
C & 0.616 & 1.69 & 0.140 & 0.0498 & 0.0647 & 0.956 & 1.0842\\
D & 0.638 & 1.61 & 0.141 & 0.0444 & 0.0564 & 0.971 & 1.0840\\
\hline
\end{tabular}
\label{table:table2}
\end{table}

\begin{figure}
\begin{subfigure}[b]{.5\textwidth}
\centering
\includegraphics[width=1\columnwidth]{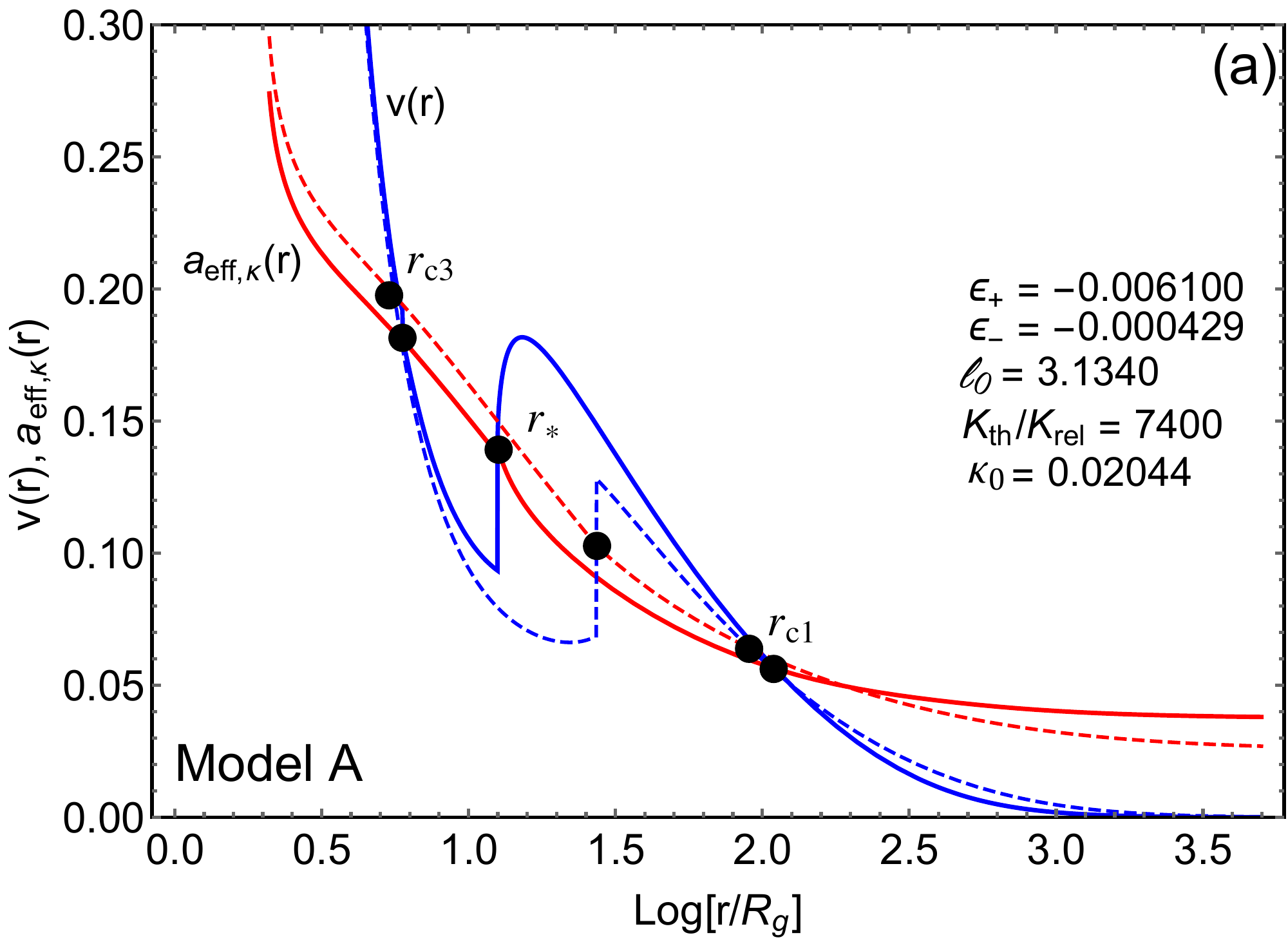}
\end{subfigure}
\quad
\begin{subfigure}[b]{.5\textwidth}
\includegraphics[width=1\columnwidth]{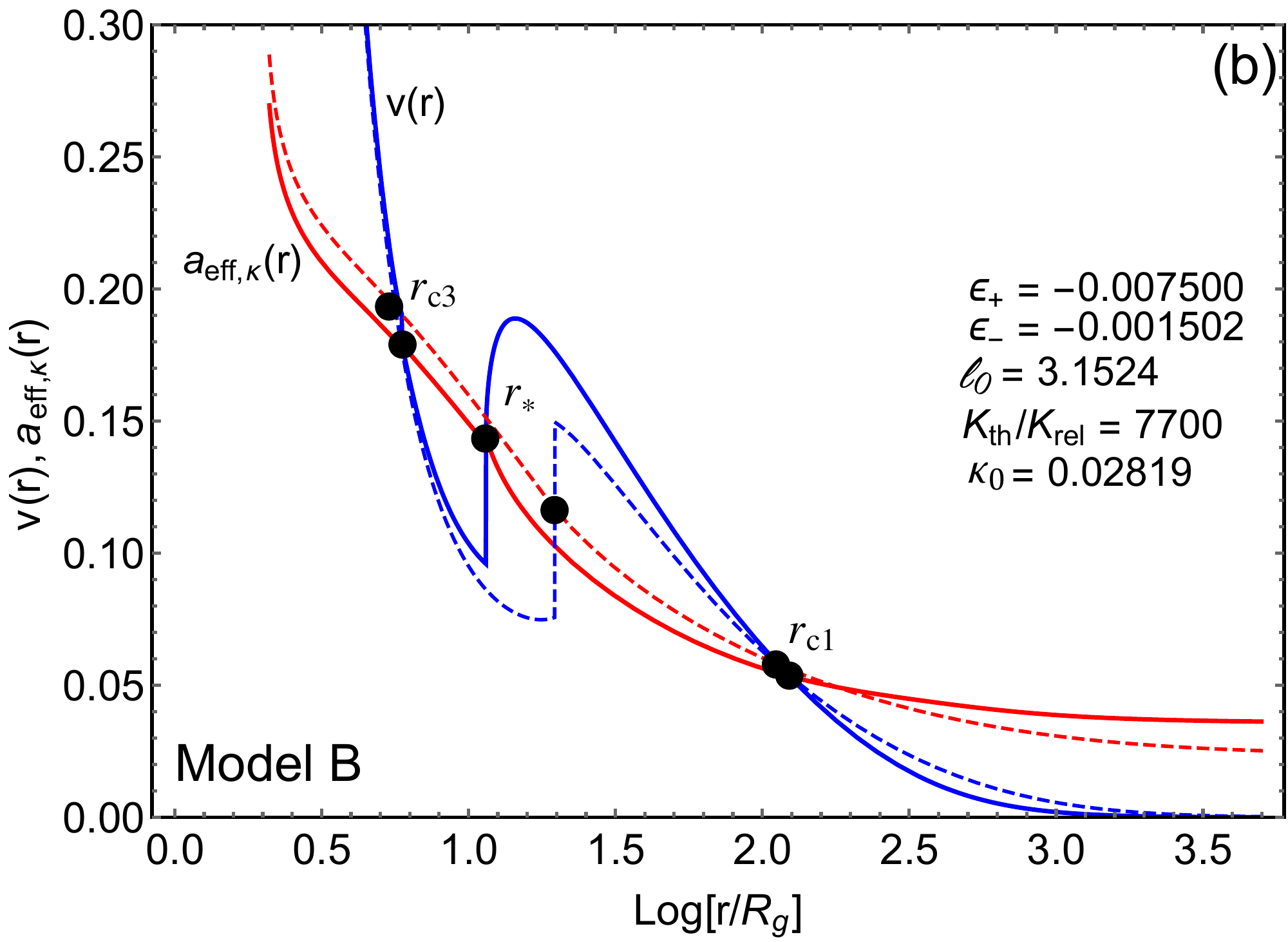}
\centering
\end{subfigure}
\caption{Velocity $\vel(r)$ (blue curves) and effective sound speed $a_{{\rm eff},\kappa}(r)$ (red curves), plotted in units of $c$, for the shocked-disc solution of (a) Model A and (b) Model B. These curves cross at the outer and inner critical points, located at radii $r_{c1}$ and $r_{c3}$, respectively. The solid lines denote the self-consistent two-fluid model developed here, which include relativistic particle pressure and diffusion. The dashed lines represent corresponding results obtained using the single-fluid model of LB05.}
\label{fig:fig4}
\end{figure}

In Figs. \ref{fig:fig4}(a) and \ref{fig:fig4}(b), we plot the inflow speed $\vel(r)$ and the effective adiabatic sound speed $a_{{\rm eff},\kappa}(r)$ (equation~\ref{eqAEffKappa}) for the shocked-disc, two-fluid solutions corresponding to Models~A and B, respectively, computed using the theory parameters listed in Table~\ref{table:table1}. In each calculation, we set $\gamma_{\rm th}=3/2$ and $\gamma_{\rm rel}=4/3$. The profiles for the flow velocity $\vel(r)$ and the sound speeds $a_{\rm th}(r)$ and $a_{\rm rel}(r)$ are computed by numerically integrating equation~(\ref{eqDAgDr}) for the thermal sound speed derivative and equation~(\ref{eqEps3alt}) for the relativistic sound speed derivative, combined with equation~(\ref{eqVfromKg}) for the inflow velocity. Figs. \ref{fig:fig4}(a) and \ref{fig:fig4}(b) also include a comparison of our two-fluid results with models 2 and 5 from the one-fluid model of LB05. In the isothermal shock model, the thermal sound speed $a_{\rm th}(r)$ is continuous at the shock location, although the particle sound speed $a_{\rm rel}(r)$ experiences a discontinuous jump. A distinctive feature of the two-fluid model is the deceleration precursor visible in the velocity profile, which is also seen in the analogous cosmic-ray modified shock model (Axford et al. 1977). This feature is completely absent in the LB05 dynamical profile, and it clearly indicates the dynamical effect of relativistic particle acceleration in the new model, as relativistic particles diffuse into the upstream region and decelerate the flow before it crosses the discontinuous shock. The results indicate that when the deceleration precursor is included in the dynamical structure, the shock is wider than in the one-fluid model, as expected.

In our model, there are two methods that can be used to compute the width of the shock, $\Delta x$, and we can compare the two results as a check on the self-consistency of our model (see Appendix~\ref{AppenShockWidth}). In the first method, we determine the shock width, $\Delta x$, including the deceleration precursor, by measuring the velocity profiles plotted in Fig. \ref{fig:fig4}, and we compute the associated dimensionless shock-width parameter, $\eta = \eta_s$, by writing
\begin{equation}
\Delta x = \eta_s \lambda_{\rm mag} \ ,
\label{eqDeltax}
\end{equation}
where $\lambda_{\rm mag}$ is the magnetic coherence length at the shock location. By using equation~(\ref{eqReif}) for the spatial diffusion coefficient to substitute for $\lambda_{\rm mag}$ in equation~(\ref{eqDeltax}), we can obtain the alternative form
\begin{equation}
\eta_s = \frac{c \Delta x}{3 \kappa_*} \ ,
\label{eqDeltax2}
\end{equation}
where $\kappa_* \equiv \left(\kappa_+ + \kappa_-\right)/2$ is the mean spatial diffusion coefficient at the shock radius. Equation~(\ref{eqDeltax2}) can be used to compute the shock-width parameter $\eta_s$ based on measurement of the velocity profiles plotted in Fig. \ref{fig:fig4}.

The second method for determining the shock-width parameter is based on utilization of a simple model for the three-dimensional random walk of the relativistic particles, as they diffuse and escape through the upper and lower surfaces of the disc at the shock location. This process is discussed in detail in Appendix~\ref{AppenShockWidth}. The result obtained for the shock-width parameter in this case is (see equation~\ref{eqB5})
\begin{equation}
\eta = \frac{A_0}{2} \left(\frac{c H_*}{3\kappa_*}\right)^2 \ ,
\label{eqB5text}
\end{equation}
where $A_0$ is the dimensionless escape parameter, and $H_*$ denotes the disc half-thickness at the shock radius. Equations~(\ref{eqDeltax2}) and (\ref{eqB5text}) represent fairly crude estimates, probably only reliable to within about a factor of two, but a comparison of these values provides a useful means for evaluating the self-consistency of our diffusive two-fluid model. The values for $\eta_s$ and $\eta$ computed using equations~(\ref{eqDeltax2}) and (\ref{eqB5text}) are reported in Table~\ref{table:table3}, and the fact that they agree reasonably well helps to support the validity of our formalism.

\begin{figure}
\centering
\includegraphics[width=4in]{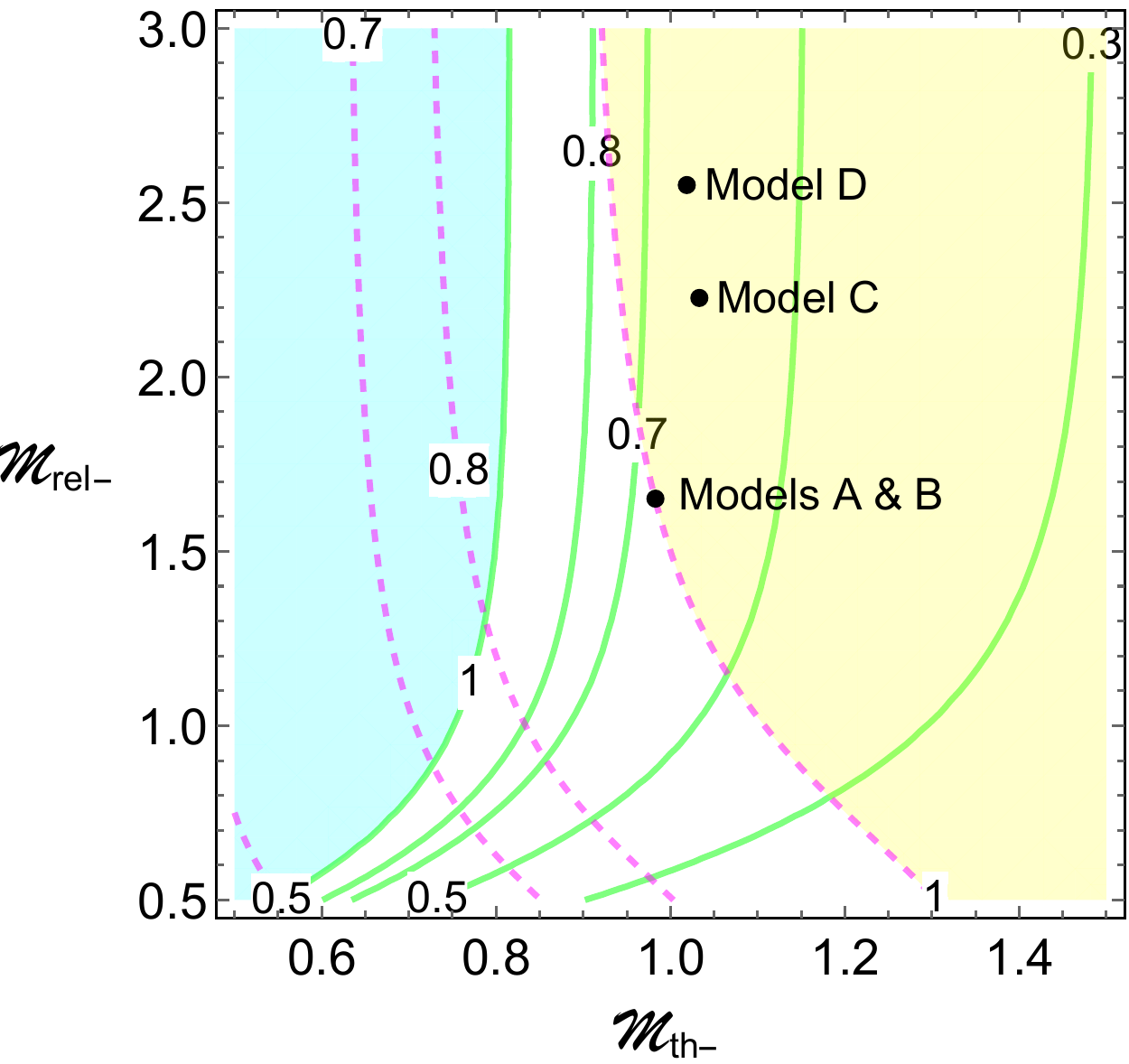}
\caption{Contour plots of the isothermal shock velocity jump ratio $Q_3 = \vel_ + /\vel_-$ (equation \ref{eqQ1}, solid lines) and the effective Mach number $\mathscr{M}_{{\rm eff},\kappa}$ (equation \ref{eqMeff}, dashed lines) as functions of $\mathscr{M}_{\rm th-}$ and $\mathscr{M}_{\rm rel-}$. The labeled points indicate the location in the parameter space for each of our four models. The region to the left of the $Q_3=1$ contour is unphysical since we must have $Q_3 <1$. Each of our models lies to the right of the $\mathscr{M}_{{\rm eff},\kappa}=1$ contour, which means that $\mathscr{M}_{{\rm eff},\kappa}>1$ for all of our models.}
\label{fig:fig5}
\end{figure}

The velocity jump ratio $Q_3$ is plotted as a function of the upstream Mach numbers $\mathscr{M}_{\rm th-}$ and $\mathscr{M}_{\rm rel-}$ in Fig.~\ref{fig:fig5}, along with the effective Mach number, $\mathscr{M}_{{\rm eff},\kappa}$, given by equation~(\ref{eqMeff}). Fig.~\ref{fig:fig5} also includes dots indicating the locations in the parameter space of the four models treated in detail here (see Section~\ref{sec:astrapp}). We note that the effective Mach number exceeds unity, barely, in the upstream region for each of our models (see Table~\ref{table:table2}). At first glance, this seems to suggest that shock acceleration is likely to be very inefficient in our two-fluid model. However, in the two-fluid model, much of the particle acceleration occurs in the extended, precursor flow deceleration region that can be clearly seen in Figs. \ref{fig:fig4}(a) and \ref{fig:fig4}(b), and is analogous to the velocity profiles seen in the cosmic-ray modified shock model (e.g., Axford et al. 1977).

\begin{figure}
\begin{subfigure}[b]{.5\textwidth}
\centering
\includegraphics[width=1\columnwidth]{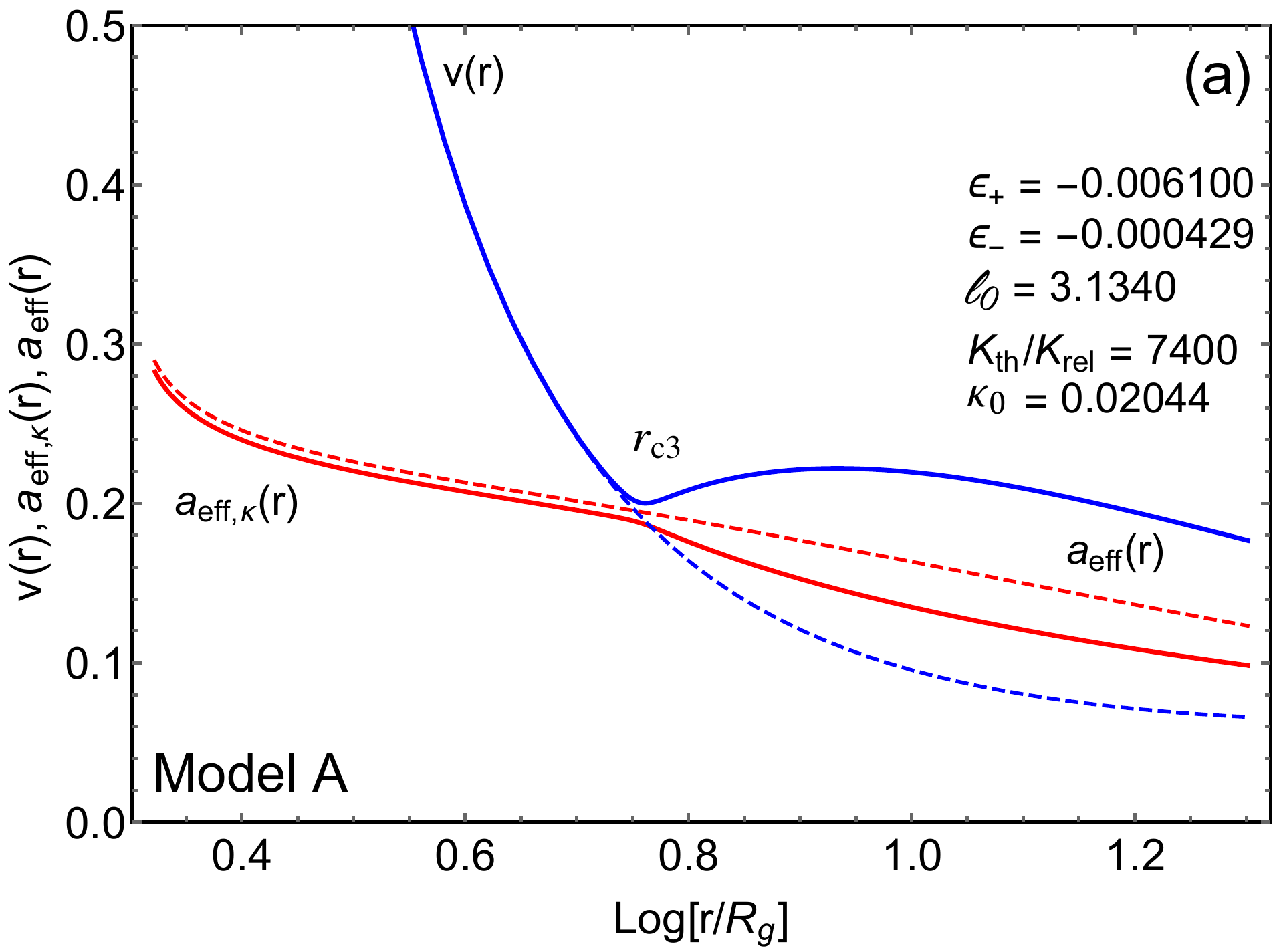}
\end{subfigure}
\quad
\begin{subfigure}[b]{.5\textwidth}
\includegraphics[width=1\columnwidth]{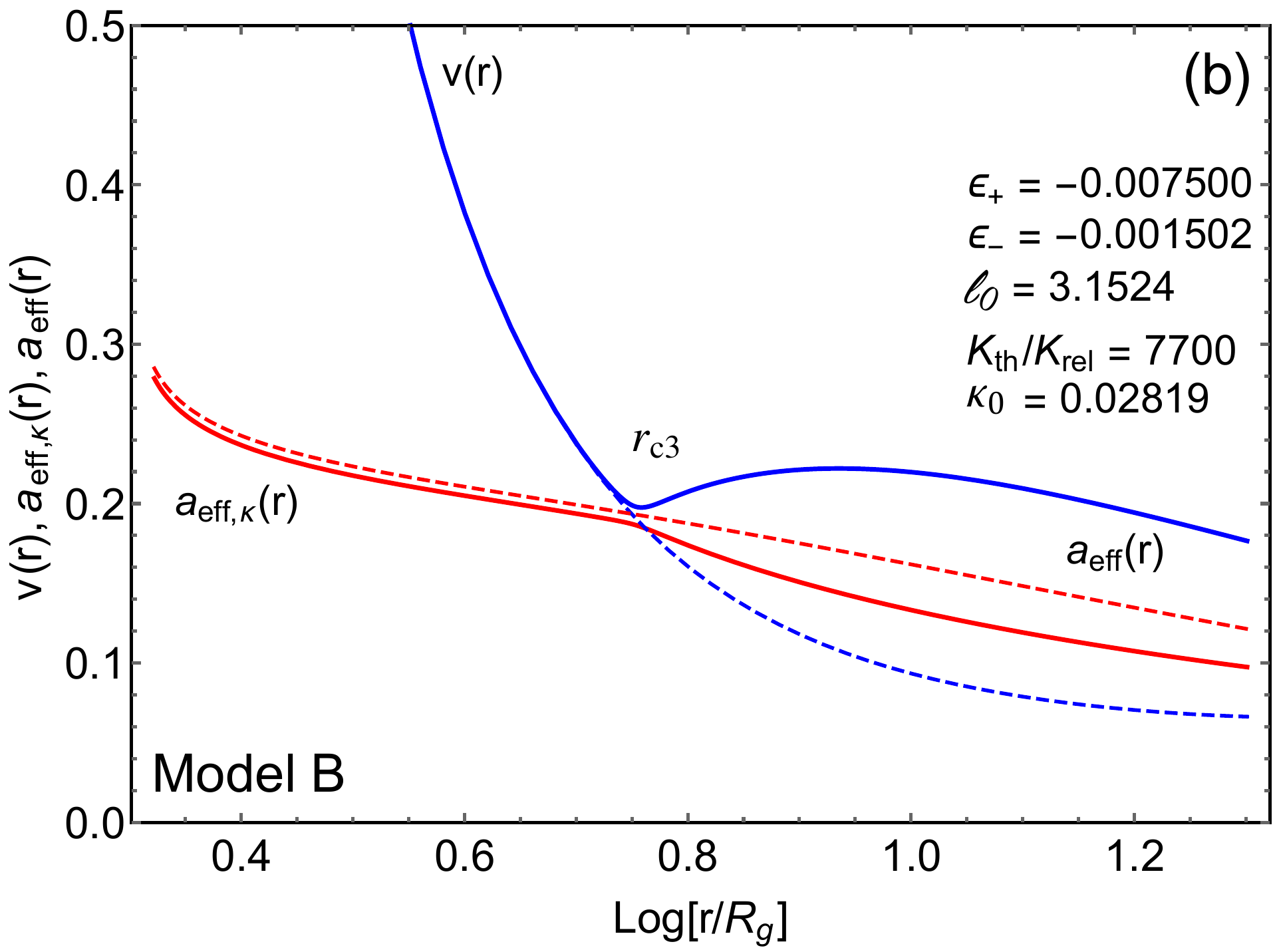}
\centering
\end{subfigure}
\caption{Velocity $\vel(r)$ (blue curves) and effective sound speed $a_{{\rm eff},\kappa}(r)$ (red curves), plotted in units of $c$, for the globally smooth (shock-free) solutions with $\epsilon_+=\epsilon_-$. The solid lines were computed using the diffusive model ($\kappa_0\neq 0$), with the parameters for (a) Model A and (b) Model B. Also plotted are the corresponding results for $\vel(r)$ and $a_{{\rm eff},\kappa}(r)$ obtained using the double-adiabatic model ($\kappa_0=0$, dashed lines). We note that a smooth, shock-free global solution is possible in the double-adiabatic case, but not in the diffusive case.}
\label{fig:fig6}
\end{figure}

In the one-fluid model of LB05, it is always possible to obtain a smooth velocity profile that corresponds to any shocked-disc solution. However, the dynamical model of LB05 did not include either relativistic particle pressure or diffusion, and therefore we must reexamine the possible existence of globally smooth flows within the context of our new two-fluid model. Figs. \ref{fig:fig6}(a) and \ref{fig:fig6}(b) depict the dynamical profiles for Model A and Model B, respectively, for globally smooth flow in the diffusive (thick lines) and double-adiabatic (dashed lines) cases. In the case of the double-adiabatic model, no diffusion is allowed ($\kappa=0$), and the dynamical structure is determined via a simple root-finding procedure based on the double-adiabatic energy equation~(\ref{eqEpsMod2Ag}). We expect the two profiles to resemble one another near the horizon, where the disc becomes purely adiabatic, and this is indeed the case. However, the globally smooth diffusive model fails to pass through the inner critical point displayed by the double-adiabatic model, and therefore it is unphysical. After an extensive exploration of the parameter space, we find that in fact it is impossible to obtain any globally smooth solutions when diffusion is included, regardless of the values for the specific angular momentum $\ell_0$, the entropy ratio $K_{\rm th}/K_{\rm rel}$, and the energy transport rate per unit mass $\epsilon_-$. On the other hand, it is always possible to obtain a globally smooth flow in the double-adiabatic case. {\it Hence we conclude that the inclusion of diffusion ($\kappa\ne 0$) in a two-fluid model invariably leads to the formation of a standing shock in the accretion flow}. 

\begin{figure}
\begin{subfigure}[b]{.4\textwidth}
\centering
\includegraphics[width=1\columnwidth]{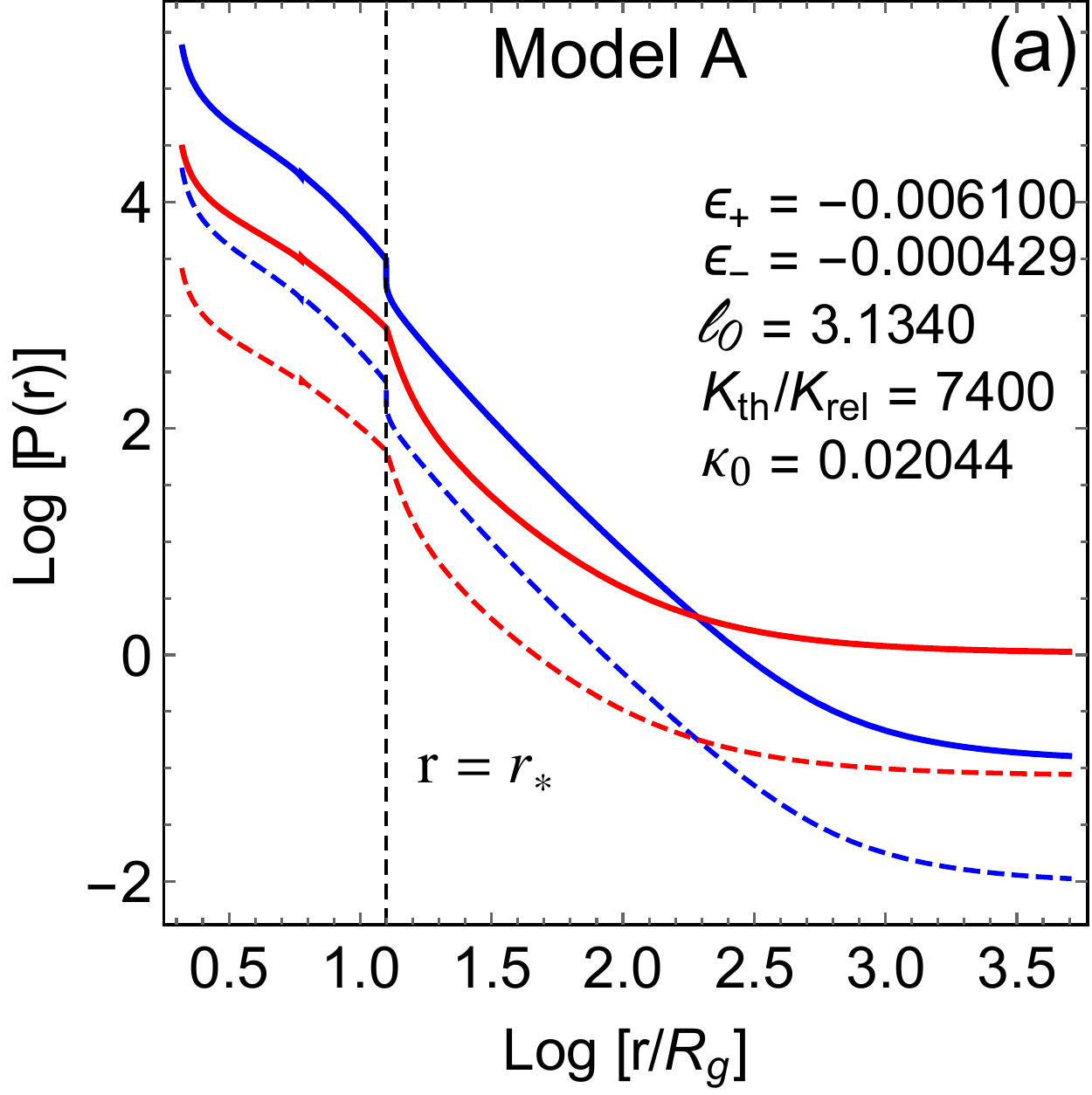}
\end{subfigure}
\quad
\begin{subfigure}[b]{.4\textwidth}
\includegraphics[width=1\columnwidth]{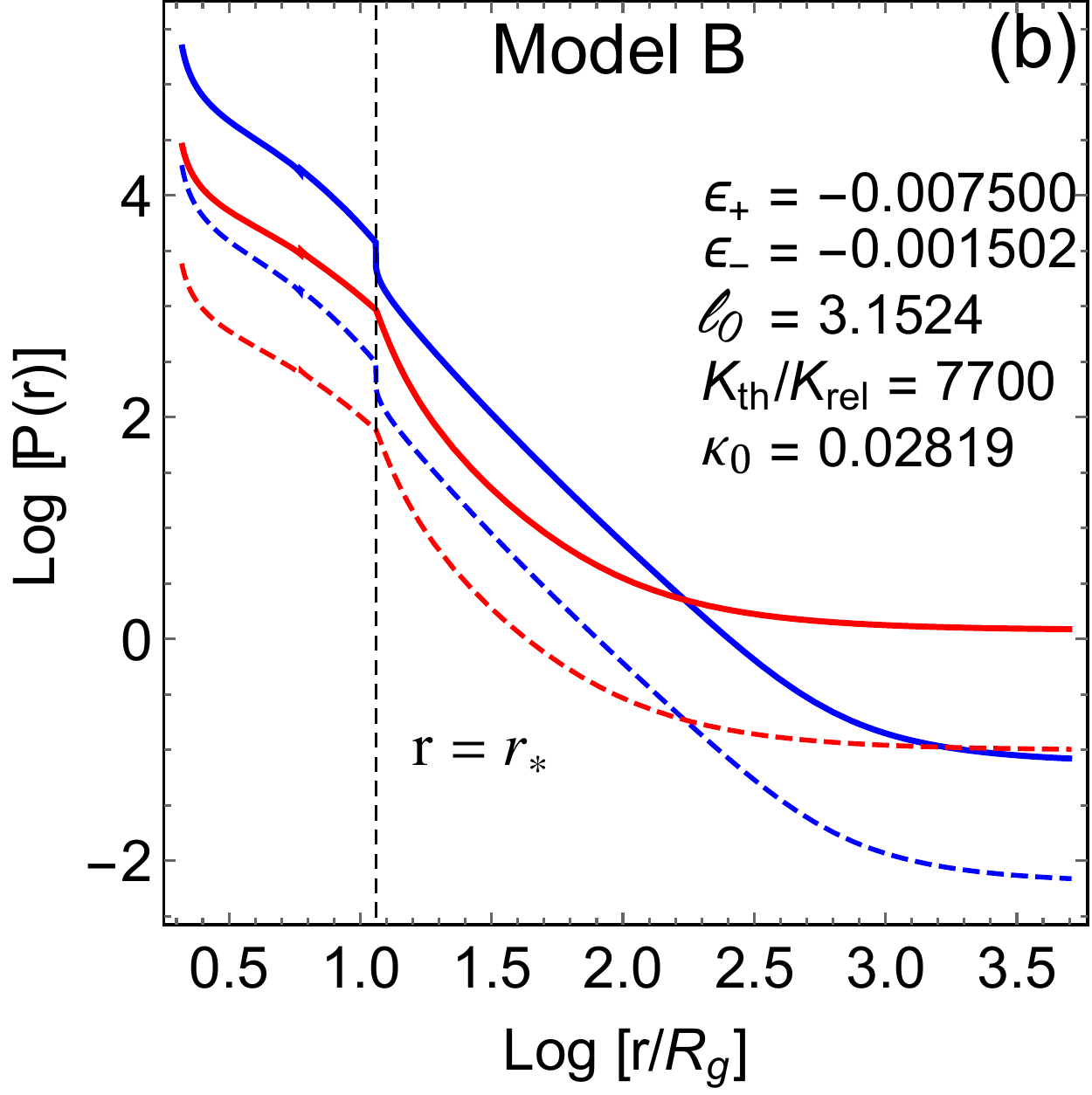}
\centering
\end{subfigure}
\caption{Hydrodynamical profiles for the thermal pressure $P_{\rm th}$ (blue curves), and the relativistic particle pressure $P_{\rm rel}$ (red curves), plotted as functions of $r$ in cgs units for (a) Model A and (b) Model B. The thick and dashed lines represent the results obtained for \sgr and M87, respectively. Note that the particle pressure is comparable to the thermal pressure at the shock.}
\label{fig:fig7}
\end{figure}

Next we study the solutions obtained for the thermal gas pressure $P_{\rm th}(r)$ and the relativistic particle pressure $P_{\rm rel}(r)$ in the disc based on the flow structures for Models A and B. Once the profiles for $\vel(r)$, $a_{\rm th}(r)$, and $a_{\rm rel}(r)$ have been obtained, the corresponding pressure distributions are computed using equations~(\ref{eqThermalP1}) and (\ref{eqRelP1}) for the thermal and relativistic particle pressures, respectively. We plot the global pressure profiles obtained in Models~A and B in Figs. \ref{fig:fig7}(a) and \ref{fig:fig7}(b), respectively, for \sgr (thick lines) and M87 (dashed lines). We observe that the pressures decrease monotonically with increasing radius. The increase in the pressures near the horizon is a consequence of strong adiabatic compression, whereas the leveling off as $r \to \infty$ reflects the dominance of diffusion far from the black hole, where conditions in the disc approach those in the surrounding medium. These results confirm that the relativistic particle pressure is comparable to the gas pressure in both \sgr and M87, in agreement with the findings of Becker et al. (2011). This validates the inclusion of the relativistic particle pressure in our computation of the disc structure in the two-fluid model.

\begin{figure}
\begin{subfigure}[b]{.4\textwidth}
\centering
\includegraphics[width=1\columnwidth]{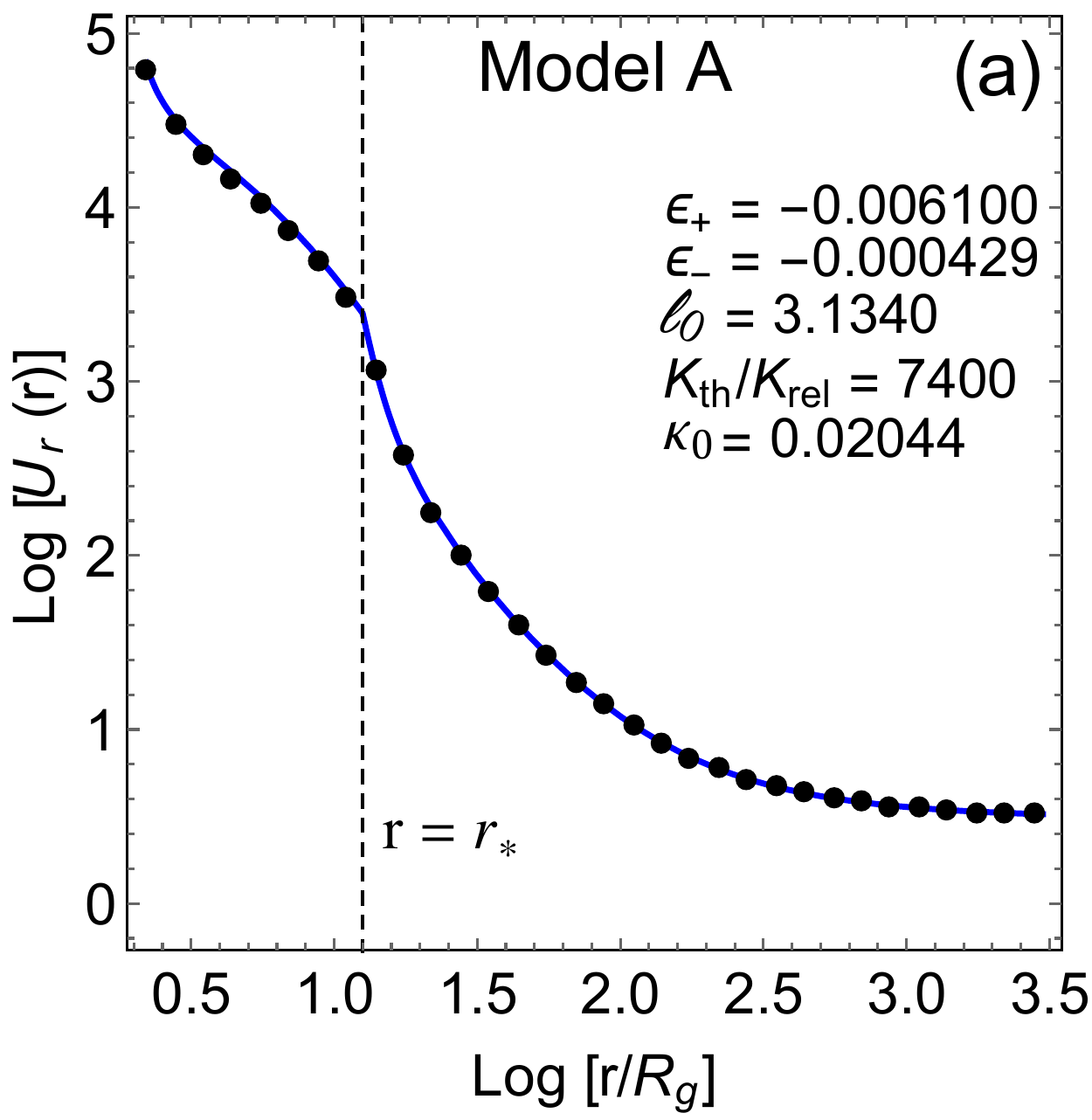}
\end{subfigure}
\quad
\begin{subfigure}[b]{.4\textwidth}
\includegraphics[width=1\columnwidth]{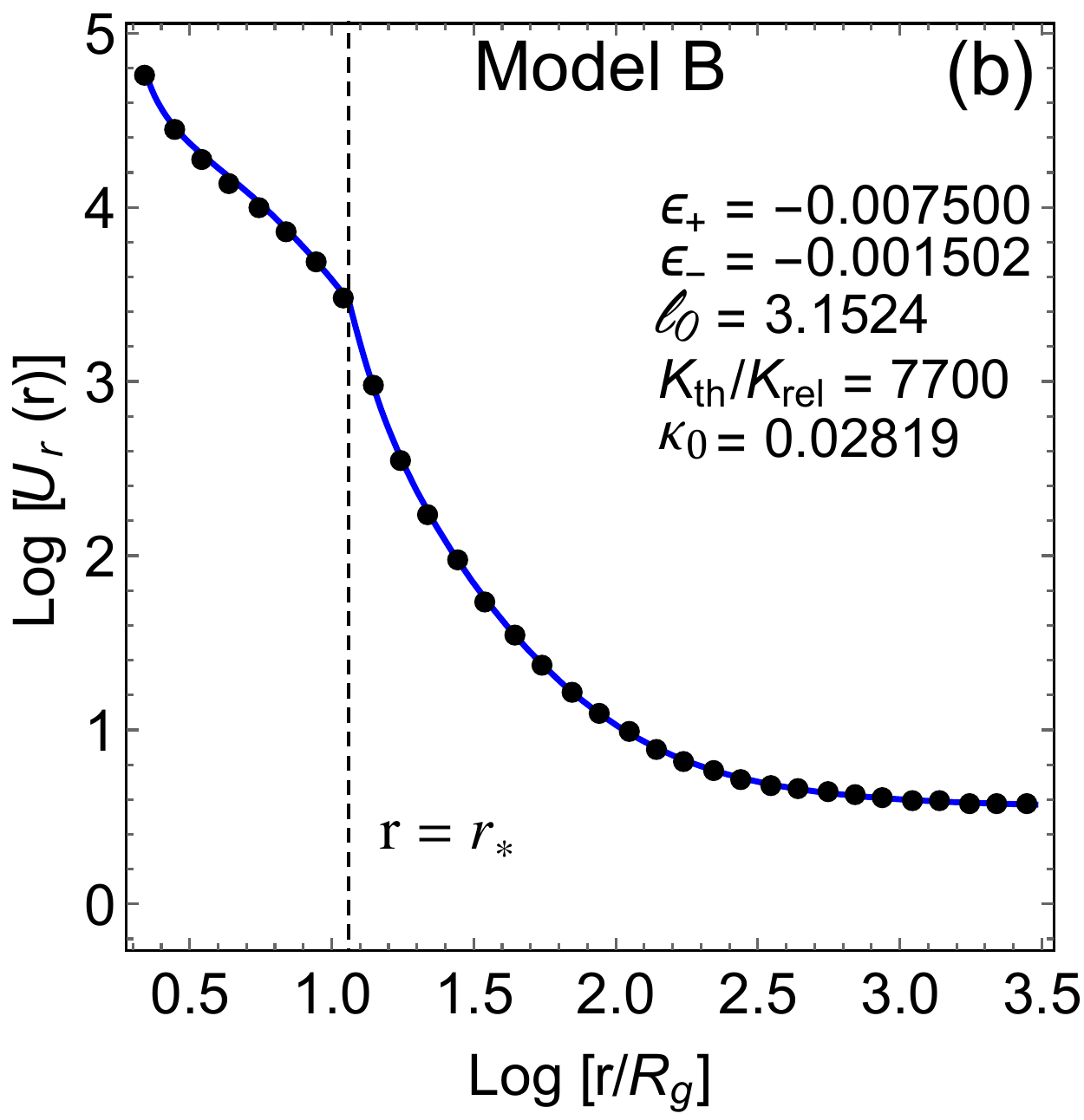}
\centering
\end{subfigure}
\caption{Global solutions for the relativistic particle energy density $U_{\rm rel}$, obtained using the particle transport equation (equation~\ref{eqGlobalUr}, solid lines) and the dynamical solution (equation~\ref{eqRelP1}, filled circles).}
\label{fig:fig8}
\end{figure}

Another means for evaluating the self-consistency of our model is provided by comparing the solution for the relativistic particle energy density $U_{\rm rel}(r)$ obtained using the transport equation method (equation~\ref{eqGlobalUr}) with that computed using the dynamical solution (equation~\ref{eqRelP1}). In Figs. \ref{fig:fig8}(a) and \ref{fig:fig8}(b), we plot the relativistic particle energy density profiles in the disc for Models~A and B, respectively. The kinks that appear in the energy density distributions at the shock radius $r=r_*$ reflect the derivative jump condition given by equation~(\ref{eqDeltaErNEW}). The overlap between the transport equation solution (solid lines) and the dynamical solution (filled circles) for $U_{\rm rel}(r)$ in Fig.~\ref{fig:fig8} demonstrates the self-consistency of our calculation of the dynamical structure of the disc-shock-outflow system.

\begin{table}
\centering
\caption{Transport equation parameters.}
\begin{tabular}{lcccccccccc}
\hline\hline
Model & $\kappa_*$ & $\lambda_{\rm mag}$ & $A_0$ & $\eta_s$ & $\eta$ & $\dot N_{\rm I}/\dot N_{\rm II}$ & $\dot N_{\rm esc}/\dot N_0$ & $E_{\rm esc}/E_0$ &$\dot M_{\rm esc}/\dot M$ & $\Gamma_\infty$\\
\hline
A & 0.134 & 0.402 & 0.050 & 6.63 & 5.95 & -0.005 & 0.386 & 2.61 & 1.64$\times10^{-3}$ & 3.48\\
B & 0.153 & 0.459 & 0.052 & 6.41 & 3.65 & -0.022 & 0.388 & 2.60 & 1.74$\times10^{-3}$ & 3.47\\
C & 0.285 & 0.855 & 0.100 & 3.56 & 3.84 & -0.140 & 0.573 & 1.78 & 2.75$\times10^{-3}$ & 2.38\\
D & 0.478 & 1.434 & 0.125 & 1.42 & 1.45 & -0.803 & 0.547 & 1.84 & 2.50$\times10^{-3}$ & 2.46\\
\hline
\end{tabular}
\label{table:table3}
\end{table}

\begin{table}
\centering
\caption{Auxiliary parameters.}
\resizebox{1\textwidth}{!}{
\begin{tabular}{lccccccccccc}
\hline\hline
& & \multicolumn{2}{c}{$L_{\rm jet}\,\left({\rm ergs \ s}^{-1}\right)$} & \multicolumn{2}{c}{$\dot{N}_0\,\left({\rm s}^{-1}\right)$} & \multicolumn{2}{c}{$\dot M\,\left(\msun {\rm yr}^{-1}\right)$} & \multicolumn{2}{c}{$n_*\,({\rm cm}^{-3})$} & \multicolumn{2}{c}{$U_*\,\left({\rm ergs \ cm}^{-3}\right)$}\\
Model & $\Delta\epsilon$ & \sgr & M87 & \sgr & M87 & \sgr & M87 & \sgr & M87 & \sgr & M87\\
\hline
A & -0.005671 & 5.0$\times10^{38}$ & 5.5$\times10^{43}$ & $2.5 \times 10^{41}$ & $2.75 \times 10^{46}$ &1.56$\times10^{-6}$ & 1.71$\times10^{-1}$ & 4.46$\times10^{5}$ & 3.66$\times10^{4}$ & 2.31$\times10^{3}$ & 1.91$\times10^{2}$\\
B & -0.005998 & 5.0$\times10^{38}$ & 5.5$\times10^{43}$ & $2.5 \times 10^{41}$ & $2.75 \times 10^{46}$ &1.47$\times10^{-6}$ & 1.62$\times10^{-1}$ & 5.40$\times10^{5}$ & 4.43$\times10^{4}$ & 2.79$\times10^{3}$ & 2.30$\times10^{2}$\\
C & -0.006427 & 5.0$\times10^{38}$ & 5.5$\times10^{43}$ & $2.5 \times 10^{41}$ & $2.75 \times 10^{46}$ &1.37$\times10^{-6}$ & 1.51$\times10^{-1}$ & 2.32$\times10^{5}$ & 1.88$\times10^{4}$ & 8.12$\times10^{2}$ & 6.71$\times10^{1}$\\
D & -0.006116 & 5.0$\times10^{38}$ & 5.5$\times10^{43}$ & $2.5 \times 10^{41}$ & $2.75 \times 10^{46}$ &1.44$\times10^{-6}$ & 1.59$\times10^{-1}$ & 2.01$\times10^5$ & 1.65$\times10^4$ & 7.38$\times10^2$ & 6.09$\times10^1$\\
\hline
\end{tabular}}
\label{table:table4}
\end{table}

\subsection{Jet formation in M87 and \sgr}

\begin{figure}
\begin{subfigure}[b]{.41\textwidth}
\centering
\includegraphics[width=1\columnwidth]{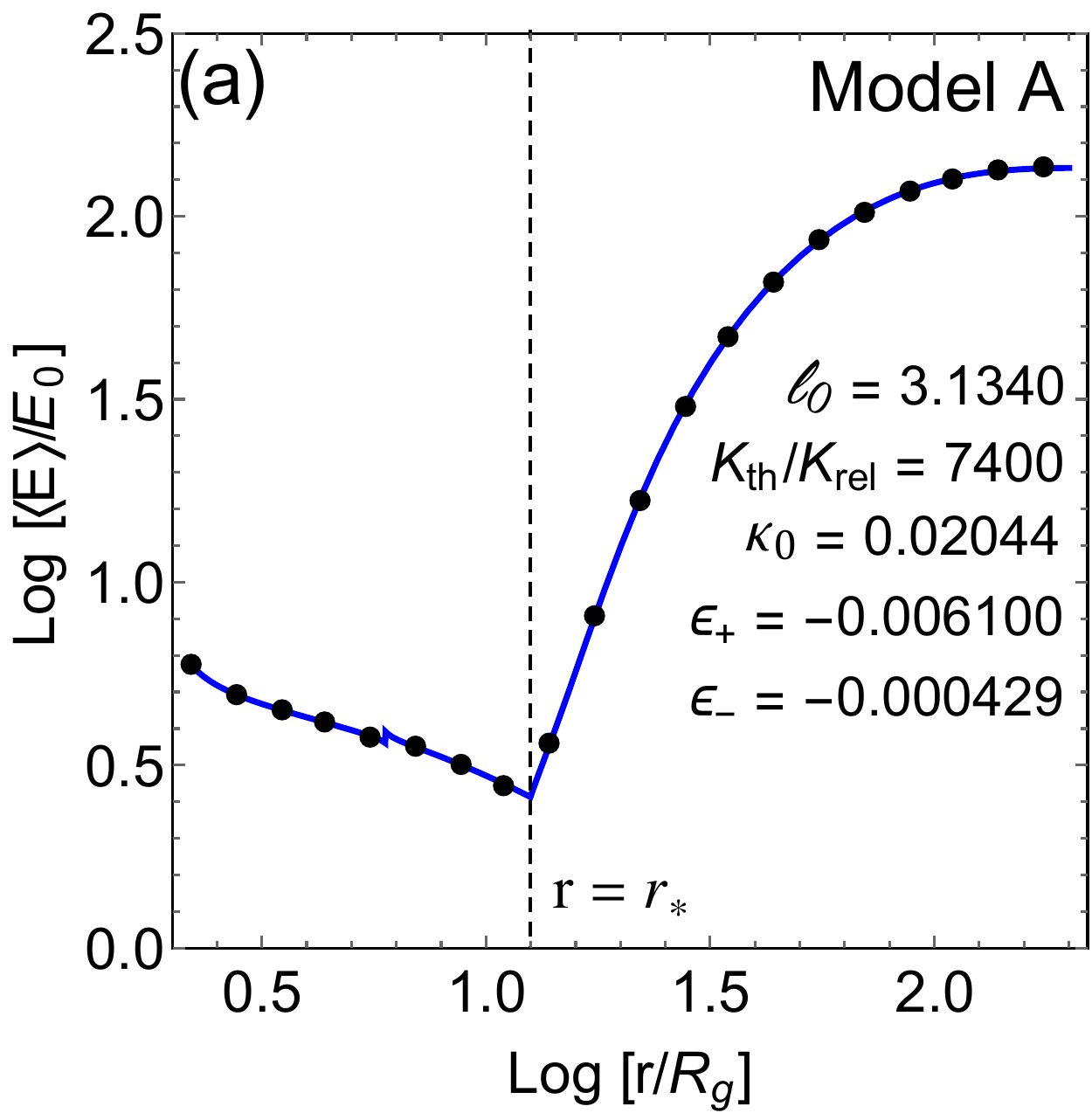}
\end{subfigure}
\quad
\begin{subfigure}[b]{.41\textwidth}
\includegraphics[width=1\columnwidth]{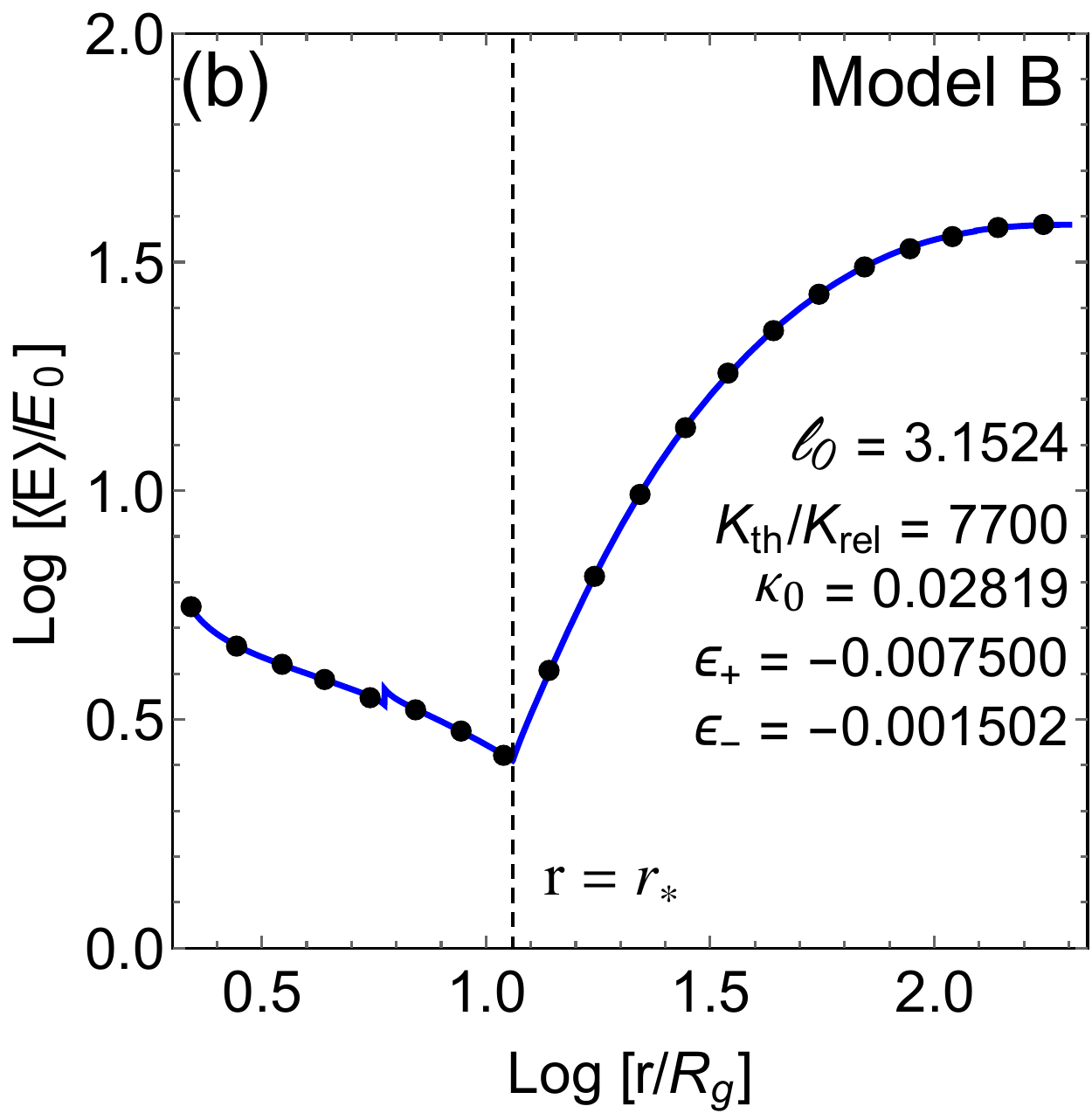}
\centering
\end{subfigure}
\caption{Mean energy of the relativistic particles in the disc, $\left<E\right>\equiv U_{\rm rel}(r)/n_{\rm rel}(r)$ (equation~\ref{eqEmean}), for Model A (a) and Model B (b), plotted in units of the injection energy $E_0$.}
\label{fig:fig9}
\end{figure}

The mean energy of the relativistic particles at radius $r$ in the disc is given by
\begin{equation}
\left<E\right>\equiv \frac{U_{\rm rel}(r)}{n_{\rm rel}(r)} \ ,
\label{eqEmean}
\end{equation}
so that $\left<E\right> = E_{\rm esc}$ at the shock location, $r=r_*$ (see equation~\ref{eqEesc}). The mean relativistic particle energy $\left<E\right>$ is plotted as a function of radius in Figs. \ref{fig:fig9}(a) and \ref{fig:fig9}(b) for Models A and B, respectively. When a shock is present in the flow, the results demonstrate that the mean particle energy is boosted, as expected. In our self-consistent two-fluid model, shock acceleration boosts the mean particle energy by a factor of $\sim 2.5$, which is less than the factor of $\sim 5-6$ found by LB05 for the same parameters. This reflects the fact that shock acceleration is weaker in the self-consistent model, since the compression ratio is reduced by the particle pressure. The analogous behaviour is observed in the models for cosmic-ray mediated shocks (e.g. Axford et al. 1977). However, even within the context of the self-consistent two-fluid model developed here, the acceleration of the relativistic particles is efficient enough to account for the outflows observed in \sgr and M87.

\subsection{Radiative losses from the jet and the disc}

While it remains unclear whether the outflows observed in many radio-loud systems containing black holes are composed of an electron-proton plasma or electron-positron pairs, or a mixture of both, the particles must provide sufficient energy to power the observed radio emission, although this requirement can be mitigated if the particles are reaccelerated by shocks propagating along the jet (e.g., Atoyan \& Dermer 2004). In our model, it is assumed that the outflows are composed of fully-ionized electron-proton plasma, which enhances the efficiency of the energy transport in the jet because the ions carry most of the kinetic energy, they don't radiate much, and they are not strongly coupled to the electrons under the typical conditions in a jet (e.g., Felten 1968; Felten et al. 1970; Anyakoha et al. 1987; Aharonian 2002).

Starting with the premise that the observed outflows are proton-driven, LB05 explored two means by which the ions in the jet lose energy: (1) via the production of synchrotron and inverse-Compton emission, or (2) via the production of indirect radiation due to Coulomb coupling with the electrons. These two possibilities were evaluated by computing the corresponding cooling timescales in the outflows. LB05 concluded that synchrotron and inverse-Compton losses have virtually no effect on the energy of the protons in either the M87 jet or the \sgr jet, but we need to reevaluate this conclusion in the context of our new two-fluid model. The energy loss timescale for the combination of inverse-Compton and synchrotron emission is computed using (see equation~112 of LB05)
\begin{equation}
t_{\rm rad} \equiv \frac{3 m_p c}{4\sigma_{_{\rm T}}\Gamma_\infty}
\left(\frac{m_p}{m_e}\right)^2\left(U_B + U_{\rm ph}\right)^{-1} \ ,
\label{eqSynch}
\end{equation}
where $U_{\rm ph}$ is the incident photon energy density and $U_B = B^2/(8\pi)$ is the magnetic energy density for a field of strength $B$. Setting $B \sim 0.1$\,G for M87 and $B \sim10$\,G for \sgr based on estimates from Biretta et al. (1999) and Atoyan \& Dermer (2004), respectively, we confirm that inverse-Compton and synchrotron losses are negligible for the jet protons since $t_{\rm rad} \sim 10^{12}\,$yr.

The ions in the jet also lose energy via Coulomb coupling with the thermal electrons, which in turn radiate efficiently. The energy loss timescale for Coulomb coupling is given by (see equation~114 of LB05)
\begin{equation}
t_{_{\rm Coul}} \equiv \frac{\Gamma_\infty m_p c^2}{\left(dE/dt\right)\Big|_{_{\rm Coul}}}
= \frac{\Gamma_\infty m_p}{30 n_e \sigma_{_{\rm T}} c m_e} \ ,
\label{eqCoulomb}
\end{equation}
where $n_e$ is the electron number density. We assume that $n_e$ decreases rapidly as the jet expands from the disc into the external medium, and therefore the strongest Coulomb coupling occurs at the base of the jet, where $n_e$ achieves its maximum value. We shall adopt this maximum value for $n_e$ in the subsequent analysis.

The value for $n_e$ at the base of the jet can be estimated by deriving two separate expressions for the proton escape rate $\dot N_{\rm esc}$. In the first method, based on the three-dimensional random walk of the protons in the disc plasma, we use equation~(\ref{eqB4}) to eliminate $A_0$ in equation~(\ref{eqNesc}) for the magnitude of the jump in the particle transport rate, which yields
\begin{equation}
\dot N_{\rm esc} \equiv \frac{8 \pi r_* \eta \lambda_{\rm mag}^2 c n_*}{H_*} \ ,
\label{eqNesc1}
\end{equation}
where $r_*$, $n_*$, $H_*$, and $\lambda_{\rm mag}$ represent the radius, the proton number density, the disc half-thickness, and the magnetic coherence length at the shock location, respectively, and $\eta$ denotes the shock width parameter. In the second method, we write the relativistic proton escape rate as the product of the annulus area (including both the upper and lower surfaces of the disc) $4 \pi r_* \Delta x$ multiplied by the escaping proton flux, $c n_p$, where $\Delta x = \eta \lambda_{\rm mag}$ is the shock width and $n_p$ is the proton number density at the base of the jet. The result obtained is
\begin{equation}
\dot N_{\rm esc} = 4 \pi r_* c n_p \Delta x = 4 \pi r_* c n_p \eta \lambda_{\rm mag} \ .
\label{eqNesc2}
\end{equation}
Eliminating $\dot N_{\rm esc}$ between equations~(\ref{eqNesc1}) and (\ref{eqNesc2}) yields
\begin{equation}
\frac{n_p}{n_*} = 2 \frac{\lambda_{\rm mag}}{H_*} \ ,
\label{}
\end{equation}
and since the electron-proton jet is charge neutral ($n_e=n_p$), we find that
\begin{equation}
n_e = 2 \frac{\lambda_{\rm mag}}{H_*} n_* \ .
\label{}
\end{equation}
Using the values for $H_*$, $\lambda_{\rm mag}$, and $n_*$ listed in Tables~\ref{table:table1}, \ref{table:table3} and \ref{table:table4}, respectively, we find that at the base of the jet, $n_e \sim 10^4\,{\rm cm}^{-3}$ for \sgr, and $n_e \sim 10^3\,{\rm cm}^{-3}$ for M87. Substituting these results for $n_e$ in equation~(\ref{eqCoulomb}), and setting $\Gamma_\infty \sim 3$, yields for the electron-proton Coulomb coupling timescale for \sgr $t_{_{\rm Coul}} \sim 10^4\,{\rm yr}$, and for M87 $t_{_{\rm Coul}} \sim 10^5\,{\rm yr}$. These results confirm that Coulomb losses are negligible in the outflowing jet, in agreement with the findings of LB05. Hence we conclude that shock acceleration of the protons in the disc is sufficient to power the observed outflows, without requiring additional energization in the jets.

The importance of radiative losses in the disc can be estimated by computing the total bremsstrahlung X-ray luminosity via integration of equation~(5.15b) from Rybicki \& Lightman (1979) over the disc volume. The result obtained for pure, fully ionized hydrogen is
\begin{equation}
L_{\rm rad} = \int^\infty_{\rs} 1.4 \times 10^{-27} T^{1/2}_e \rho^2 m^{-2}_p \, dV \ ,
\label{}
\end{equation}
where $T_e$ represents the electron temperature, and $dV=4\pi rHdr$ denotes the differential volume element in cylindrical coordinates. Based on the assumption that the electron temperature is equal to the ion temperature $T_*$, we find that $L_{\rm rad}/L_{\rm jet} \sim 10^{-4}$ and $L_{\rm rad}/L_{\rm jet} \sim 10^{-2}$ for \sgr and M87, respectively. We emphasize that these are the most unfavorable possible scenarios, since in an actual ADAF disc, the electron temperature is likely to be at least two orders of magnitude lower than the proton temperature, which will greatly reduce the bremsstrahlung luminosity below the estimate obtained here. Hence we are fully justified in neglecting radiative losses.

\section{CONCLUSION}
\label{sec:conclusion}

In this paper we have developed the first self-consistent, two-fluid model for the accretion hydrodynamics and the associated particle acceleration occurring in an inviscid, advection-dominated accretion disc. In particular, this is the first time that the test particle approximation has been relaxed in studies of black hole accretion. Our results demonstrate that particle acceleration at a standing, isothermal shock in an ADAF accretion disc can provide relativistic protons with the energy required to unbind them from the disc and thereby power the outflows observed from radio-loud sources containing black holes. The work presented here is a modified, improved version of the model developed by LB04 and LB05, which now includes relativistic particle pressure and diffusion, and is self-consistent with the dynamical results. The new diffusive, two-fluid model we have developed allows us to study in detail transonic relativistic accretion discs around both stellar-mass black holes and supermassive black holes in the cores of AGNs.

The existence of shocks in black-hole accretion disc remains a controversial issue, although a preponderance of the most recent studies seem to support the existence of shocks in discs (e.g., Chattopadhyay \& Kumar 2016). Our work lends further support to that conclusion, since we find that smooth (shock-free) solutions cannot occur in diffusive, two-fluid discs. Our results for the predicted shock/jet location and the asymptotic Lorentz factor are consistent with the findings of other models, and with the observations of M87 and \sgr. In particular, we find that our model for M87 agrees with the findings of Biretta et al. (2002), who concluded that the M87 jet forms in a region no farther than $r_* \sim 30\,GM/c^2$ from the black hole. Likewise, our results for \sgr demonstrate that the shock/outflow forms at radius $r_* \sim 11-14\,GM/c^2$, in agreement with the conclusions of Yuan (2000), who estimated that the disc truncates at radius $r \sim 10\,GM/c^2$.

In relation to earlier work, the new model developed here is analogous to the two-fluid model for cosmic-ray modified shocks, describing the acceleration of cosmic rays at supernova-driven shock waves, in which the back-reaction due to the pressure of the accelerated particles influences the structure of the shock. We find that within the context of our diffusive, two-fluid model, the pressure of the accelerated particles is comparable to that of the thermal background gas, in agreement with the transition from the earliest test-particle models for cosmic-ray acceleration (e.g., Blandford \& Ostriker 1978), to the two-fluid, cosmic-ray modified shock model (e.g., Axford et al. 1977). We also find that the incorporartion of relativistic particle pressure into the dynamical model increases the width of the shock, allowing for the development of a distinctive deceleration precursor, similar to that observed in the two-fluid model for cosmic-ray modified shocks (e.g., Becker \& Kazanas 2001).

The values we obtain for the asymptotic Lorentz factor, $\Gamma_\infty$, using our new two-fluid model are slightly lower than the values obtained using the single-fluid model of LB05, as expected, since the back-reaction due to the relativistic particle pressure tends to decrease the compression ratio. However, even with this effect included, we obtain asymptotic Lorentz factors that agree reasonably well with the observational estimates for both M87 and \sgr.

In future work, we plan to incorporate viscosity in order to explore shock formation and particle acceleration within the context of a more realistic dynamical model. However, based on the findings of B11, we anticipate that the inclusion of viscosity will not significantly alter the conclusions reached in this work because significant particle acceleration will occur regardless of the level of viscosity when a shock is present in the disc. We plan to use the viscous model to reexamine the question of whether smooth flow is possible when particle diffusion and viscosity are both included. We conclude that the diffusive, two-fluid model developed here provides for the first time a completely self-consistent explanation for the outflows observed in many radio-loud systems containing black holes.

\section*{Acknowledgements}

The authors are grateful to the anonymous referee, who provided a number of useful comments that led to improvements in the presentation.

\appendix

\section{REVERSE SHOCK JUMP CONDITION}
\label{AppendixReverseJump}

The numerical integration procedure required to determine the disc structure begins near the event horizon and proceeds in an outward direction. It is therefore convenient to have available ``reverse'' jump conditions that can be used to cross over the shock from the downstream side to the upstream side. We emphasize that the shock itself is physically still a ``normal'' (compressive) shock, rather than an unphysical anti-shock. The difference here is that the integration procedure requires us to compute the upstream quantities (just outside the shock radius $r_*$) in terms of the downstream quantities (just inside $r_*$). We begin by noting that equation~(\ref{eqJumpArplus1}) for the downstream relativistic particle Mach number and equation~(\ref{eqShockJump1}) for the relativistic particle sound speed jump are each symmetrical with respect to the interchange of the upstream and downstream quantities, since they are based on the conservation of the mass and momentum transport rates across the shock. Hence we can immediately write the equivalent relations
\begin{equation}
\mathscr{M}_{\rm rel-}^{-2} Q_*^2= \frac{\gamma_{\rm rel}}{\gamma_{\rm th}}\left(Q_*-1\right) \mathscr{M}_{\rm th+}^{-2}
+ Q_* \, \mathscr{M}_{\rm rel+}^{-2} + \gamma_{\rm rel} \, Q_* \left(1-Q_*\right) \ ,
\label{eqJumpArplus1Reverse}
\end{equation}
and
\begin{equation}
\mathscr{M}^{-4}_{\rm rel-} Q_*^4 = \mathscr{M}^{-4}_{\rm rel+} Q_*^2 \left(\frac{Q_*^2 \frac{\gamma_{\rm th}}{\gamma_{\rm rel}}
\mathscr{M}^{-2}_{\rm rel-} + \mathscr{M}^{-2}_{\rm th+}}
{\frac{\gamma_{\rm th}}{\gamma_{\rm rel}} \mathscr{M}^{-2}_{\rm rel+} + \mathscr{M}^{-2}_{\rm th+}}\right) \ ,
\label{eqShockJump2Reverse}
\end{equation}
where
\begin{equation}
Q_* \equiv \frac{\vel_-}{\vel_+} = \frac{1}{Q} \ .
\label{eqJumpRatReverse}
\end{equation}

By using equation~(\ref{eqJumpArplus1Reverse}) to substitute for $\mathscr{M}_{\rm rel-}$ in equation~(\ref{eqShockJump2Reverse}), we can obtain a quartic equation for $Q_*$ in terms of coefficients that depend only on the downstream Mach numbers $\mathscr{M}_{\rm rel+}$ and $\mathscr{M}_{\rm th+}$. The trivial upstream root, $Q_*=1$, can be factored out by dividing the quartic equation by $(Q_*-1)$. After some algebra, we obtain the cubic equation
\begin{equation}
Q_*^3{\cal F}+Q_*^2{\cal H}+Q_*{\cal I}+{\cal J}=0 \ ,
\label{eqQJumpCubicb}
\end{equation}
where
\begin{equation}
\begin{split}
{\cal F} & =1+\gamma_{\rm rel}^{-1}\mathscr{M}_{\rm rel+}^{-2}-\left(\gamma_{\rm rel}\mathscr{M}_{\rm rel+}^2
+ \gamma_{\rm th}\mathscr{M}_{\rm th+}^2\right)^{-1} \ ,\\
{\cal H} & =-2\gamma_{\rm th}^{-1}\mathscr{M}_{\rm th+}^{-2}-\gamma_{\rm rel}^{-2}\mathscr{M}_{\rm rel+}^{-4}
\left(1+\gamma_{\rm rel}\mathscr{M}_{\rm rel+}^2\right)^2 \ ,\\
{\cal I} & =\gamma_{\rm th}^{-2}\mathscr{M}_{\rm th+}^{-4}\left[2\gamma_{\rm th}\mathscr{M}_{\rm th+}^2
\left(1+\gamma_{\rm rel}^{-1}\mathscr{M}_{\rm rel+}^{-2}\right) +1\right] \ ,\\
{\cal J} & =-\gamma_{\rm th}^{-2}\mathscr{M}_{\rm th+}^{-4} \ ,
\label{eqQjumpCubic1b}
\end{split}
\end{equation}
Only one of the three solutions is physically valid. 

The three solutions to the cubic equation are given by,
\begin{equation}
\begin{split}
Q_{*1}& =S+T-\frac{1}{3}\frac{\cal H}{\cal F} \ ,\\
Q_{*2}& =-\frac{1}{2}\left(S+T\right)-\frac{1}{3}\frac{\cal H}{\cal F}+\frac{1}{2}i\sqrt{3}\left(S-T\right) \ ,\\
Q_{*3}& =-\frac{1}{2}\left(S+T\right)-\frac{1}{3}\frac{\cal H}{\cal F}-\frac{1}{2}i\sqrt{3}\left(S-T\right) \ ,
\label{eqQ1b}
\end{split}
\end{equation}
where 
\begin{equation}
\begin{split}
S& = \left(X+\sqrt{W^3+X^2}\right)^{1/3}\\
T& = \left(X-\sqrt{W^3+X^2}\right)^{1/3}
\end{split}
\end{equation}
and
\begin{equation}
\begin{split}
W &= \frac{1}{9}\left(\frac{3\cal I}{\cal F} - \frac{{\cal H}^2}{{\cal F}^2}\right) \ ,\\
X &=\frac{1}{54}\left(\frac{9{\cal H}{\cal I}}{{\cal F}^2} - \frac{27\cal J}{\cal F} - \frac{2{\cal H}^3}{{\cal F}^3}\right) \ .
\end{split}
\end{equation}
By analogy with the discussion in Section~\ref{VelJump}, we find that the root $Q_{*1}$ is unacceptable because $Q_{*1} < 1$, which implies the existence of an unphysical ``anti-shock'' with $\vel_+ > \vel_-$. Likewise, utilization of the root $Q_{*2}$ yields the unphysical result $a^2_{\rm rel-} < 0$. The only physically acceptable root is therefore $Q_{*3}$, computed using the final relation in equations~(\ref{eqQ1b}).

\section{ENERGY MOMENT EQUATIONS}
\label{AppendixEnergyMoments}

The relativistic particle Green's function, $\greens(E_0,E,r_*,r)$, represents the particle distribution resulting from the injection of $\dot N_0$ particles per second, with energy $E_0$, from a source located at the shock radius, $r=r_*$. The Green's function is related to the relativistic particle number density $n_{\rm rel}$ and energy density $U_{\rm rel}$ via the expressions
\begin{equation}
n_{\rm rel}(r) \equiv \int_0^\infty 4\pi E^2 f_{\rm G} \, dE \ , \qquad
U_{\rm rel}(r) \equiv \int_0^\infty 4\pi E^3 f_{\rm G} \, dE \ .
\label{eqBB2}
\end{equation}
These two relations can be generalized in terms of the energy moments of the Green's function, $I_n(r)$, defined by
\begin{equation}
I_n(r) \equiv \int_0^\infty 4\pi E^n f_{\rm G} \, dE \ ,
\label{eqBB3}
\end{equation}
where $n_{\rm rel}(r)=I_2(r)$ and $U_{\rm rel}(r)=I_3(r)$. The Green's function satisfies the vertically-integrated transport equation (cf. equation~A9 from LB05)
\begin{equation}
H \vel_r \frac{\partial f_{\rm G}}{\partial r} = \frac{1}{3r}\frac{d}{dr}
(rH\vel_r)E\frac{\partial f_{\rm G}}{\partial E} + \frac{1}{r}\frac{\partial}{\partial r}
\left(rH\kappa\frac{\partial f_{\rm G}}{\partial r}\right) + \frac{\dot N_0\delta(E-E_0)
\delta(r-r_*)}{\left(4\pi E_0\right)^2 r_*}-A_0 cH_* \delta(r-r_*)f_{\rm G} \ ,
\label{eqVertIntTrans}
\end{equation}
where $\vel=-\vel_r < 0$, the quantities $\greens$, $\vel_r$, $H$, and $\kappa$ are considered vertically averaged, and the constant $A_0$ is a dimensionless parameter that determines the rate at which particles escape through the surface of the disc at the shock location (see Appendix~\ref{AppenShockWidth}).

We can derive the differential equation satisfied by the energy moment $I_n(r)$ by operating on equation~(\ref{eqVertIntTrans}) with $\int_0^\infty 4\pi E^n dE$ and integrating by parts, which yields
\begin{equation}
H \vel_r \frac{dI_n}{dr} = -\left(\frac{n+1}{3}\right)\frac{I_n}{r}\frac{d}{dr}
(rH\vel_r) + \frac{1}{r}\frac{d}{dr}\left(rH\kappa\frac{dI_n}{dr}\right)
+\frac{\dot N_0 E_0^{n-2}\delta(r-r_*)}{4\pi r_*}-A_0 cH_* \delta(r-r_*) I_n \ .
\label{eqVertIntTransIn}
\end{equation}
Setting $n=2$ in equation~(\ref{eqVertIntTransIn}) yields the differential equation satisfied by the total relativistic particle number density, $n_{\rm rel}(r)=I_2(r)$, given by
\begin{equation}
H \vel_r \frac{dn_{\rm rel}}{dr} = -\frac{n_{\rm rel}}{r}\frac{d}{dr}(rH\vel_r)+\frac{1}{r}\frac{d}{dr}
\left(rH\kappa\frac{dn_{\rm rel}}{dr}\right)+\frac{\dot N_0 \delta(r-r_*)}{4\pi r_*}
- A_0 c H_* \delta(r-r_*) n_{\rm rel} \ .
\label{eqVertIntTransNr}
\end{equation}
Likewise, setting $n=3$ in equation~(\ref{eqVertIntTransIn}) yields the differential equation satisfied by the total relativistic particle energy density, $U_{\rm rel}(r)=I_3(r)$, which can be written as
\begin{equation}
H \vel_r \frac{dU_{\rm rel}}{dr} = -\frac{\gamma_{\rm rel} U_{\rm rel}}{r}\frac{d}{dr}(rH\vel_r)+\frac{1}{r}\frac{d}{dr}\left(rH\kappa\frac{dU_{\rm rel}}{dr}\right) + \frac{\dot N_0 E_0 \delta(r-r_*)}{4\pi r_*} - A_0 c H_* \delta(r-r_*) U_{\rm rel} \ .
\label{eqVertIntTransUr}
\end{equation}

Equation~(\ref{eqVertIntTransIn}) can be rewritten in the flux-conservation form,
\begin{equation}
\frac{dG_n}{dr} = 4\pi rH\left[\left(\frac{2-n}{3}\right) \vel \frac{dI_n}{dr} + \frac{\dot N_0 E_0^{n-2}
\delta(r-r_*)}{4\pi r_*H_*} - A_0 c \delta(r-r_*) I_n \right] \ ,
\label{eqFluxConsForm}
\end{equation}
where $\vel = - \vel_r > 0$, and the transport rate for the $n$th moment is defined by
\begin{equation}
G_n \equiv 4\pi r H F_n \ ,
\label{eqFn0}
\end{equation}
with $F_n$ representing the associated flux, computed using
\begin{equation}
F_n \equiv - \left(\frac{n+1}{3}\right) \vel I_n-\kappa\frac{dI_n}{dr} \ .
\label{eqFn}
\end{equation}
Integrating equation~(\ref{eqFluxConsForm}) with respect to $r$ in a small region surrounding the shock location $r=r_*$ yields the jump condition
\begin{equation}
\Delta G_n = - \dot N_0 E_0^{n-2} + 4 \pi r_* H_* A_0 c I_n(r_*) \ ,
\label{eqDeltaGN1}
\end{equation}
where we remind the reader that $\Delta f \equiv f_+ - f_-$. The energy moment $I_n$ is continuous across the shock ($\Delta I_n=0$), and therefore the jump in its derivative $dI_n/dr$ can be computed by combining equations~(\ref{eqFn0}) and (\ref{eqFn}) to show that
\begin{equation}
\Delta \left(\kappa\frac{dI_n}{dr} \right)
= - \left(\frac{n+1}{3}\right) I_n(r_*) \Delta \left(H \vel \right)
+ \frac{\dot N_0 E_0^{n-2}}{4 \pi r_*} - H_* A_0 c I_n(r_*) \ .
\label{eqDeltaGN3}
\end{equation}

The global solution for the energy moments $I_n(r)$ is given by
\begin{equation}
I_n(r) = \begin{cases}
AQ_{\rm I}(r) \ , & r > r_* \ , \\
BQ_{\rm II}(r) \ , & r < r_* \ ,
\end{cases}
\label{eqGlobalIn}
\end{equation}
where $A$ and $B$ are normalization constants, and the functions $Q_{\rm I}(r)$ and $Q_{\rm II}(r)$ satisfy the homogeneous differential equation (cf. equation~\ref{eqVertIntTransIn}),
\begin{equation}
H \vel_r \frac{dQ}{dr} = -\left(\frac{n+1}{3}\right)\frac{Q}{r}\frac{d}{dr}(rH\vel_r)
+ \frac{1}{r}\frac{d}{dr}\left(rH\kappa\frac{dQ}{dr}\right) \ ,
\label{eqHomoDifEq}
\end{equation}
along with the boundary conditions (see equations~\ref{eqAsympIn} and \ref{eqEigenH30})
\begin{equation}
Q_{\rm I}(r_{\rm out}) = \frac{C_1}{r_{\rm out}} + 1 \ , \qquad
Q_{\rm II}(r_{\rm in}) = \left(\frac{r_{\rm in}}{\rs}-1\right)^{-(n+1)/(3\gamma_{\rm th}+3)} \ ,
\label{eqBoundaryQ}
\end{equation} 
where $C_1$ is a constant determined with reference to the relativistic particle energy density solution (see equation~\ref{eqUrAsympInf}), and $r_{\rm in}$ and $r_{\rm out}$ denote the inner and outer radii for the computational domain, respectively.

The normalization constants $A$ and $B$ are derived by applying the continuity and derivative jump conditions at the shock location, $r = r_*$. The function $I_n$ is continuous across the shock ($\Delta I_n=0$), and its derivative jump is given by equation~(\ref{eqDeltaGN3}). Combining relations yields, after some algebra,
\begin{equation}
A = B\frac{Q_{\rm II}}{Q_{\rm I}}\Bigg|_{r=r_*} \ ,
\label{eqAIn}
\end{equation}
\begin{equation}
B = \frac{\dot N_0 E_0^{n-2}}{4\pi r_*}Q_{\rm I}\left[ \frac{n+1}{3}
(H_+ \vel_+ - H_ - \vel_-)Q_{\rm I}Q_{\rm II} - H_- \kappa_ - Q_{\rm II}Q'_{\rm I}
+ H_+ \kappa_+ Q_{\rm I}Q'_{\rm II}
+ H_* A_0 c Q_{\rm II}Q_{\rm I}\right] ^{-1}\Bigg|_{r=r_*} \ ,
\label{eqBIn}
\end{equation}
where the primes denote differentiation with respect to radius. The solutions for the functions $Q_{\rm I}(r)$ and $Q_{\rm II}(r)$ are obtained by numerically integrating equation~(\ref{eqHomoDifEq}), subject to the boundary conditions given by equations~(\ref{eqBoundaryQ}). Once $A$ and $B$ are computed, the global solution for $I_n (r)$ is evaluated using equation~(\ref{eqGlobalIn}). The solutions for $n_{\rm rel}=I_2$ and $U_{\rm rel}=I_3$ are obtained by setting $n=2$ and $n=3$, respectively. This completes the formal solution procedure for the energy moments.

\section{ASYMPTOTIC RELATIONS}
\label{AppendixAsymptoticRelations}

In this section, we consider the asymptotic variation of the physical quantities $\vel$, $a_{\rm th}$, $a_{\rm rel}$, and $H$ near the event horizon, and at a large distance from the black hole. We also use these results to generate suitable boundary conditions for the integration of the differential equations governing the relativistic particle number density $n_{\rm rel}$ and the relativistic particle energy density $U_{\rm rel}$.

\subsection{Asymptotic dynamical behaviour as $r\to\rs$}
\label{sec:AsymHoriz}

As discussed in Section~\ref{sec:asympHor}, near the event horizon, the radial velocity $\vel$ approaches the free-fall velocity $\vel^2_{\rm ff}(r) \equiv 2GM/(r-\rs)$, and therefore
\begin{equation}
\vel^2(r) \propto (r-\rs)^{-1} \ , \qquad r \to \rs \ .
\label{eqEigenAsyRS1a}
\end{equation}
Spatial diffusion is overwhelmed by advection in any region in which the flow velocity approaches $c$ (Weinberg 1972). Hence in the asymptotic domain $r \to \rs$, spatial diffusion of the relativistic particles becomes negligible compared with advection, and this in turn implies that the relativistic particle sound speed $a_{\rm rel}$ becomes adiabatically related to the thermal sound speed $a_{\rm th}$, as expressed by equation~(\ref{eqArAgfunMod2}). By combining equations (\ref{eqEigenAsyRS1a}), (\ref{eqKg}), and (\ref{eqArAgfunMod2}), we can write the thermal entropy parameter $K_{\rm th}$ in the form
\begin{equation}
K_{\rm th} \propto (r-\rs)^{1/2} \, a_{\rm th}^{2/(\gamma_{\rm th}-1)} \left[\frac{\gamma_{\rm th}}{\gamma_{\rm rel}} a_{\rm rel,c}^2
\left(\frac{a_{\rm th}}{a_{\rm th,c}}\right)^{2(\gamma_{\rm rel}-1)/(\gamma_{\rm th}-1)} + a_{\rm th}^2 \right]^{1/2}
\ , \qquad r \to \rs \ .
\label{eqKgAsymp1}
\end{equation}
Incorporating the values $\gamma_{\rm th}=3/2$ and $\gamma_{\rm rel}=4/3$ yields the equivalent result
\begin{equation}
K_{\rm th} \propto (r-\rs)^{1/2} \, a_{\rm th}^4 \left[\frac{9}{8} \, a_{\rm rel,c}^2
\left(\frac{a_{\rm th}}{a_{\rm th,c}}\right)^{4/3} + a_{\rm th}^2 \right]^{1/2}
\ , \qquad r \to \rs \ .
\label{eqEigenKgRSMod3}
\end{equation}

In our ADAF disc model, the thermal gas pressure $P_{\rm th}$ varies adiabatically throughout the disc (except as the gas crosses the shock), and therefore $K_{\rm th}=$constant. In this case, $a_{\rm th}$ increases without bound as $r \to \rs$, and the second term inside the square brackets in equation~(\ref{eqEigenKgRSMod3}) dominates near the horizon. We therefore conclude that the asymptotic variation of the gas sound speed $a_{\rm th}$ is given by
\begin{equation}
a_{\rm th}^2(r) \propto (r-\rs)^{(1-\gamma_{\rm th})/(1+\gamma_{\rm th})}
\propto (r-\rs)^{-1/5} \ ,
\qquad r \to \rs \ ,
\label{eqEigenAsyRS1b}
\end{equation}
which can also be substituted into equation (\ref{eqArAgfunMod2}) to show that the asymptotic variation of the relativistic particle sound speed $a_{\rm rel}$ is given by
\begin{equation}
\arsq(r) \propto (r-\rs)^{(1-\gamma_{\rm rel})/(1+\gamma_{\rm th})}
\propto (r-\rs)^{-1/7} \ ,
\qquad r \to \rs \ .
\label{eqEigenAsyRS1c}
\end{equation}
Based on equations~(\ref{eqEigenAsyRS1b}) and (\ref{eqEigenAsyRS1c}), we conclude that $a_{\rm th}$ diverges faster than $a_{\rm rel}$ as $r \to \rs$, and therefore the gas pressure dominates over the relativistic particle pressure. This in turn implies that near the horizon, equation~(\ref{eqH}) for the disc half-thickness $H$ reduces to the standard single-fluid relation (e.g., LB05)
\begin{equation}
H(r) = \frac{a_{\rm th}}{\OmegaK} \ ,
\qquad r \to \rs \ .
\label{eqHappC}
\end{equation}
By combining equations~(\ref{eqOmegaK}), (\ref{eqEigenAsyRS1b}) and (\ref{eqHappC}), we can show that the explicit radial dependence of $H$ near the event horizon is given by
\begin{equation}
H(r) \propto (r-\rs)^{(\gamma_{\rm th}+3)/(2\gamma_{\rm th}+2)} \ , \qquad r \to \rs \ .
\label{eqEigenAsyRS2}
\end{equation}
Likewise, we can combine equations~(\ref{eqTranM}), (\ref{eqEigenAsyRS1a}), and (\ref{eqEigenAsyRS2}) to show that the asymptotic variation of the gas density $\rho$ is given by
\begin{equation}
\rho(r) \propto (r-\rs)^{-1/(\gamma_{\rm th}+1)} \ , \qquad r \to \rs \ .
\label{eqEigenAsyRS3}
\end{equation}
Equations~(\ref{eqEigenAsyRS2}) and (\ref{eqEigenAsyRS3}) are identical to the single-fluid results obtained by LB05, which is expected since in our model the gas pressure dominates over the relativistic particle pressure as $r \to \rs$.

\subsection{Asymptotic dynamical behaviour as $r \to \infty$}
\label{sec:AsymInf}

At very large radii, advection is negligible, and the particle transport in the disc is dominated by outward-bound diffusion. Furthermore, in the asymptotic upstream limit, $r\to\infty$, the thermal and relativistic particle sound speeds approach constant values, denoted by $a_{\rm th,\infty}$ and $a_{\rm rel,\infty}$, respectively. We also note that the gas entropy parameter, $K_{\rm th}$, is a global constant, except at the location of a discontinuous shock (see equation \ref{eqKg}). These facts imply that the asymptotic variation of the flow velocity $\vel$ at large radii is the same as that observed in the one-fluid model (LB05),
\begin{equation}
\vel \propto r^{-5/2} \ , \qquad r \to \infty \ ,
\label{eqEigenAsyInf2}
\end{equation}
resulting in the variation of the disc half-thickness (equation \ref{eqH})
\begin{equation}
H \propto r^{3/2} \ , \qquad r\to\infty \ ,
\label{eqEigenAsyInf3}
\end{equation}
as well as the density (equation \ref{eqTranM})
\begin{equation}
\rho \approx {\rm constant} \ , \qquad r \to \infty \ .
\label{eqEigenAsyInf4}
\end{equation}

\subsection{Asymptotic behaviour of the energy moments as $r \to \rs$}

Next we employ the relations developed in Secs.~\ref{sec:AsymHoriz} and \ref{sec:AsymInf} to derive suitable asymptotic boundary conditions for the energy moments $I_n$ appearing in equation~(\ref{eqVertIntTransIn}). First we substitute for $\kappa(r)$ using equation (\ref{eqKappa}) to obtain, after some algebra,
\begin{equation}
\frac{d^2 I_n}{dr^2} + \left[\frac{\rs}{\kappa_0(r-\rs)^2} + \frac{2}{r-\rs}
+ \frac{d\ln(rH\vel)}{dr}\right]\frac{dI_n}{dr} + \left(\frac{n+1}{3}\right)\frac{\rs}{\kappa_0
(r-\rs)^2}\frac{d\ln(rH\vel)}{dr} \, I_n = 0 \ ,
\label{eqEigenH1}
\end{equation}
where $\vel=-\vel_r$, and $I_2=n_{\rm rel}$, $I_3=U_{\rm rel}$ (see equations \ref{eqBB2} and \ref{eqBB3}).

The asymptotic relations given by equations~(\ref{eqEigenAsyRS1a}) and (\ref{eqEigenAsyRS2}) can be combined to show that
\begin{equation}
\frac{d\ln\left(rH\vel\right)}{dr} \approx \frac{1}{\gamma_{\rm th}+1} \, \frac{1}{r-\rs} \ ,
\qquad r \to \rs \ ,
\label{eqEigenH5}
\end{equation}
which can be used to reduce equation (\ref{eqEigenH1}) to the asymptotic form
\begin{equation}
\frac{d^2 I_n}{dr^2} + \left[\frac{\rs}{\kappa_0 (r-\rs)^2}
+ \left(2+\frac{1}{\gamma_{\rm th}+1}\right)\frac{1}{r-\rs}\right] \frac{dI_n}{dr}
+ \left(\frac{n+1}{3}\right)\frac{\rs}{\kappa_0(\gamma_{\rm th}+1)(r-\rs)^3} \, I_n \approx 0 \ ,
\qquad r \to \rs \ .
\label{eqEigenH1b}
\end{equation}
The asymptotic solution for the energy moment $I_n$ is found by applying the Frobenius method to equation (\ref{eqEigenH1b}), which results in two values for the power-law index $\alpha$, where
\begin{equation}
I_n(r) \propto (r-\rs)^{-\alpha} \ , \qquad r \to \rs \ .
\label{eqasympPower}
\end{equation}

We can focus on the physical root by balancing the dominant terms in equation (\ref{eqEigenH1b}), obtaining
\begin{equation}
\frac{\rs}{\kappa_0 (r-\rs)^2}\frac{dI_n}{dr} + \left(\frac{n+1}{3}\right)\frac{\rs}{\kappa_0
(\gamma_{\rm th} + 1)(r-\rs)^3} \, I_n \approx 0 \ , \qquad r \to \rs \ .
\label{eqEigenH1b1}
\end{equation}
The solution obtained for the power-law index is
\begin{equation}
\alpha = \frac{n+1}{3(\gamma_{\rm th}+1)} \ ,
\label{eqasympPower2}
\end{equation}
and therefore we find that 
\begin{equation}
I_n(r) \propto (r-\rs)^{-(n+1)/(3\gamma_{\rm th}+3)} \ , \qquad r \to \rs \ .
\label{eqAsympIn}
\end{equation}

The explicit asymptotic form for the particle number density ($n=2$) is given by
\begin{equation}
n_{\rm rel}(r) \propto (r-\rs)^{-1/(\gamma_{\rm th}+1)} \ , \qquad r \to \rs \ ,
\label{}
\end{equation}
and the corresponding result for the energy density $(n=3)$ is
\begin{equation}
U_{\rm rel}(r) \propto (r-\rs)^{-4/(3\gamma_{\rm th}+3)} \ , \qquad r \to \rs \ .
\label{}
\end{equation}
These results agree with the asymptotic relations derived by LB05 in the context of their one-fluid model, which is expected since in our model the thermal gas pressure dominates over the relativistic particle pressure near the event horizon (see equations~\ref{eqEigenAsyRS1b} and \ref{eqEigenAsyRS1c}).

\subsection{Asymptotic behaviour of the energy moments as $r \to \infty$}

Next we employ the relations developed in Secs.~\ref{sec:AsymHoriz} and \ref{sec:AsymInf} to derive suitable asymptotic boundary conditions for the energy moments $I_n$ appearing in equation~(\ref{eqVertIntTransIn}). First we substitute for $\kappa(r)$ using equation (\ref{eqKappa}) to obtain, after some algebra,
\begin{equation}
\frac{d^2 I_n}{dr^2} + \left[\frac{\rs}{\kappa_0(r-\rs)^2} + \frac{2}{r-\rs}
+ \frac{d\ln(rH\vel)}{dr}\right]\frac{dI_n}{dr} + \left(\frac{n+1}{3}\right)\frac{\rs}{\kappa_0
(r-\rs)^2}\frac{d\ln(rH\vel)}{dr}I_n = 0 \ ,
\label{eqEigenH2}
\end{equation}
where $\vel=-\vel_r$, and $I_2=n_{\rm rel}$, $I_3=U_{\rm rel}$ (see equations \ref{eqBB2} and \ref{eqBB3}).

The asymptotic relations given by equations~(\ref{eqEigenAsyInf2}) and (\ref{eqEigenAsyInf3}) can be combined to show that
\begin{equation}
\frac{d\ln(rH\vel)}{dr} \approx 0 \ , \qquad r \to \infty \ .
\label{eqEigenH36}
\end{equation}
Incorporating this result into equation~(\ref{eqEigenH1}) yields the new asymptotic form
\begin{equation}
\frac{d^2 I_n}{dr^2} + \left[\frac{\rs}{\kappa_0 r^2}
+ \frac{2}{r}\right]\frac{dI_n}{dr} \to 0 \ , \qquad r \to \infty \ .
\label{eqEigenH1c}
\end{equation}
The second term inside the square brackets dominates as $r \to \infty$, and therefore equation~(\ref{eqEigenH1c}) reduces to
\begin{equation}
\frac{d^2 I_n}{dr^2} = -\frac{2}{r}\frac{dI_n}{dr} \ , \qquad r \to \infty \ .
\label{eqEigenH1d}
\end{equation}
This equation can be immediately integrated to obtain the asymptotic solution for $I_n(r)$, given by
\begin{equation}
I_n(r) \approx I_{n,\infty} \left(\frac{C_1}{r} + 1\right) \ , \qquad r \to \infty \ ,
\label{eqEigenH30}
\end{equation}
where $C_1$ is an integration constant, and $I_{n,\infty}$ is the asymptotic value of $I_n$ at an infinite distance from the black hole. Equation~(\ref{eqEigenH30}) provides the spatial boundary conditions (equations \ref{eqHomoDifEq}) required in order to integrate equation~(\ref{eqBoundaryQ}) to determine the spatial variation of the energy moment function $Q$.

In our model, the spatial diffusion coefficient $\kappa$ described by equation~(\ref{eqKappa}) is independent of the energy of the relativistic particles, and therefore in the asymptotic domain $r \to \infty$, we expect that the relativistic particle number and energy densities, $n_{\rm rel}$ and $U_{\rm rel}$, respectively, will vary in proportion to each other. We can therefore express the explicit asymptotic forms for the relativistic particle number and energy densities by writing
\begin{equation}
n_{\rm rel} \approx n_{\rm rel,\infty} \left(\frac{C_1}{r} + 1\right) \ , \qquad r \to \infty \ ,
\label{eqnrAsympInf}
\end{equation}
and
\begin{equation}
U_{\rm rel} \approx U_{\rm rel,\infty} \left(\frac{C_1}{r} + 1\right) \ , \qquad r \to \infty \ ,
\label{eqUrAsympInf}
\end{equation}
where $n_{\rm rel,\infty}$ and $U_{\rm rel,\infty}$ denote values measured at infinity. The constant $C_1$ is determined by requiring that the numerical solution for $U_{\rm rel}$ (equation~\ref{eqUrAsympInf}) agrees with the dynamical solution for the relativistic particle energy density (equation~\ref{eqRelP1}) far from the black hole.

\section{SHOCK WIDTH AND PARTICLE ESCAPE}
\label{AppenShockWidth}

In the vertically-integrated model considered here, the rate at which particles escape through the surface of the disc at the shock location is quantified by the value of the dimensionless parameter $A_0$ introduced in equation~(\ref{eqVertIntTrans}). We can relate the value of $A_0$ to the physical parameters describing the structure of the accretion disc by focusing on the ``leaky pipe'' analogy utilized by LB05, where it was assumed that the shock width is exactly equal to the magnetic coherence length, $\lambda_{\rm mag}$. In the present application, the pressure of the accelerated relativistic particles is included, and we expect that this will lead to a broadening of the shock, as is seen in the case of cosmic-ray modified shocks (e.g., Axford et al. 1977; B11). We will therefore set the shock thickness using $\Delta x = \eta \lambda_{\rm mag}$, where $\eta$ is a dimensionless quantity. The case treated by LB05 corresponds to $\eta = 1$. In the case under consideration here, the relativistic particle pressure creates a precursor deceleration which increases the effective width of the shock. Observation of the velocity profiles plotted in Fig. \ref{fig:fig4} suggests that $\eta \sim 2-6$.

We also can estimate the fraction of particles that escape from the pipe over the shock thickness by using a simple model for the three-dimensional random walk executed by the relativistic particles as they diffuse through the plasma and escape from the upper and lower surfaces of the disc at the shock location. Based on equations (B1)-(B4) of LB05, we can write 
\begin{equation}
f_{\rm esc} = 1 - \frac{n_{\rm rel}}{n_0}\Bigg|_{x=\eta\lambda_{\rm mag}}
= \frac{\eta\lambda_{\rm mag}}{\vel_x t_{\rm esc}} \ ,
\label{eqB1}
\end{equation}
where $\vel_x$, $n_{\rm rel}$, $n_0$, and $t_{\rm esc}$, respectively, represent the flow velocity, the relativistic particle number density, the incident number density as the flow encounters the exit in the pipe (at $x=0$), and the average time for the particles to ``leak'' through the pipe via diffusion.

Working from the transport equation representation of the process, the fraction of particles that escape as the gas crosses the shock is given by (cf. equation B5 of LB05)
\begin{equation}
f_{\rm esc} = A_0\frac{c}{\vel_*} \ ,
\label{eqB2}
\end{equation} 
where the mean velocity at the shock is defined as $\vel_*\equiv\left(\vel_+ + \vel_-\right)/2$. By setting $\vel_x=\vel_*$, we can combine equations~(\ref{eqB1}) and (\ref{eqB2}) to show that
\begin{equation}
A_0 = \frac{\eta\lambda_{\rm mag}}{c t_{\rm esc}} \ .
\label{eqB3}
\end{equation}

In order to proceed, we need to derive an expression for the mean diffusive escape time, $t_{\rm esc}$. The diffusion velocity for particles experiencing a three-dimensional random walk between magnetic scattering centers in the vicinity of the shock is given by
\begin{equation}
\vel_{\rm diff} = \frac{c \lambda_{\rm mag}}{H_*/2} \ ,
\label{eqVdiff}
\end{equation}
where $H_*$ is the disc half-thickness at the shock location. The diffusive escape timescale can now be written as
\begin{equation}
t_{\rm esc} = \frac{H_*}{\vel_{\rm diff}} = \frac{H_*^2}{2 c \lambda_{\rm mag}} \ .
\label{eqVdiff2}
\end{equation}
Using equation~(\ref{eqVdiff2}) to substitute for $t_{\rm esc}$ in equation~(\ref{eqB3}) yields
\begin{equation}
A_0 = 2 \, \eta \left(\frac{\lambda_{\rm mag}}{H_*}\right)^2 \ .
\label{eqB4}
\end{equation}

We can derive an equation that can be used to evaluate the shock-width parameter $\eta$ by starting with the standard formula for the spatial diffusion coefficient (e.g., Reif 1965), which can be written as
\begin{equation}
\kappa_* = \frac{c\lambda_{\rm mag}}{3} \ ,
\label{eqReif}
\end{equation}
where $\kappa_* \equiv \left(\kappa_+ + \kappa_-\right)/2$ is the average of the upstream and downstream values of the diffusion coefficient (equation~\ref{eqKappa}) on either side of the shock. Using equation~(\ref{eqReif}) to substitute for $\lambda_{\rm mag}$ at the shock location in equation~(\ref{eqB4}) yields
\begin{equation}
\eta = \frac{A_0}{2} \left(\frac{c H_*}{3\kappa_*}\right)^2 \ .
\label{eqB5}
\end{equation}
We can use equation~(\ref{eqB5}) to compute $\eta$ for given values of the parameters $A_0$, $\kappa_*$, and $H_*$.

\bsp	
\label{lastpage}

\end{document}